\documentclass[twocolumn]{aa} 
\usepackage[varg]{txfonts}
\usepackage{listings}
\usepackage{amsmath,amssymb}
\usepackage[normalem]{ulem}
\usepackage{yhmath}
\usepackage{mathtools}
\usepackage{cancel}
\usepackage{mathrsfs}  
\newcommand{\be}{\boldsymbol{e}}
\newcommand{\br}{\boldsymbol{r}}

\newcommand{\buu}{\boldsymbol{u}}

\newcommand{\bnabla}{\boldsymbol{\nabla}}
\newcommand{\bg}{\boldsymbol{g}}

\newcommand{\dd}{{\rm d}}

\DeclareMathOperator{\Tr}{Tr}

\newcommand{\msun}{\ensuremath{M_\odot}}
\newcommand{\rsun}{\ensuremath{R_\odot}}
\newcommand{\menv}{\ensuremath{M_\text{env}}}

\newsavebox\myboxA
\newsavebox\myboxB
\newlength\mylenA
\usepackage{hyperref}
\hypersetup{
  unicode=true,      
  pdftoolbar=true,    
  pdfmenubar=true,    
  pdffitwindow=true,   
  pdftitle={Post-dynamical spiral-in phase of common envelope evolution},
  pdfauthor={Gagnier \& Pejcha},
  pdfkeywords={},     
  pdfnewwindow=true,   
  colorlinks=true,    
  linkcolor=blue,     
  citecolor=blue,     
  filecolor=gray,     
  urlcolor=blue      
}

\makeatletter
\renewcommand*\aa@pageof{, page \thepage{} of \pageref*{LastPage}}
\makeatother

\newcommand*\xoverline[2][0.75]{%
    \sbox{\myboxA}{$\m@th#2$}%
    \setbox\myboxB\null
    \ht\myboxB=\ht\myboxA%
    \dp\myboxB=\dp\myboxA%
    \wd\myboxB=#1\wd\myboxA
    \sbox\myboxB{$\m@th\overline{\copy\myboxB}$}
    \setlength\mylenA{\the\wd\myboxA}
    \addtolength\mylenA{-\the\wd\myboxB}%
    \ifdim\wd\myboxB<\wd\myboxA%
       \rlap{\hskip 0.5\mylenA\usebox\myboxB}{\usebox\myboxA}%
    \else
        \hskip -0.5\mylenA\rlap{\usebox\myboxA}{\hskip 0.5\mylenA\usebox\myboxB}%
    \fi}

\begin{document}

\title{Post-dynamical inspiral phase of common envelope evolution:}
\subtitle{Binary orbit evolution and angular momentum transport}
  \titlerunning{Post-dynamical inspiral of common envelope}
  \authorrunning{Gagnier \& Pejcha} 

\author{Damien Gagnier
\and Ond\v{r}ej Pejcha}

\institute{Institute of Theoretical Physics, Faculty of Mathematics and Physics, Charles University, V Hole\v{s}ovi\v{c}k\'{a}ch 2, Praha 8, 180 00, Czech Republic, \email{damien.gagnier@matfyz.cuni.cz}}

\date{Received 2023}

\abstract{
After the companion dynamically plunges through the primary's envelope, the two cores remain surrounded by a common envelope and the decrease of the orbital period $P_\text{orb}$ stalls. The subsequent evolution has never been systematically explored with multidimensional simulations. For this study, we performed 3D hydrodynamical simulations of an envelope evolving under the influence of a central binary star using an adaptively refined spherical grid. We followed the evolution over hundreds of orbits of the central binary to characterize the transport of angular momentum by advection, gravitational torques, turbulence, and viscosity. We find that local advective torques from the mean flow and Reynolds stresses associated with the turbulent flow dominate the angular momentum transport, which occurs outward in a disk-like structure about the orbital plane and inward along the polar axis. Turbulent transport is less efficient, but can locally significantly damp or enhance the net angular momentum radial transport and may even reverse its direction. Short-term variability in the envelope is remarkably similar to circumbinary disks, including the formation and destruction of lump-like overdensities, which enhance mass accretion and contribute to the outward transport of eccentricity generated in the vicinity of the binary. If the accretion onto the binary is allowed, the orbital decay timescale settles to a nearly constant value $\tau_\text{b} \sim 10^3$ to $10^4\,P_\text{orb}$, while preventing accretion leads to a slowly increasing $\tau_\text{b} \sim 10^5\,P_\text{orb}$ at the end of our simulations. Our results suggest that the post-dynamical orbital contraction and envelope ejection will slowly continue while the binary is surrounded by gas and that $\tau_\text{b}$ is often much shorter than the thermal timescale of the envelope.
}

\maketitle
\section{Introduction}

Common envelope evolution (hereafter CEE) of a binary star system occurs when one of the stars engulfs its companion, which then rapidly spirals in through the envelope \citep{Paczynski1976}. The drag experienced by the companion moving in the noncorotating envelope leads to energy and angular momentum deposition in the surrounding gas. One possible outcome of CEE is that the companion star is dissolved inside the primary and the two stars merge into one. Alternatively, the dynamical inspiral slows down and a quasi-steady spiral-in phase ensues. The reasons for the stalling of the inspiral are not completely understood, but the reduction of the drag does occur when the density decreases either due to envelope expansion or heating, or when the gas starts to locally corotate with the companion \citep{Roepke22}. Simulations show that this stalled inspiral phase lasts for at least hundreds of orbits and that the two cores remain surrounded by a shared envelope \citep[e.g.,][]{Ricker2012,Passy2012,Ohlmann2016,Ivanova2016}. It is believed that a self-regulating feedback loop of the weak local drag and its associated energy dissipation slowly brings the central binary together and eventually ejects the envelope on its thermal timescale leaving behind a post-CEE binary \citep[e.g.,][]{Ivanova2013,Clayton2017,Glanz2018}. 

Common envelope evolution is responsible for a wide variety of binary systems such as cataclysmic variables \citep[][]{Paczynski1976}, X-ray binaries \citep[e.g.,][]{Kalogera1998,Taam2010,Chen2020}, progenitors of Type Ia supernovae \citep[e.g.,][]{Iben1984,Belczynski2005,Ablimit2016}, or planetary nebulae nuclei \citep[e.g.,][]{DeMarco2009,Jones_Boffin2017}.
CEE might be responsible for a substantial fraction of gravitational wave progenitors \citep[e.g.,][]{Dominik2012,Klencki2021,Marchant2021}, but CEE is also expected to emit gravitational waves  on its own that are likely to be detected by space-based gravitational-wave detectors such as LISA \citep[][]{Baker2019} or TianQin \citep[][]{Huang2020} during the post-dynamical in-spiral CEE stage \citep[][]{Renzo2021}. Binaries that do not survive CEE and merge can be observed as luminous red novae  \citep[e.g.,][]{Soker2006,Ivanova2013Sci,Kochanek2014,Pejcha2016b,Blagorodnova2021}.

Despite its importance, CEE is far from being fully understood. Great efforts have been made to confront numerical simulations' outcomes to observational constraints over the last few decades. 
Three-dimensional hydrodynamical simulations have provided insight into the physical processes important in the dynamical inspiral \citep[e.g.,][]{Passy2012,Ohlmann2016,Macleod2018,Chamandy2020,Sand2020}, the CEE ejecta dynamics and thermodynamics \citep[e.g.,][]{Glanz2018,Iaconi2019,Iaconi2020}, or radiation hydrodynamics of the ejecta and the associated transients \citep[e.g.,][]{Pejcha2016b,Pejcha2016,Pejcha2017,Metzger2017,MacLeod2017_M31,Matsumoto2022}. However, many ab initio simulations fail to eject the common envelope during dynamical plunge-in when only orbital energy injection by the secondary star is considered, and the obtained post-dynamical inspiral orbital separations are often larger than that observed in post-CE systems \citep[e.g.,][]{GM2011,Iaconi_deMarco2019,Politano2021,Passy2012,Kruckow2021}. A more realistic equation of state that takes ionization states into account seems to facilitate mass ejection, but might not affect the final separation  \citep[e.g.,][]{Nandez2015,Reichardt2020,Lau2022a,Lau2022b}.  Because of the wide range of temporal and spatial scales that need to be resolved and the associated high numerical cost, 3D hydrodynamical simulations are often stopped soon after the end of the dynamical inspiral phase.  The consecutive slow contraction of the orbit on a thermal timescale necessitates reverting to 1D models that cannot capture the multidimensional features \citep[][]{Taam1978,Meyer1979,Fragos2019}.

To facilitate rapid prediction of outcomes, the binary configurations preceding and following CEE are often linked using energy conservation with one free parameter, $\alpha_\text{CEE}$ \citep[e.g.,][]{Webbink1984,Livio1988}. The value of $\alpha_\text{CEE}$ can be estimated from simulations or from various observed binary populations. Yet, it is not clear whether the $\alpha_\text{CEE}$ formalism can truly encompass the complicated physics of CEE. In particular, if thermal timescale processes such as predynamical nonconservative mass transfer or post-dynamical self-regulated inspiral are important, then adiabatic energy conservation is violated. Of course, it is often possible to select a value of $\alpha_\text{CEE}$ to explain an observed population even if some assumptions of the formalism are not satisfied. This displeasing situation has motivated the development of alternative formalisms based on the conservation of angular momentum \citep{Nelemans2000,DiStefano2022} or a two-step prescription combining energy and angular momentum \citep{Hirai2022}.

The post-dynamical self-regulating inspiral plays an important role in many of the unsolved aspects of CEE. Due to numerical difficulties in studying late stages of CEE when the envelope has expanded and the central binary has tightened, very little is known about this phase, especially the duration, mechanism of orbital contraction and angular momentum transfer, and whether the thermal-timescale self-regulation is actually established. Clearly, even if the central binary orbits in a locally corotating gas and the gravitational drag, which is the prevailing source of orbital tightening during dynamical in-spiral, becomes very weak \citep[e.g.,][]{Ostriker1999,Ricker2012,MacLeod2015,MacLeod2017,Reichardt2019,Chamandy2019b,De2020}, the corotation cannot be maintained over arbitrary distances. As a result, the complex interaction between the binary and the gravitationally perturbed shared envelope can take over and drive the orbital separation evolution on a shorter timescale than the thermal one. An additional issue is the possibility that mass and angular momentum can reaccrete onto the binary. 

The configuration of the post-dynamical inspiral resembles a very thick circumbinary disk (CBD), where a binary is embedded in a low density cavity surrounded by a disk with which it interacts by gravitational, advective, and viscous torques, mass accretion onto the central binary, and by binary eccentricity evolution \citep{Sandquist1998}. In the case of thin CBDs, such intricate interactions can lead to either orbital expansion or contraction and to excitation of the binary eccentricity  \citep[e.g.,][]{Artymowicz1994,Macfayden2008,Shi2012,Tang2017,Miranda2017,Munoz2019,Munoz2020,Duffell2020,DOrazio2021,Dittmann2021,Penzlin2022}. These results suggest an exciting possibility that the CEE post-dynamical evolution does not have to proceed as a simple monotonic contraction of a circular orbit, but there can be phases of orbital expansion or an eventual formation of an eccentric post-CEE binary. Our connection between post-dynamical CEE and CBDs is different from previous explorations of fallback CBDs around post-CEE binaries \citep{DeMarco2011,Soker2011}.

There are also important differences between CBDs and the post-dynamical phase of CEE. CEE might not result in the formation of a cavity around the central binary, instead, the central binary could virialize the gas in its vicinity, which would provide pressure support of the envelope and prevent accretion. Therefore, we can identify two extreme regimes of zero or maximum accretion onto the binary. Which of the two regimes of accretion occurs depends on the absence or presence of a ``pressure valve'' inside the orbit, which allows the material to accrete onto the binary \citep[][]{Chamandy2018}. An example of such a pressure valve could be jets \citep[e.g.,][]{Soker1994,Enrique2017,Shiber2019,Camara2019,Camara2022}. A realistic situation probably lies between these two extreme regimes of accretion.

In this paper, we aim to clarify the nature and dynamics of the post-dynamical inspiral of CEE by performing the first dedicated series of 3D hydrodynamical numerical simulations. To establish a well-controlled numerical experiment, we mimic the outcome of the dynamical inspiral phase by artificially injecting angular momentum in the primary envelope following the procedure of \cite{Morris2006}. To follow the evolution of the system over long timescales, we excise an inner sphere containing the binary, but study the gravitational influence of the orbiting binary on the surrounding envelope by prescribing time-changing gravitational potential. The inner boundary condition at the excised sphere allows us to control the accretion on the central binary. To analyze our results, we employ techniques and diagnostics inspired by those commonly used in the context of CBDs \citep[e.g.,][]{Shi2012,Miranda2017,Munoz2019,Penzlin2022}. 


This work follows the following structure: in Sect.~\ref{sec:model}, we introduce our physical model and describe the numerical setup used in our common envelope simulations. In Sect.~\ref{sec:results}, we present the results of our simulations. In particular, we measure the timescale of binary separation evolution resulting from the various torques acting on the system, when accretion is turned on or off. We measure the typical frequencies associated with the short-term variability of mass accretion onto the binary, and we compare them with that from CBDs. We then study the formation of overdensities, the excitation of eccentricity, and the convective stability of the envelope. Finally, we analyze the angular momentum transport within the envelope. In Sect.~\ref{sec:discussions}, we discuss implications of our findings for CBDs and CEE. In Sect.~\ref{sec:conclusions}, we summarize our results.  

\section{Physical model and numerical setup}\label{sec:model}

We construct our post-dynamical inspiral model in the inertial frame at rest with the center of mass of the binary. We do not follow the previous evolution of the inspiraling binary, instead, we mimic its outcome following a procedure similar to \cite{Morris2006,Morris2007,Morris2009} and \cite{Hirai2021}, where the envelope is spun-up until a satisfactory amount of total angular momentum is injected. This mimics angular momentum transfer from  the secondary's orbit into the envelope during the dynamical plunge-in (see Sect.~\ref{sec:spinup}).
We set the gravitational constant $G$, the total binary mass $M = M_1 + M_2$, the  primary's  initial radius $R$, and thus the angular velocity $\sqrt{GM/R^3}$ to unity. The orbital velocity is fixed to $\Omega_{\rm orb} = \sqrt{GM/a_\text{b}^3}$, where $a_\text{b} = r_1 + r_2$ is the fixed binary separation,  $M_1$ is the mass of the primary's core located at  $\{r_1,\theta_1,\varphi_1\}$, and $M_2$ is the mass of the secondary object (either a main-sequence star or a compact object) located at $\{r_2,\theta_2,\varphi_2\}$. Orbital period is $P_\text{orb} = 2\pi/\Omega_\text{orb}$. The two objects are not resolved and are considered as constant point masses. To simplify our model, we consider an equal mass binary ($q \equiv M_2/M_1 = 1$) on a fixed circular orbit. The mass of the envelope is $M_\text{env} = 2$ in our units. Because we are most concerned with the angular momentum transport within the common envelope in the two extreme regimes of mass and angular momentum accretion onto the binary rather than the specific details of the individual cores, we excise a central region  encompassing the binary, which has a radius $r_{\rm in} = 0.625~a_\text{b} = R/10$. This excised region represents the gas bubble virialized by the orbiting binary and the enforced conditions at its boundary determine whether the binary is accreting or not.

We compare our setup to several ab initio simulations of CEE in Table \ref{tab:comp}. The key quantity is the ratio of final separation to the initial radius of the primary, which we set in our model to $a_\text{b}/R = 0.16$. 
This comparison suggests that our choice of initial parameters to the binary and envelope broadly represents results of ab initio simulations across for a range of progenitor types.

\begin{table*}
\caption{Setup comparison to several ab initio simulations of CEE.} \label{tab:comp}
\begin{center}
\begin{tabular}{lcccccc}
\hline\hline
{Reference} &  {$M_1$} &  {$M_{\rm env}$}  &  {$M_2$} & {$q$}  &  {$R$} &   {$a_\text{b}/R$} \\
\hline
\citet{Passy2012} (SPH3) & $0.392$ & $0.488$ & $0.3$ & $0.77$ & $83$ & $0.14$ \\
\citet{Ohlmann2016} & $0.38$ & $1.6$ & $0.99$ & $2.6$ & $49$ & $0.09$ \\
\citet{Sand2020} & $0.545$ & $0.425$ & $0.485$ & $0.89$ & $173$ & $0.2$--$0.24$ \\
\citet{Lau2022a} & $3.84$ & $8.16$ & $3$ & $0.78$ & $619$ & $0.05$--$0.07$ \\
This work & ... & $2(M_1+M_2)$ & ... & $1$ & & $0.16$ \\

\hline
\end{tabular}
\tablefoot{$M_1$, $\menv$, and $M_2$ are expressed in $\msun$, $R$ is expressed in $\rsun$. The primary star is a red giant in \citet{Passy2012} and \citet{Ohlmann2016}, an AGB star in  \citet{Sand2020}, and a red supergiant in \citet{Lau2022a}.}
\end{center}
\end{table*}

In the rest of this Section, we describe the equations used for solving the problem (Sect.~\ref{sec:equations}), boundary conditions (Sect.~\ref{sec:boundary}), initial conditions (Sect.~\ref{sec:IC}), and initial deposition of angular momentum (Sect.~\ref{sec:spinup}). We then present our numerical setup for the averaging of the polar zones (Sect.~\ref{sec:polar}), mesh refinement (Sect.~\ref{sec:mesh}), and equatorial symmetry of the simulations (Sect.~\ref{sec:sym}).

\subsection{Equations of hydrodynamics}
\label{sec:equations}

We use Athena++ \citep[][]{Stone2020} to solve the equations of hydrodynamics
    \begin{align}\label{eq:eq}
        \frac{\partial \rho}{\partial t} + \bnabla\cdot \rho \buu &= 0\ , \\
        \frac{\partial \rho \buu}{\partial t} + \bnabla \cdot (\rho \buu \buu + P \boldsymbol{I} + \boldsymbol{T}) &= - \rho \bnabla \Phi\ , \\
        \frac{\partial E}{\partial t} + \bnabla \cdot \left[(E+P \boldsymbol{I})\buu + \boldsymbol{T} \cdot \buu \right] &= - \rho \bnabla \Phi \cdot \buu \ ,
    \end{align}
where $E = e +\rho u^2/2$, $e$ is the internal energy density, $P = (\Gamma - 1)e $, $\Gamma = 5/3$ is the adiabatic index, $\Phi(\br)$ is the gravitational potential of the binary,
\begin{equation}\label{eq:fullphi}
    \Phi(\br) = -\sum_{i=1}^2 \frac{GM_i}{|\br - \br_i|}\ ,
\end{equation}
and $T_{ij}$ is the symmetric viscous stress tensor,
\begin{equation}
    T_{ij} = -\rho \nu \left( \partial_i u_j + \partial_j u_i - \frac{2}{3} (\bnabla \cdot \buu) \delta_{ij} \right)\ ,
\end{equation}
which is nonzero when we prescribe a kinematic viscosity $\nu$. 

For runs with nonzero viscosity, we prescribe an isotropic effective kinematic viscosity of turbulent nature, $\nu(r,\theta,\varphi) = \frac{1}{3} v l$, where $v$ is the velocity of the turbulent eddies, and $l$ is their vertical mean free path or the mixing length. We further assume that the mixing length is proportional to the local pressure scale height, $l = \alpha_1 H_P$ \citep[e.g.,][]{Vitense1953,Zahn89}, and that the characteristic eddy velocity is a fraction of the local sound speed, $v = \alpha_2  c_s$. Following \citet{SS73}, we obtain $\nu(r,\theta,\varphi) = \alpha_\nu c_s H_P$, where $\alpha_\nu = \alpha_1 \alpha_2/3$ is a free parameter. Assuming a typical  effective kinematic viscosity of $\mathcal{O}(10^{15}~ \rm cm^2~s^{-1})$, we take $ \alpha_\nu = 10^{-3}$ in this work. The nature of such effective viscosity is unknown and has been debated lively in the context of astrophysical accretion flows.
We further impose zero kinematic viscosity radial gradient in ghost cells at both boundaries. In our viscous simulation, we use the Runge–Kutta–Legendre super-time-stepping algorithm in Athena++ \citep[][]{MEYER2014,Stone2020}, which integrates diffusive terms forward with hyperbolic timesteps. Although this algorithm dramatically reduces the timestep constraints, we still could not evolve our viscous runs for as long as we could when $\alpha_\nu =0$.

\subsection{Boundary conditions}
\label{sec:boundary}

\subsubsection{At the inner edge of the domain}

During the initial spin-up of the star (see Sect~\ref{sec:spinup}), we assume that the inner boundary supports the weight of the primary's envelope and we forbid the fluid to flow through it. 
To achieve that, we assume that $\rho$ is constant in ghost cells, which we initialize with the value of the density in the first interior cell $i$ and  which we assume to be in equilibrium with ghost cells. Considering a simple first-order finite volume integration algorithm with a number of ghost cells $N_g = 2$, the pressure and density boundary conditions in the inner ghost cell of index $j$ read 
    \begin{align}
    &\rho_{i-j} =  \rho_i\ , \\
    &P_{i-j} = P_{i-j+1} + \Delta r \hat{\rho} \left.\Bigg(\frac{\dd \langle \Phi \rangle_\theta }{ \dd r }- \frac{\hat{u}_\varphi^2}{r}\Bigg)\right\vert_{r=\hat{r}} \ , \label{eq:BC_in}
    \end{align}
where $\hat{f} $ is the cell-face value of the variable $f$ evaluated at $ \hat{r} = r_{\rm in} - (j-1)\Delta r $, $\Delta r = r_j - r_{j+1}$, and $\langle \Phi \rangle_\theta$ is the time and latitude average of the binary gravitational potential (see Sect.~\ref{sec:IC}). We deal with the horizontal velocity by applying zero radial gradient in the adjacent ghost cells and  we impose reflecting radial velocity to enforce that $u_r = 0$ at the inner boundary,
    \begin{align}
    u_{r,i-j}  &= - u_{r,i+j-1}\ , \\ 
    u_{\theta,i-j} &=  u_{\theta,i+j-1}\ , \\
    u_{\varphi,i-j} &=u_{\varphi,i+j-1} \ .
    \end{align}
We note that in some cases Eq.~(\ref{eq:BC_in}) gives negative $P$ in ghost zones. When that is the case, equilibrium cannot be enforced at the boundary, and we impose a zero pressure gradient instead.
 
Once the spin-up phase is terminated, we either allow or forbid accretion onto the central binary. When accretion is allowed, we open the inner boundary to angular momentum and mass flow by imposing zero radial gradient of $\rho$, $P$, $u_\theta$, and angular momentum in ghost cells, and a diode-type radial infall only condition, $u_{r,i-j}  =  \min{(u_{r,i}, - u_{r,i+j-1})}$. When accretion is forbidden, we impose purely reflecting boundary conditions.

\subsubsection{At the outer edge of the domain}\label{sec:BC_out}

We use diode-type boundary conditions at the outer edge of the domain and we impose zero density and pressure gradient in the outer ghost zones. However, this condition implies that the ambient medium of our initial model is out of equilibrium and there is an inflow near the outer boundary during the initial spin-up phase. In this region, $\rho$ and $P$ are initially very low (see Sect.~\ref{sec:IC}) and  thus there is negligible influx of mass. 

\subsection{Initial conditions and outer low-density medium}\label{sec:IC}

We assume that the gas in the envelope is initially in hydrostatic equilibrium and that it can be described by a polytropic equation of state, as is often done in stellar physics \citep[e.g.,][]{maeder09,jones_etal09,Gagnier2020}. Ignoring the gas self-gravity and considering purely radial initial profiles, the equations governing the envelope initial structure read
\begin{equation}\label{eq:HE}
    \frac{\dd P}{\dd r} = - \rho \frac{\dd \Phi}{\dd r}  \quad {\rm and} \quad P = K \rho^\Gamma  \ ,
\end{equation}
where $K$ is a constant related to the thermal conditions at the inner boundary. The Green's function for Eq.~(\ref{eq:fullphi}) satisfies \citep{Jackson1975}
\begin{equation}
   G(\br, \br_i) =  \frac{1}{|\br - \br_i|} = 4 \pi \sum_{\ell = 0}^\infty \sum_{m=-\ell}^{\ell} \frac{1}{2\ell +1} \frac{r_i^\ell}{r^{\ell +1}} (Y_\ell^m (\theta_i,\phi_i))^\ast Y_\ell^m (\theta,\phi),
\end{equation}
for $r \ge r_{\rm in} \ge \max(r_1,r_2)$, where $Y_\ell^m$ are the usual normalized scalar spherical harmonic functions of degree $\ell$ and order $m$. The parity properties of the spherical harmonic function and of the time and latitude average of the binary potential, that is $\langle \Phi(\br) \rangle_\theta = \langle \Phi(-\br) \rangle_\theta$, imply that
\begin{equation}
   \langle \Phi \rangle_\theta =  -4\pi \sum_{i=1}^2 \sum_{k = 0}^\infty \sum_{m=-2k}^{2k} \frac{G M_i}{4k +1} \frac{r_i^{2k}}{r^{2k +1}} \langle (Y_{2k}^m (\theta_i,\phi_i))^\ast Y_{2k}^m (\theta,\phi) \rangle_\theta \ ,
\end{equation}
where $\langle \cdot \rangle_\theta$ indicates a time and latitude average. The $\ell \leq 2$ latitude and time averaged binary potential finally reads
\begin{equation}\label{eq:phi}
    \langle \Phi \rangle_\theta = - \frac{GM}{r} \left[ 1 - \frac{a_\text{b}^2 q}{8(1+q)^2r^2} \right] \ .
\end{equation}
We use this latitude- and time-averaged potential in the momentum equation during spin-up. We replace the averaged potential by its time-dependent expression (Eq.~(\ref{eq:fullphi})) after the spin-up. Our choice of the initial potential facilitates the transition from the initial spin-up by preventing a large injection or removal of gravitational energy, which  could lead to a sudden envelope ejection or collapse. By combining Eqs.~(\ref{eq:HE}) and (\ref{eq:phi}) and by specifying the ratio between $\rho$ at  the stellar surface and at the inner boundary surface, $\kappa^n = \rho(R)/\rho(r_{\rm in})$, we obtain the initial density and pressure profiles
\begin{align}\label{eq:rho_sh}
    \frac{\rho(r)}{\rho(r_{\rm in})} &=  \left[ 1 + C \left(\frac{B}{3} \left(\frac{1}{r^3} - \frac{1}{r_{\rm in}^3} \right) - A \left(\frac{1}{r} - \frac{1}{r_{\rm in}} \right) \right) \right]^n,\\
    P(r) &= K \rho^\Gamma \ ,
    \end{align}    
where $n = 1/(\Gamma -1)$ is the polytropic index and
\begin{equation}
\begin{aligned}
  A &= GM \ , &    B &= \frac{3 A a^2 q}{8(1+q)^2} \ ,\\
  A' &= A \left( \frac{1}{R} - \frac{1}{r_{\rm in}} \right) \ ,    &   B' &= \frac{B}{3} \left ( \frac{1}{R^3} - \frac{1}{r_{\rm in}^3} \right) \ , \\
  C &=   \frac{\kappa  - 1}{B' - A'} \ , \ {\rm and}  &  K &= \frac{(1-\Gamma)(B' - A')}{\Gamma (\kappa -1) \rho(r_{\rm in})^{1/n}} \ .
\end{aligned}
\end{equation}
Density at the inner boundary $\rho(r_{\rm in})$ can be calculated from the prescribed total mass of the envelope. 

\begin{figure}[t]
\centering
      \includegraphics[width=0.49\textwidth]{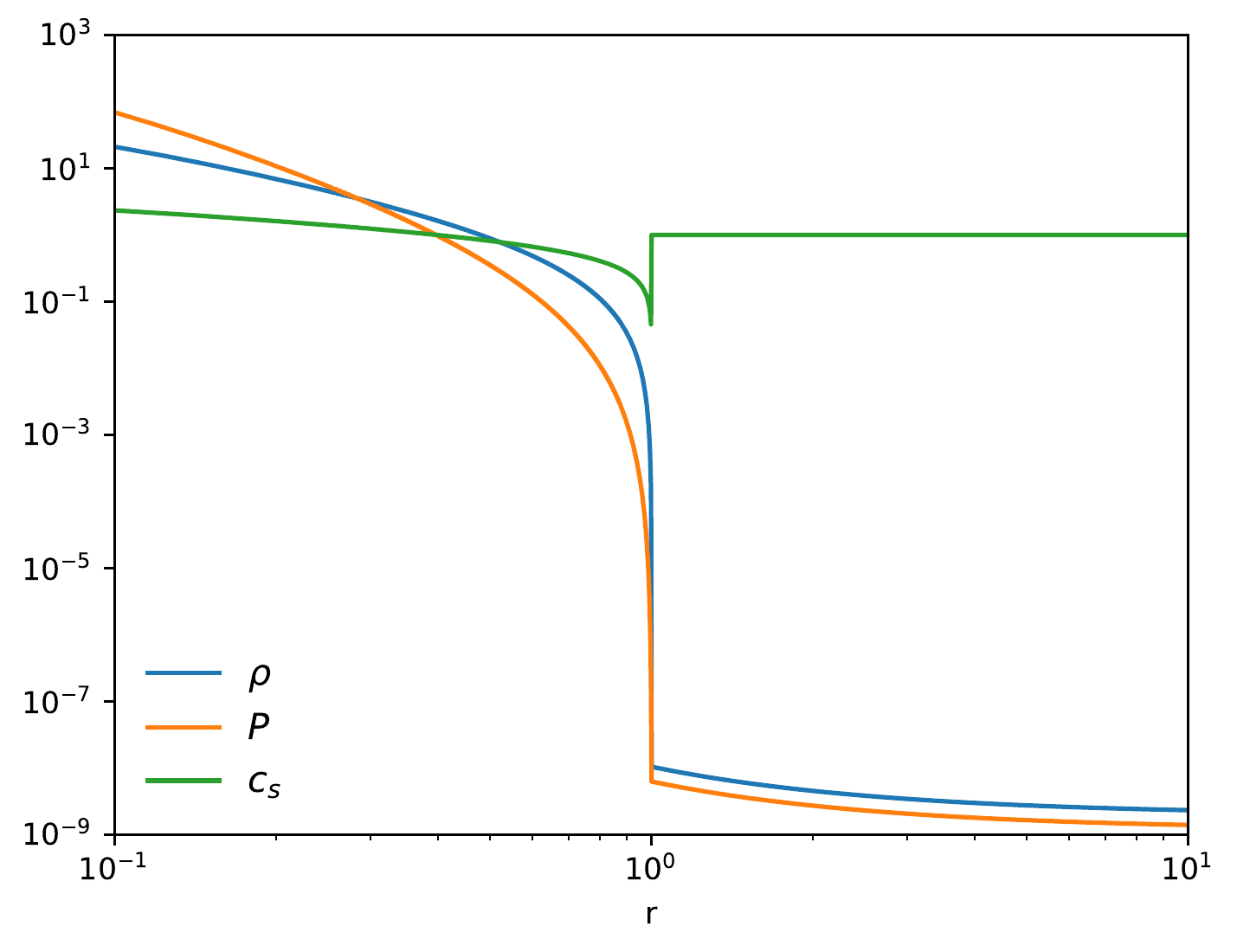}\\
   \caption{Initial density, pressure, and sound speed profiles used in all our simulation runs.}
\label{fig:ambient}
\end{figure}

The primary star is initially embedded in a low-density medium to which we apply our outer boundary conditions and in which the envelope will expand later on. To model this low-density medium, we consider an atmosphere in hydrostatic equilibrium with constant ambient sound speed $c_{s, \rm amb}$ \citep[see also][]{Macleod2018} and we obtain analytical $\rho$ and $P$ profiles assuming $P = \rho c_{s, \rm amb}^2/\Gamma$, which yields
 \begin{equation}
     \rho_{\rm ext} = C_1 \exp \left( \frac{\Gamma}{c_{s, \rm amb}^2}\left(\frac{A}{r} - \frac{B}{3r^3}\right) \right).
 \end{equation}
We note that it is not possible to transition from the envelope to the ambient region without a discontinuity, either in $\rho$, $P$, or both. To accommodate this discontinuity, we derive the constant $C_1$ such that the stellar surface is in hydrostatic equilibrium with the low-density atmosphere,  which gives
 \begin{equation}
     C_1 = \frac{  P^- - 0.5\Delta r(R) \rho^- \dd \langle \Phi \rangle_\theta/\dd r }{0.5 \Delta r (R) \dd \langle \Phi \rangle_\theta/\dd r + c_{s, \rm amb}^2 / \Gamma} \exp \left( -\frac{AC'}{r^+} + \frac{BC'}{3 r^{+ 3}}\right).
 \end{equation}
Here, $P^-$ and $\rho^-$ are the pressure and density in the last radial cell of the envelope from Eq.~(\ref{eq:rho_sh}), $r^+$ is the radius of the first cell of the ambient medium, $\Delta r(R)$ is the difference between $r^+$ and the radius of the last  radial cell of the envelope, and $C'= \Gamma/c_{s, \rm amb}^2$. 
We illustrate our initial conditions in Fig.~\ref{fig:ambient}. 

In order to minimize the effects of the nonexact numerical hydrostatic equilibrium resulting from the finite grid resolution, we use Gauss-Legendre quadrature to map the initial profiles onto the mesh as volume averaged variables at the volume averaged center of each cell, which is different from geometric center in polar-spherical coordinates, especially near the polar axis because of the converging grid geometry. 

\subsection{Initial spin-up}\label{sec:spinup}

We aim to construct a model with an initial total angular momentum that is consistent with what is available in the system, 
\begin{equation}
J_z = \frac{GM_2(M_1+M_{\rm env})( 1+\beta)}{M+M_{\rm env}}\sqrt{G(M + M_{\rm env})a_\text{i}}  - J_{z,\rm b} \ , 
\label{eq:angmom}
\end{equation}
where $a_\text{i}$ is the initial binary separation\footnote{We ignore potential mass and angular momentum loss from the outer Lagrange point ($L_2$) preceding common envelope \citep[e.g.,][]{Shu1979,Pejcha2014,Pejcha2016,Hubova2019}. Such additional angular momentum loss can be mimicked by lowering the value of $\beta$.}, $a_\text{b}$ is the enforced separation at the end of the dynamical plunge-in, $M_{\rm env} = 2$ is the total mass of the envelope, and $\beta$ is the ratio between the primary's envelope angular momentum and the orbital angular momentum before the plunge-in. $J_{z,\text{b}} =   \mu \sqrt{GM a_\text{b}}$ is  the orbital angular momentum of the binary at the end of the dynamical plunge-in, which coincides with the beginning of our simulations, and $\mu = M_1M_2/(M_1+M_2)$ is the reduced mass. We require  $\beta \leq 1/3$ to ensure Darwin stability \citep[e.g.,][]{Hut1980}. To impart angular momentum to the envelope, we use the procedure of \cite{Morris2006,Morris2007,Morris2009} and we apply a fixed spin-up rate to all cells in which the angular velocity  is  sub-Keplerian, $u_\varphi^2 < |\langle \Phi \rangle_\theta |$. Simultaneously, the structure of the envelope slowly restructures. 

We stop the spin-up once a satisfactory amount of total angular momentum is injected in the envelope. After a short adjustment phase, we replace the latitude and time averaged potential $\langle \Phi \rangle_\theta$ with its real expression (Eq.~(\ref{eq:fullphi})). As a result, there is a small bump of internal energy that is exclusively due to the increase of the gravitational energy density in the inner envelope. Though it has no physical origin beyond the sudden anisotropy of the gravitational potential and despite the fact that its amplitude cannot be easily constrained, it has the benefit of mimicking a small gravitational energy deposition by the spiral-in of the secondary star. We  discuss this more in Sect.~\ref{sec:Energy}.  

 \subsection{Polar averaging}
 \label{sec:polar}

It is well known that the use of spherical coordinates leads to strong time-step constraints resulting from the converging grid geometry and the Courant–Friedrichs–Lewy (CFL) condition. To mitigate this issue, we use a polar averaging technique based on the Ring Average technique of \citet{Zhang2019}, which is conservative and computationally inexpensive. This technique consists of a post-processing treatment of the variables in cell ``chunks'' adjacent to the polar axis, which is applied after the cells have been updated by the  Riemann solver. Hence, this technique does not involve the modification of the grid nor of the solver.  For nonuniformly spaced spherical coordinates, we compute the appropriate number of chunks  $N_c = 2^{k}$  per latitudinal ring of index $m$ in each mesh block, where 
\begin{equation}
 k =  \left[ \log_2 \left(\frac{r m \Delta \theta \Delta \varphi_{\rm block}}{\Delta r}\right)\right].
 \end{equation}
Here, square brackets indicate rounding to the nearest integer and $\Delta \varphi_{\rm block}$ is the azimuthal extent of the mesh block. Then, we average conserved variables in the azimuthal direction within each chunk of each ring and in each mesh block. We subsequently apply second-order spatial reconstruction procedure to the averaged values and we correct the minimum time-step within a mesh block to account for the coarsened effective mesh. 

\subsection{Mesh refinement}
\label{sec:mesh}

Our initial models are statically refined to properly resolve regions with strong initial gradients. These are regions close to the central binary and to the initial surface of the star. Specifically, our initial mesh is refined two levels above the base in the regions $r_{\rm in} \leq r \leq 0.25$ and  $0.95 \leq r \leq 1.05$. After the first timestep of the initial spin-up of the envelope, we switch from static to adaptive mesh refinement. We adopt a criterion based on the second derivative error norm of a function $\sigma$ of a variable \citep[]{Lohner1987}. This criterion measures the smoothness  of  the  solution for a given refinement variable. Similarly to the PLUTO code \citep{PLUTO2012}, a mesh block is tagged for refinement whenever 
\begin{equation}\label{eq:Lohner}
 \chi^2 =    \frac{\sum_d |\Delta_{d,+1/2} \sigma  - \Delta_{d,-1/2} \sigma|^2                }{\sum_d \left(|\Delta_{d,+1/2} \sigma |  + |\Delta_{d,-1/2}| + \epsilon \sigma_{d,\rm ref}  \right)^2} \ge \chi_r^2\ .
\end{equation}
Here, $\Delta_{r,\pm 1/2} = \pm (\sigma_{i \pm 1} - \sigma_i) $ and $\sigma_{r,\rm ref} =  |\sigma_{i+1}| + 2 |\sigma_i| + |\sigma_{i-1}|$. The value of the threshold $\chi_r$ is problem dependent and also depends on the chosen refinement variable $\sigma$. Finally, $\epsilon$ acts as a filter preventing refinement in regions of small ripples. We find that for our simulations, $\sigma = \rho |\buu|$ tends to capture the flow contrasts the best with $\chi_r^2 = 0.2$ and $\epsilon = 0.01$.
We note that the criterion in Eq.~(\ref{eq:Lohner}) does not include cross derivatives, unlike the original work of \citet{Lohner1987}. We find little difference when those terms are included and we opt not to include them.

\subsection{Equatorial symmetry}\label{sec:sym}

Since the orbiting cores are aligned in the equatorial plane at all times, our setup should be exactly symmetric about the equator. In practice, such symmetry can be difficult to enforce despite Athena++'s integration method being well-suited to preserving it \cite[][]{Stone2020}. For instance a mesh symmetric about the double precision rounding accuracy of $\pi/2$ in the $\theta$-direction yields asymmetric volume averaged cell colatitudes \citep[][]{Mignone2014}. For example, 
\begin{equation}
    x_{2,v} = \frac{\int_{\rm cell} \theta \dd V}{\int_{\rm cell} \dd V} = \frac{\dd (\sin \theta - \theta \cos \theta)}{\dd(-\cos \theta)}\ ,
\end{equation}
is asymmetric because of the asymmetry of the trigonometric functions about the rounded value of $\pi/2$. This introduces asymmetry in the theoretically symmetric source terms in our problem. Volume-averaged colatitudes are thus computed about the double precision rounding of $\pi/2$, hereafter noted $\widetilde{\pi/2}$. We obtain
\begin{equation*}
    x_{2,v} =  \widetilde{\pi/2}  + \frac{\int_{\rm cell} \widetilde{\phi} \dd V}{\int_{\rm cell} \dd V}  =  \widetilde{\pi/2} - \frac{\dd (-\cos \widetilde{\phi} - \widetilde{\phi}\sin \widetilde{\phi})}{\dd(-\sin \widetilde{\phi})}\ ,
\end{equation*}
where  $\widetilde{\phi}$ is the latitude measured from $\widetilde{\pi/2}$. Although this change considerably improves symmetry, face-centered cell colatitudes are, in practice, not symmetric in the last place precision. Such tiny asymmetry of $d \theta$ leads to the asymmetry of the physical and geometric sources terms and to a residual asymmetric flow that may amplify when it is linearly unstable. Furthermore, additional sources of asymmetries may amplify the problem, such as compiler value-unsafe optimizations of floating-point operations or the nonassociativity of floating-point arithmetic.
In order to control such inevitable perturbations, we impose ad hoc initial random weak seed perturbation to the initial density profile with maximum amplitude $10^{-6}~\rho(r,\theta,\varphi)$, which is orders of magnitude larger than the amplitude of perturbations resulting from grid asymmetries.  

\section{Results}\label{sec:results}

\begin{figure*}[t]
\centering
      \includegraphics[width=\textwidth]{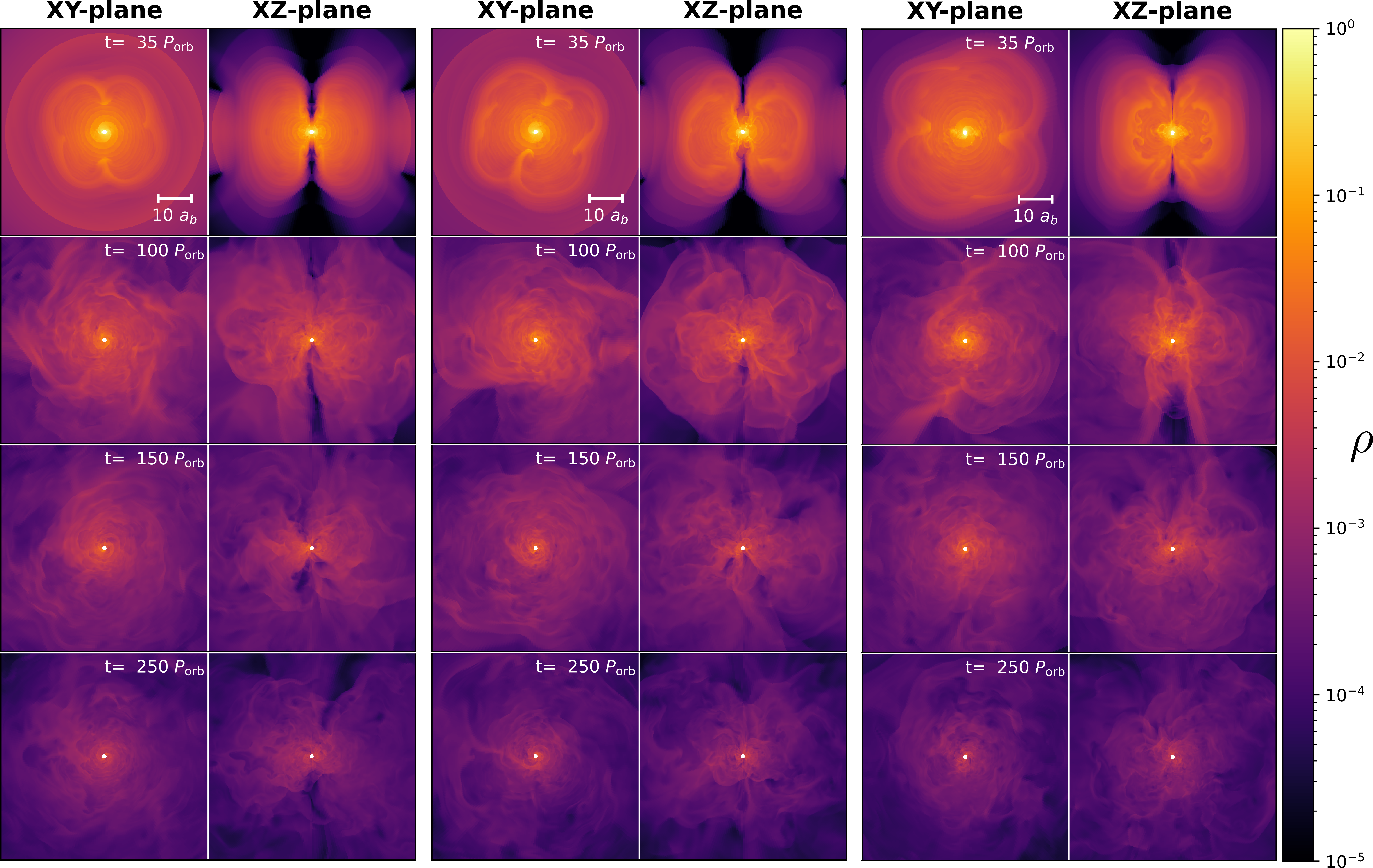} \hfill
   \caption{Zoomed-in snapshots of density cross section in the $xy$ and $xz$ planes at different times and for our three inviscid simulations runs A (left), B (middle) and C (right). The snapshots on the first line are taken shortly after the end of the initial spin-up phase.}
\label{fig:snap}
\end{figure*}

\begin{table*}
\caption{Run parameters and simulations outcome. \label{tab:runs}}
\begin{center}
\begin{tabular}{lccccccc}
\hline\hline
{Run} &  {$\alpha_\nu$} &  {$\beta$}  & {Accretion} &  {$\Lambda_{\rm I}$\textsuperscript{140--250}} &   {$\Lambda_{\rm II}$\textsuperscript{140--250}} &   {$\Lambda_{\rm III}$\textsuperscript{140--250}} \\
\hline
A &  0 & 0.3  & yes &0.608 & 1.501 & 3.880\\
A' &  0 & 0.3  & no\\
B & 0 & 0.1 & yes & 0.602 & 1.893 & 5.507 \\
C & 0 & -0.3 & yes & 0.877 & 3.000 & 5.331 \\
D & $10^{-3}$ & 0.3 & yes   \\
\hline
\end{tabular}
\tablefoot{$\Lambda_{\rm I,II,III}$ are the normalized autocorrelations of the azimuthally averaged turbulent latitudinal velocity on the orbital plane integrated over an arbitrary radial domain, expressed in units $a_\text{b}$, and interpreted as the associated typical convective eddy scale. The superscript 140--250 indicates a time average on the interval $140 \le t/P_{\rm orb} \le 250$.  More details are provided in Sect.~\ref{sec:vertical_eddy_scales}.}
\end{center}
\end{table*}

We used a total of 4.6 million CPU hours on the Karolina cluster at IT4Innovations to perform our simulation runs. In Table~\ref{tab:runs}, we summarize the parameters of the runs.  In Fig.~\ref{fig:snap}, we present zoomed-in snapshots of density cross section in the $xy$ and $xz$ planes at different times and for three inviscid simulation runs. The inviscid runs A, B, and C only differ by the initial size of the envelope's angular momentum reservoir. Run A is computed with $\beta = 0.3$, that is close to the limit of Darwin instability, run B is computed with $\beta = 0.1$, and run C  with $\beta = -0.3$. Negative value of $\beta$ implies that the total $z$ component of angular momentum is smaller than the initial orbital angular momentum of the binary orbit. Although our setup only approximates the process of angular momentum transfer from the orbit to the primary's envelope during the dynamical plunge-in, the density structure and flow morphology early in our simulations have striking resemblance with late-time snapshots from ab initio simulations of dynamical plunge-in \citep[e.g.,][]{Ohlmann2016, Chamandy2020}. 

Our simulations show that overall the envelope is destabilized by the central binary gravitationally torquing the inner envelope, exciting spiral density waves, and shearing the fluid flow. Energy from such flow is transferred to large-scale turbulence, and angular momentum is then transported by mean and turbulent flows. In the rest of this Section, we investigate these processes in detail. We address the initial jump in energy (Sect.~\ref{sec:Energy}), binary evolution and mass accretion (Sect.~\ref{sec:bin_evol}), short timescale dependence of accretion (Sect.~\ref{sec:tdepacc}), presence and origin of the lump (Sect.~\ref{sec:lump}), eccentricity of the envelope (Sect.~\ref{sec:ecc}), and convective stability and angular momentum transport (Sect.~\ref{sec:AMtrans}).

\subsection{Energy injection}\label{sec:Energy}

\begin{figure}[t]
\centering
   \includegraphics[width=0.49\textwidth]{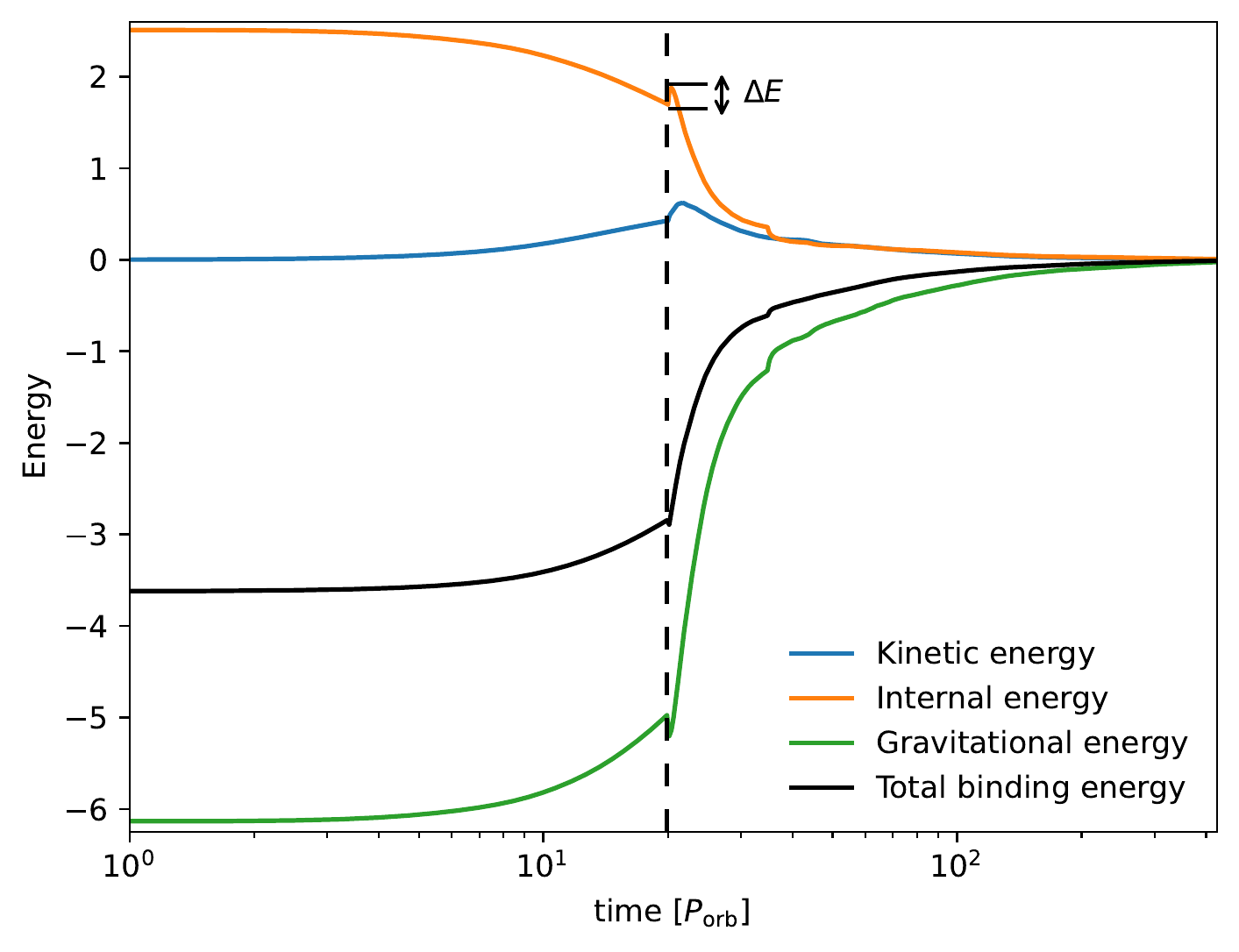} \\
   \caption{Kinetic, internal, gravitational, and total binding energy evolution in units $GM^2/R$ for model A. The vertical black dashed line indicates the replacement of the averaged binary potential with its full expression (Eq.~(\ref{eq:fullphi})), and the bump in energy is indicated with the double arrow.}
\label{fig:EnergyB}
\end{figure}

In Fig.~\ref{fig:EnergyB}, we show the kinetic, internal, gravitational, and total binding energy evolution for model A. We first discuss the bump in energy, which occurs at the end of the initial spin-up when we replace the latitude and time averaged binary potential with its real expression. 

To asses the importance of the bump, we estimate the CEE efficiency parameter $\alpha_\text{CEE}$ corresponding to injection  of internal energy $\Delta E$, 
\begin{equation}
    \alpha_\text{CEE} = \frac{2 \Delta E}{GM_2 \left( \frac{M_1}{a_\text{b}} - \frac{M_1 + M_{\rm env}}{a_\text{i}}\right)} \simeq 0.46 \ ,
\end{equation}
where $a_\text{i}$ is the initial binary orbital separation before plunge-in that we assume to be equal to $10~a_\text{b}$ \citep[e.g.,][]{Passy2012,Ohlmann2016,Chamandy2020}. That is, the amplitude of the internal energy bump corresponds to a gravitational energy deposition during plunge-in of 46\% of the difference between initial and final total orbital energy. Furthermore, because the difference between averaged and real gravitational potentials is only significant in the vicinity of the central binary, the energy is almost exclusively injected in the inner part of the envelope, which agrees with numerical simulations of \cite{Chamandy2019}. Because both amplitude and location of the energy injection are consistent with orbital energy deposition during CEE, we do not add more.

\subsection{Binary evolution and mass accretion}\label{sec:bin_evol}

Here, we address the evolution of the orbit of the central binary. So far, CEE theory has assumed that the binary separation decreases almost monotonically in time. However, recent studies of CBDs \citep[e.g.,][]{Munoz2019,Penzlin2022} suggest that for equal mass binaries there is a wide range of viscosity and disk thickness that leads to the expansion of the orbit. Therefore, finding out what actually happens to the binary separation in post-dynamical CEE inspiral is of fundamental importance. In our setup, we keep the orbital parameters fixed, but we can measure how much angular momentum was exchanged between the binary and the envelope and therefore assess what would happen to the binary if it was self-consistently coupled to the envelope. The advantage of our setup is that by excising the inner region, we can run the simulations for more orbits of the central binary.

\subsubsection{Torques and angular momentum conservation}\label{sec:torques}

  \begin{figure}[t]
\centering
      \includegraphics[width=0.49\textwidth]{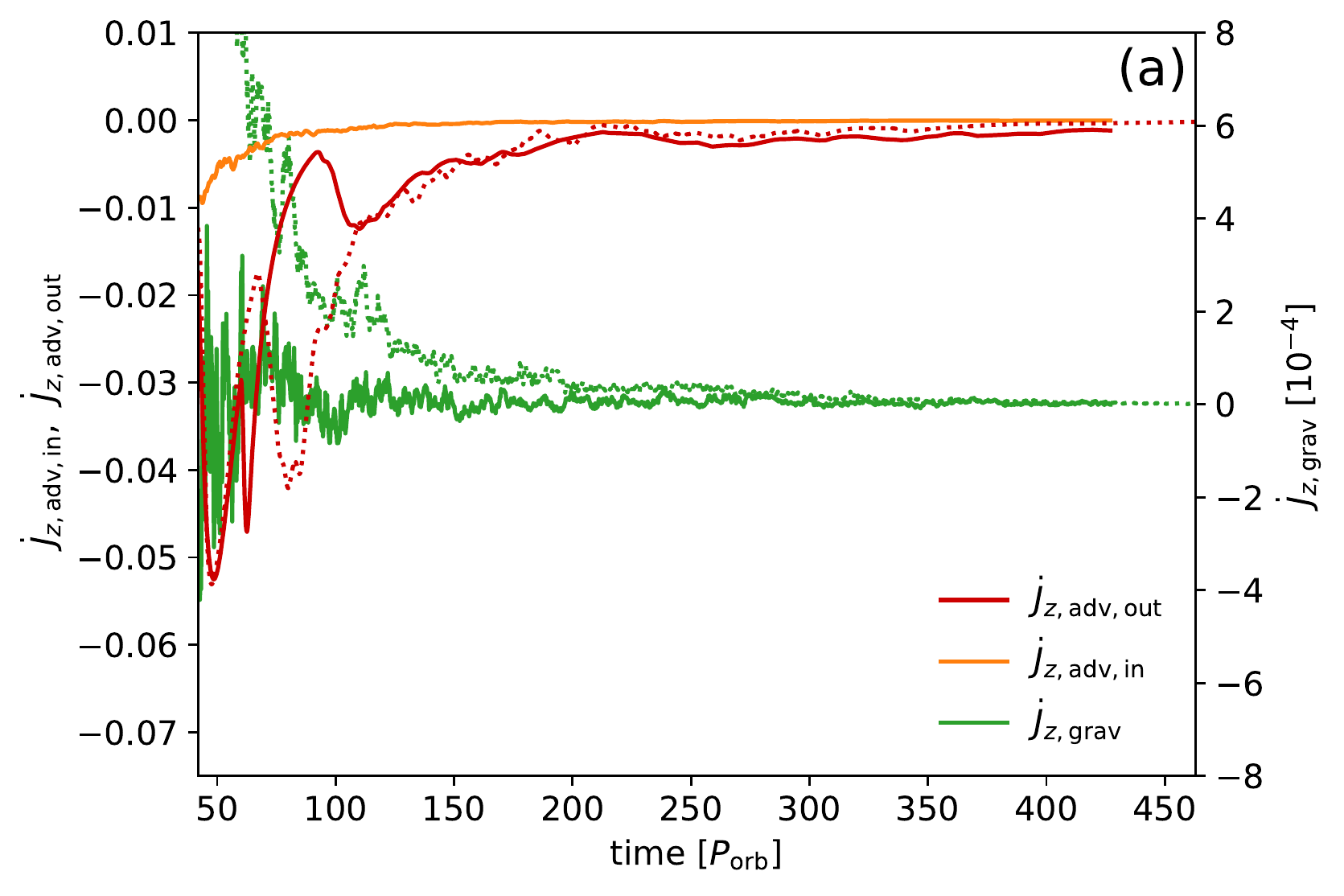} \\
    \includegraphics[width=0.49\textwidth]{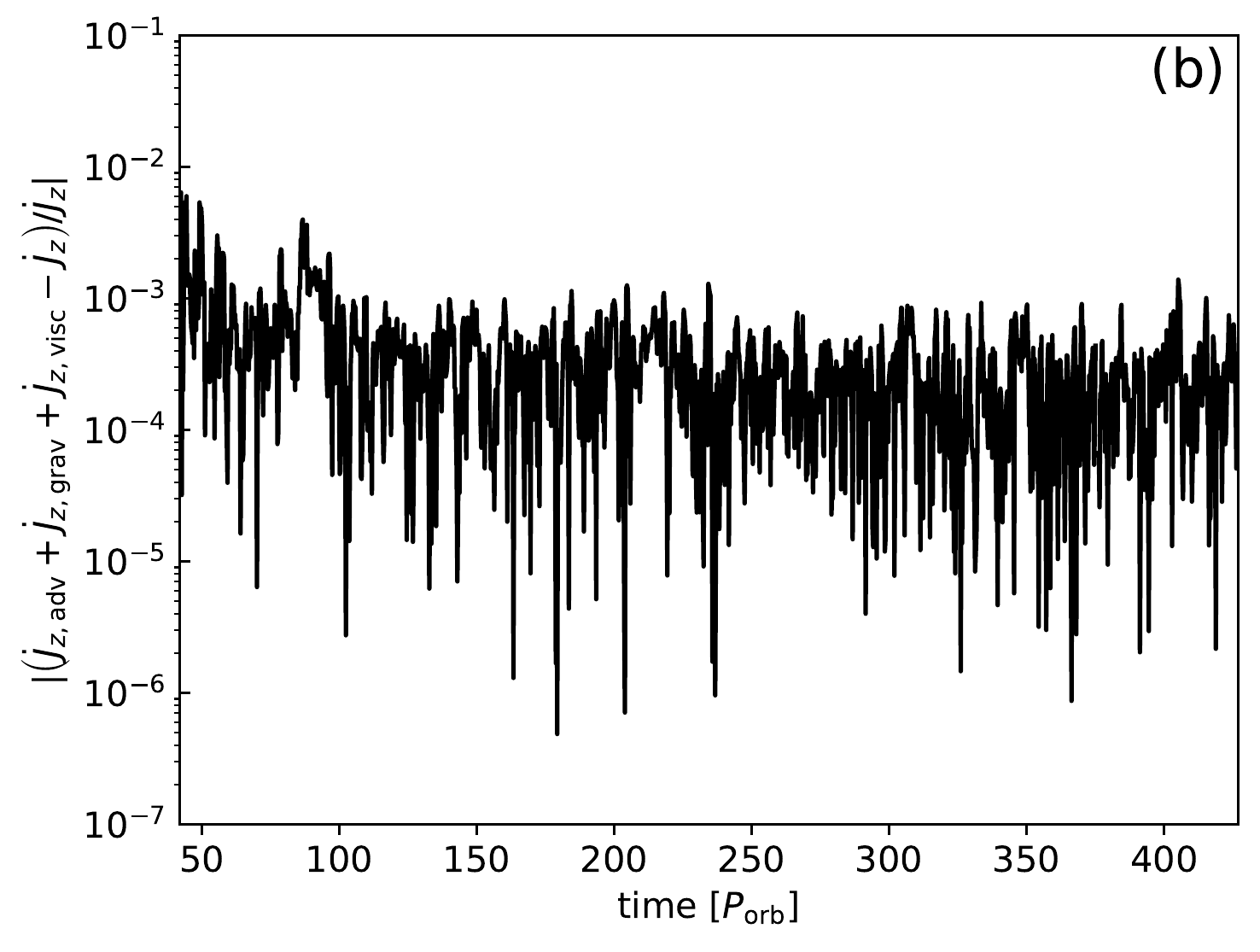} 
   \caption{Angular momentum evolution and conservation in our simulations. Panel (a): Time evolution of the advective, viscous, and gravitational torques for runs A  and, A' (dotted lines). Panel (b):  Relative difference between the sum of the torques and the measured time derivative of the total angular momentum showing the angular momentum conservation for run A.}
\label{fig:AMconsA}
\end{figure}

In order to predict the secular evolution of the binary separation, it is necessary to evaluate the torques in the common envelope. Such torques originate from the quadrupolar component of the gravitational potential as well as the advective (and perhaps viscous) angular momentum fluxes through the domain boundaries. The angular momentum conservation equation reads
\begin{equation}
    \dot{J}_z = \dot{J}_{z, \rm adv} + \dot{J}_{z, \rm grav} + \dot{J}_{z, \rm visc} \ ,
\end{equation}
where $\dot{J}_{z, \rm adv}$ is the advective torque associated with the loss of angular momentum through the boundaries,
\begin{equation}
    \dot{J}_{z, \rm adv} = - \int_{\partial R} \rho s u_\varphi\buu \cdot \boldsymbol{n}_\perp \dd S \ ,
\end{equation}
$\dot{J}_{z, \rm grav}$ is the gravitational torque exerted by the binary,
\begin{equation}
    \dot{J}_{z, \rm grav} =  -\int \rho \frac{\partial \Phi}{\partial \varphi} \dd V \ ,
\end{equation}
and $\dot{J}_{z, \rm visc}$ is the viscous torque,
\begin{equation}
    \dot{J}_{z, \rm visc} =  - \int_{\partial R} \left[\left(   \boldsymbol{\br} \times \boldsymbol{T} \right) \cdot \be_z \right] \cdot \boldsymbol{n}_\perp \dd S.
\end{equation}
Here, $\boldsymbol{n}_\perp$ is the outward-pointing unit vector at the boundaries' surface and $s = r \sin \theta$ is the radial cylindrical coordinate. We give more details in Appendix~\ref{App:dotJz}.  

In Fig.~\ref{fig:AMconsA}, panel (a), we show the evolution of these torques for runs A and A'. We also perform consistency check for angular momentum conservation by comparing the time evolution of the individual torques with the evolution of the total angular momentum budget for all of our models after the initial spin-up, and we show the result for model A in Fig.~\ref{fig:AMconsA}, panel (b). We find that the angular momentum is conserved to within about $0.1$--$1\%$  margin for all of our models. We also see that for all our models the total angular momentum evolution is dominated by the outflow at the outer boundary, which results from the expansion of the envelope and the finite radial extent of our numerical domain. When the inner boundary is open to angular momentum and mass flow toward the binary, angular momentum accretion dominates over the gravitational torque, which only weakly contributes to the injection of the angular momentum in the envelope. Because we choose zero radial gradient of angular momentum and viscosity at the inner boundary, the contribution of viscous torque remains weak even for eccentric flows in the binary close vicinity or for larger values of $\alpha_\nu$. After 140 orbital periods, we consider the various torques to be sufficiently time-steady so that we can qualitatively assume that all initial transients have decayed and that the flow properties have reached a quasi-steady state. 

\subsubsection{Binary orbital evolution}

In this work, we set the orbital eccentricity $e_\text{b}$ to zero and impose the binary mass ratio $q = 1$. We thus assume that mass and angular momentum accretion through the inner boundary distribute equally between the two cores, $\dot{q} = 0$. Furthermore, we assume that accretion does not excite orbital eccentricity, as suggested by CBD simulations, and we therefore fix $\dot{e}_\text{b} = 0$ \citep{Munoz2019,Heath2020,Penzlin2022}. The validity of this assumption will be discussed in Sect.~\ref{sec:ecc}. The time derivative of the binary's angular momentum can be written as the orbital separation evolution equation 
 \begin{equation}\label{eq:adota}
 \frac{\dot{a_\text{b}}}{a_\text{b}} = \frac{\dot{M}}{M} \left( 2 \frac{M \dot{J}_{z,\text{b}}}{\dot{M}J_{z,\text{b}}} - 3\right) \ .
 \end{equation}
If the central binary does not accrete from the shared envelope, Eq. (\ref{eq:adota}) simplifies to 
\begin{equation}
     \frac{\dot{a_\text{b}}}{a_\text{b}} = -2 \frac{\dot{J}_{z,\rm grav}}{J_{z,\text{b}}} \ ,
\end{equation}
and the binary orbit contracts ($\dot{a}/a < 0$) if $\dot{J}_{z,\rm grav} > 0$, that is if the gravitational torque transfers angular momentum from the binary orbit to the envelope. 
If the inner boundary is open to mass and angular momentum flow onto the binary, it is useful to consider the specific angular momentum transfer rate
\begin{equation}\label{eq:jjcrit}
    j \equiv \frac{\dot{J}_{z,b}}{\sqrt{GMa_\text{b}}\dot{M}} \ ,
\end{equation}
which yields a critical value $j_{\rm crit} = 3/8$ for $q=1$ \citep[e.g.,][]{Miranda2017,Moody2019,Dittmann2021,Penzlin2022}. Above this value, the binary orbit expands ($\dot{a}/a > 0$)  and below it contracts ($\dot{a}/a < 0$). In Eq.~(\ref{eq:jjcrit}), $\dot{M}$ is the measured mass accretion through the inner boundary
\begin{equation}\label{eq:movingj}
    \dot{M} = -\int_{\partial R_{\rm in}} \rho u_r \dd S \ ,
\end{equation}
and $\dot{J}_{z,b} = -\left.\dot{J}_{z, \rm adv}\right\vert_{r = r_{\rm in}} - \dot{J}_{z, \rm grav} - \left.\dot{J}_{z, \rm visc}\right\vert_{r = r_{\rm in}}$. 

\begin{figure}[t]
\centering
      \includegraphics[width=0.49\textwidth]{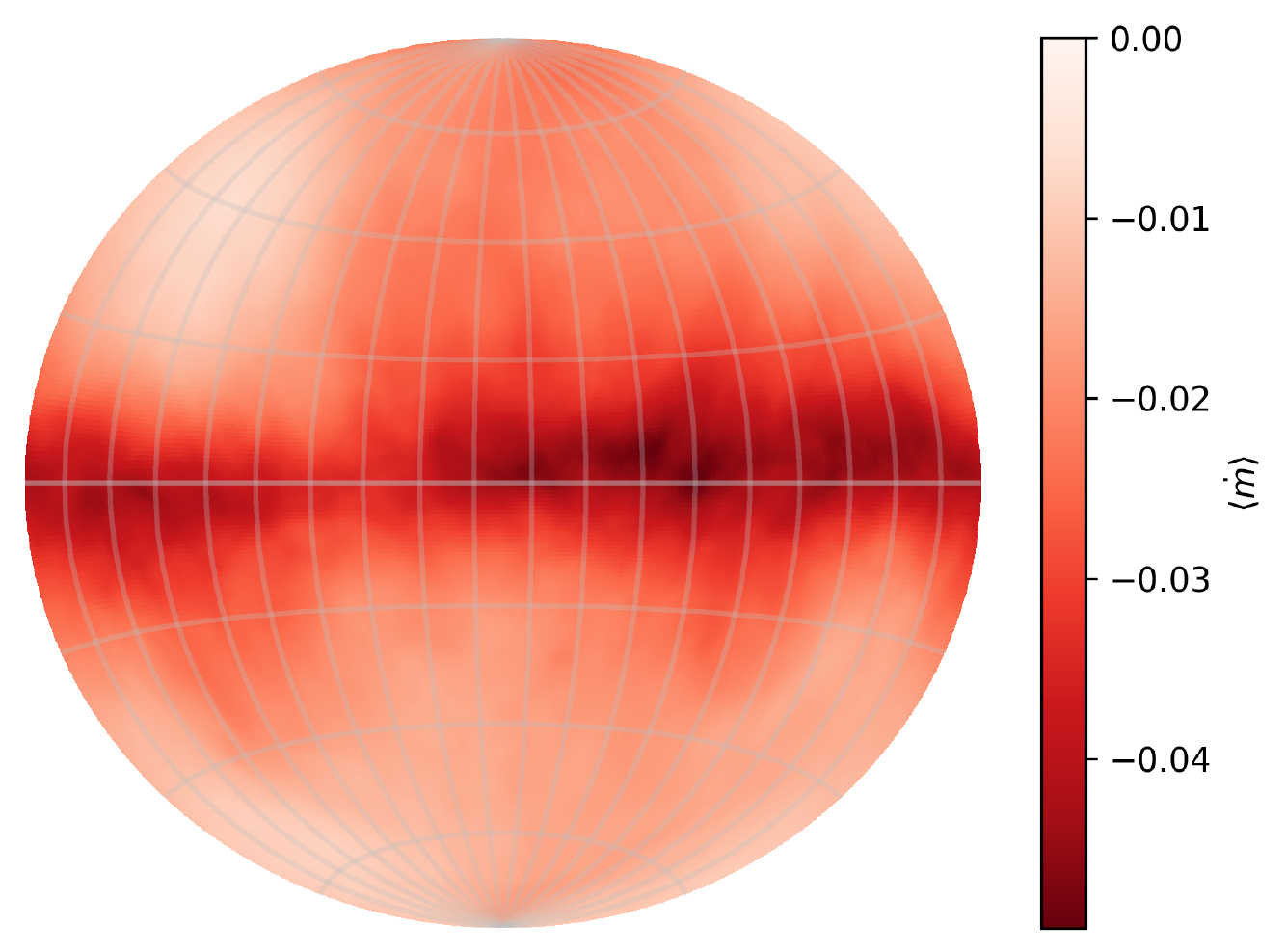} \\ 
      \includegraphics[width=0.49\textwidth]{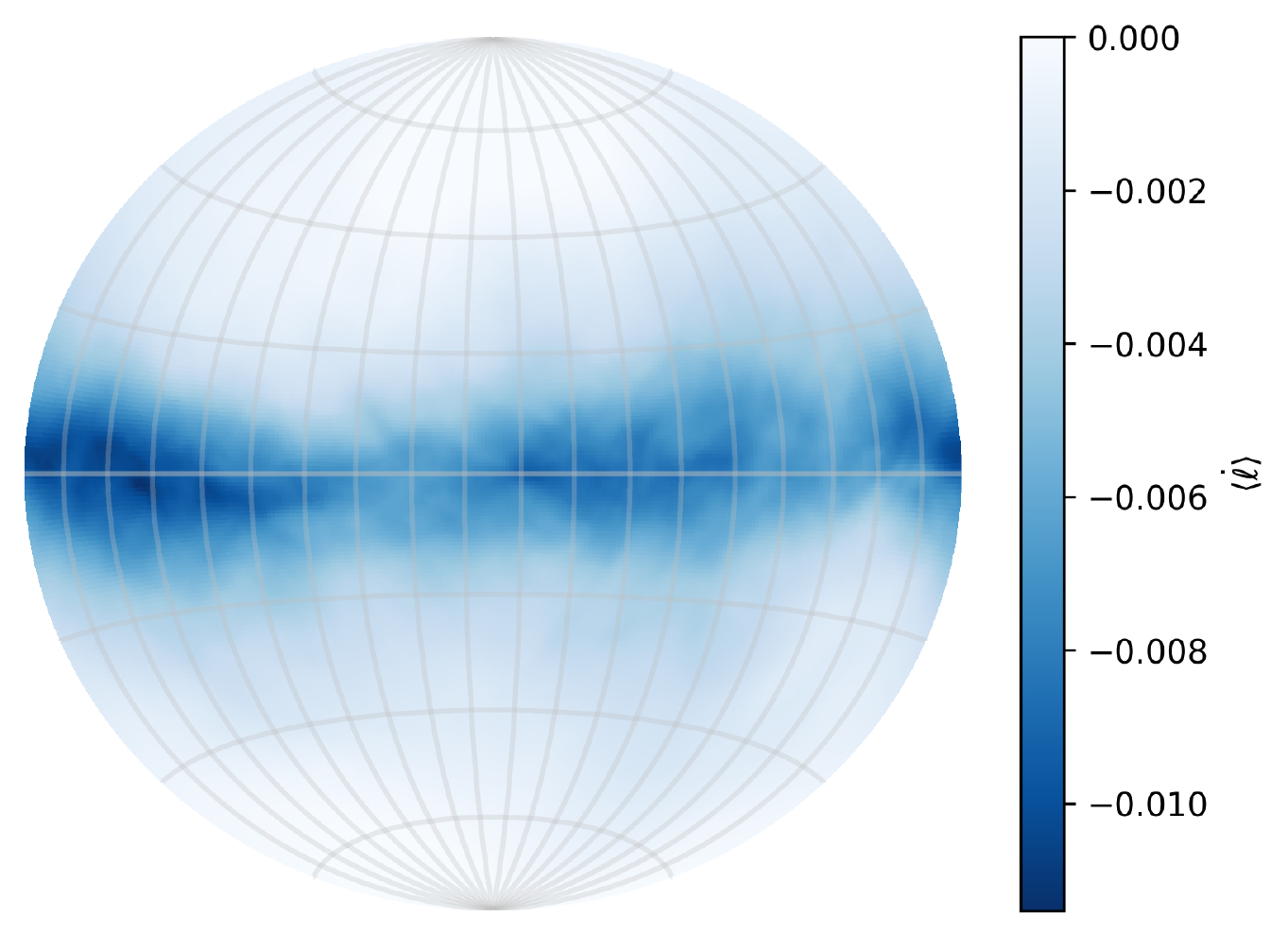} 
   \caption{Time average of the angular distribution of mass (panel a) and angular momentum (panel b) accretion fluxes through the inner boundary for ten orbital periods of model A at $148 \le t/P_{\rm orb} \le 158$.}
\label{fig:acc}
\end{figure}

Contrary to 3D simulations of CBDs where  mass  and angular momentum accretion only occur within a limited angle about the orbital plane dictated by the geometrical thickness of the disk, mass accretion could span the whole solid angle in our simulations. Hence, we may observe accretion along the polar axis with very small $j$, which could favor the contraction of the orbit according to Eq.~(\ref{eq:jjcrit}). To diagnose this issue, we show in Fig.~\ref{fig:acc} the time average over $10P_\text{orb}$ of the angular distribution of mass and angular momentum accretion fluxes through the inner boundary for model B. We see that mass and angular momentum accretion mostly occur within an annular ring centered on the orbital plane. Above and below such annular ring, mass accretion is accompanied with weak angular momentum accretion. Hence, the geometry of the CEE problem favors contraction of the binary orbit unlike what is the case for CBDs. 

  \begin{figure*}
\centering
      \includegraphics[width=0.75\textwidth]{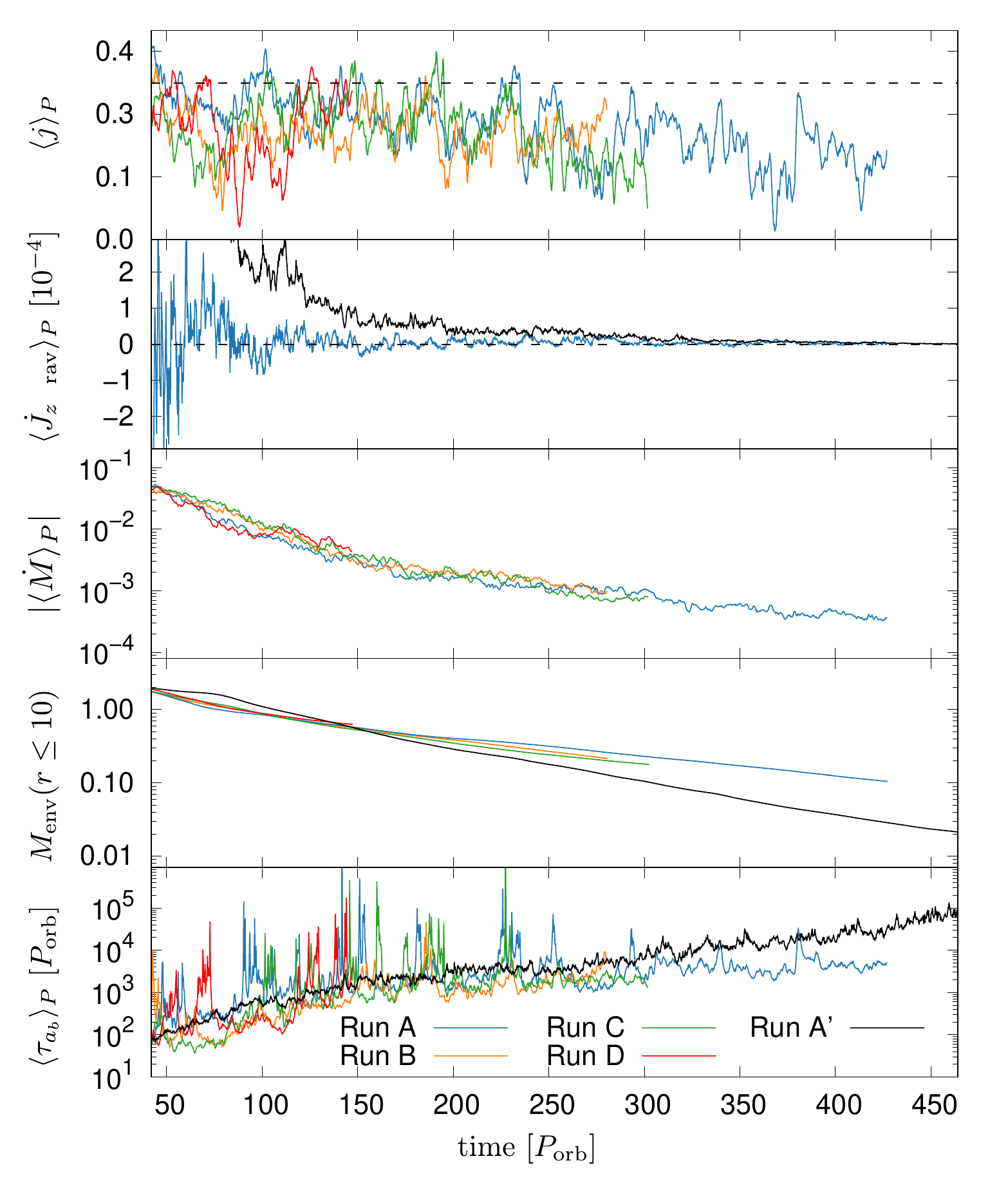} \hfill
    \caption{Evolution of key quantities relevant for the binary orbit after the initial envelope spin-up and adjustment. Panel (a) shows the moving average of the specific angular momentum transfer $\langle j \rangle_P$ (Eq.~(\ref{eq:movingj})) for models A, B, C, and D. The black dashed line indicates the critical specific angular momentum transfer $j_{\rm crit} = 3/8$. Panel (b) shows the moving average of the gravitational torque $\langle \dot{J}_{z, \rm grav} \rangle_P$ for simulation runs A and A'. The black dashed line indicates $\langle \dot{J}_{z, \rm grav} \rangle_P = 0$. Panel (c) showes the moving average of the mass accretion rate through the inner boundary $\langle \dot{M} \rangle_P$. Panel (d) shows the envelope mass in our numerical domain $M_\text{env} (r \leq 10)$. Panel (e) shows the moving average of the orbital separation evolution timescale $\langle \tau_{a_\text{b}} \rangle_P = \langle |a_\text{b}/\dot{a_\text{b}}| \rangle_P$.}
\label{fig:mmcrit}
\end{figure*}

In the top panel of Fig.~\ref{fig:mmcrit}, we show the evolution of the moving average of the specific angular momentum transfer rate over one orbital period
\begin{equation}
    \langle j \rangle_{P} = \frac{1}{P_{\rm orb}}\int_t^{t+P_{\rm orb}} j(t') \dd t'.
\end{equation}
We see that the combined effects of gravitational torque and mass and angular momentum accretion lead to the contraction of the orbit for all the considered values of $\beta$ and for both viscous and inviscid fluids. In the second panel of Fig.~\ref{fig:mmcrit}, we show the evolution of the moving average of the gravitational torque for all our models. We see that $\dot{J}_{z,\rm grav}$ rapidly settles to a value that closely oscillates around zero and thus does not contribute to the orbital evolution when the inner boundary is open to mass and angular momentum flow toward the binary. Conversely, when accretion is prevented by reflecting boundary conditions $\dot{J}_{z,\rm grav}$ decreases much slower, remains positive, and thus drives orbital contraction.  This crucial difference comes from the stabilizing effect of higher density in the vicinity of the binary when reflecting boundary conditions are enforced. This is discussed in more depth in Appendix~\ref{App:dotJzgrav}. In the bottom panel of Fig.~\ref{fig:mmcrit}, we show the orbital separation evolution timescale $ \tau_{a_\text{b}} = |a_\text{b}/\dot{a_\text{b}}|$. We find that $\tau_{a_\text{b}}$ reaches a statistically steady value of $O(10^3~P_{\rm orb})$ for all models allowing mass and angular momentum accretion onto the binary. Conversely, when accretion is forbidden (simulation run A'), the gravitational torque is exclusively responsible for the orbital contraction and the slow decrease of $\dot{J}_{z,\rm grav}$ implies a slow increase of $\tau_{a_\text{b}}$. The orbital separation evolution timescale eventually reaches a value of $O(10^5~P_{\rm orb})$ at the end of simulations run A' at $t\approx 450 P_\text{orb}$. It is possible that $\tau_{a_\text{b}}$ would continue increasing if we were able to run our model for more orbits.

Finally, we address the influence of envelope viscosity on binary evolution. Unfortunately, we could not run simulation run D for as long as the inviscid ones. Still, we can see that $\alpha_\nu = 10^{-3}$ does not significantly affect $\tau_{a_\text{b}}$. Because of the very variable nature of mass and angular momentum accretion rates, it is not clear whether the limited impact of viscosity would eventually lead to a slower or faster contraction of the orbit. Similarly, higher values of $\alpha_\nu$ should be investigated as well.

\subsection{Time variability of  mass and angular momentum accretion}\label{sec:tdepacc}

\begin{figure}[t]
 \centering
  \includegraphics[width=.49\textwidth]{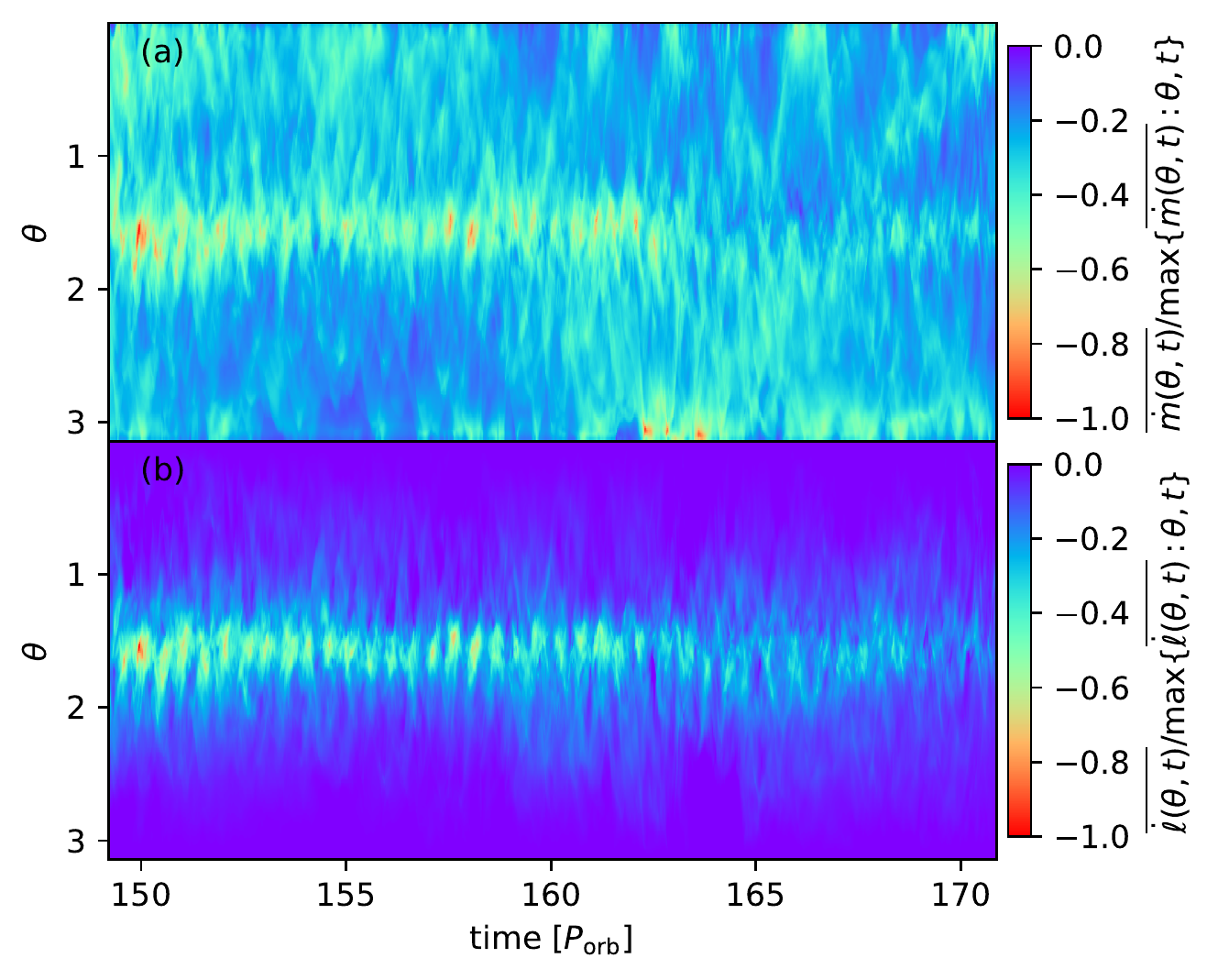}
   \caption{Space-time diagram of the mass (a) and angular momentum (b) fluxes onto the binary during 21 orbital periods for model A.}
 \label{fig:spacetimeB}
 \end{figure}

\begin{figure}[t]
\centering
        \includegraphics[width=.49\textwidth]{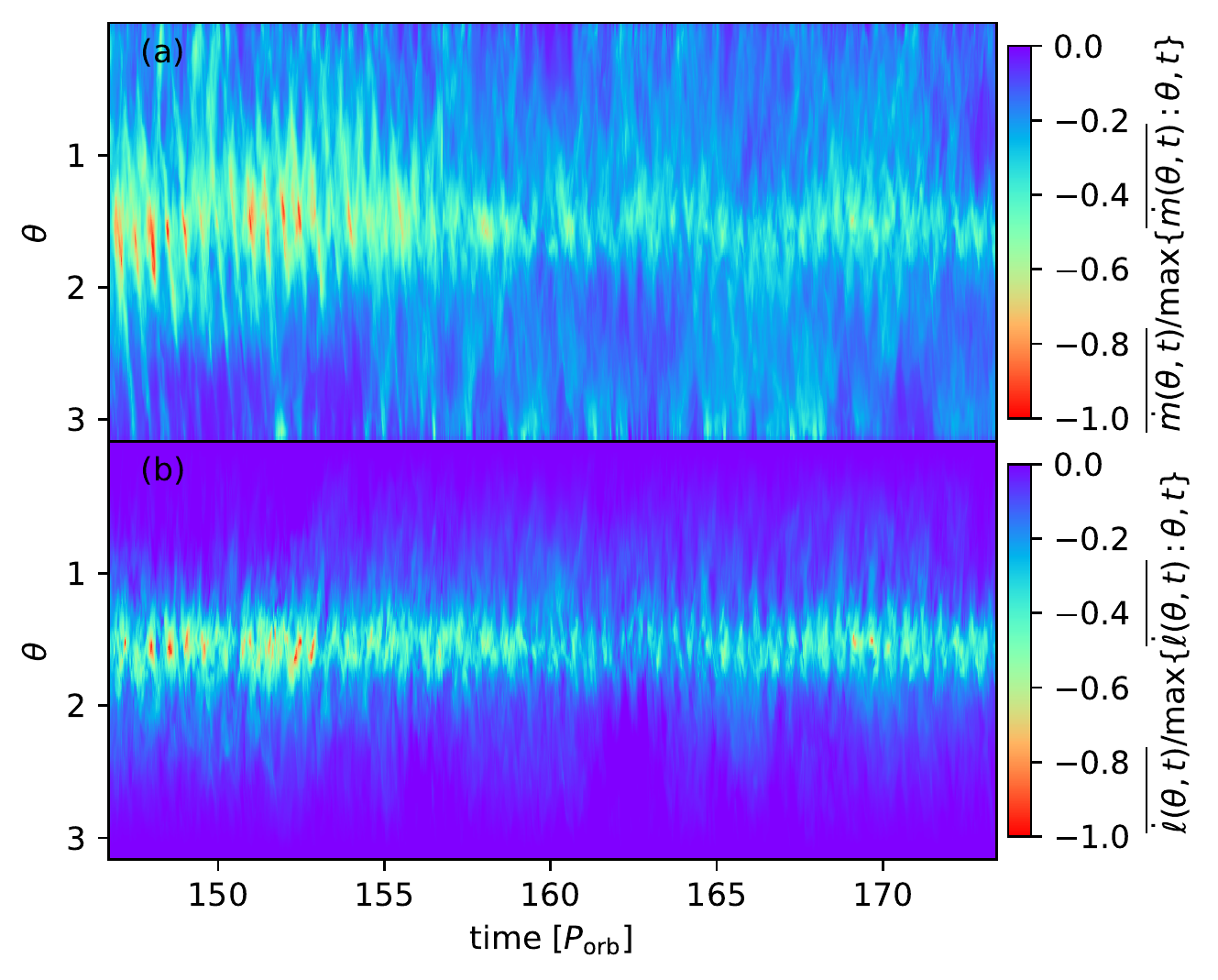}
   \caption{Same as Fig.~\ref{fig:spacetimeB} but for model B during 26 orbital periods.}
\label{fig:spacetimeA}
\end{figure}

\begin{figure}[t]
\centering
      \includegraphics[width=0.49\textwidth]{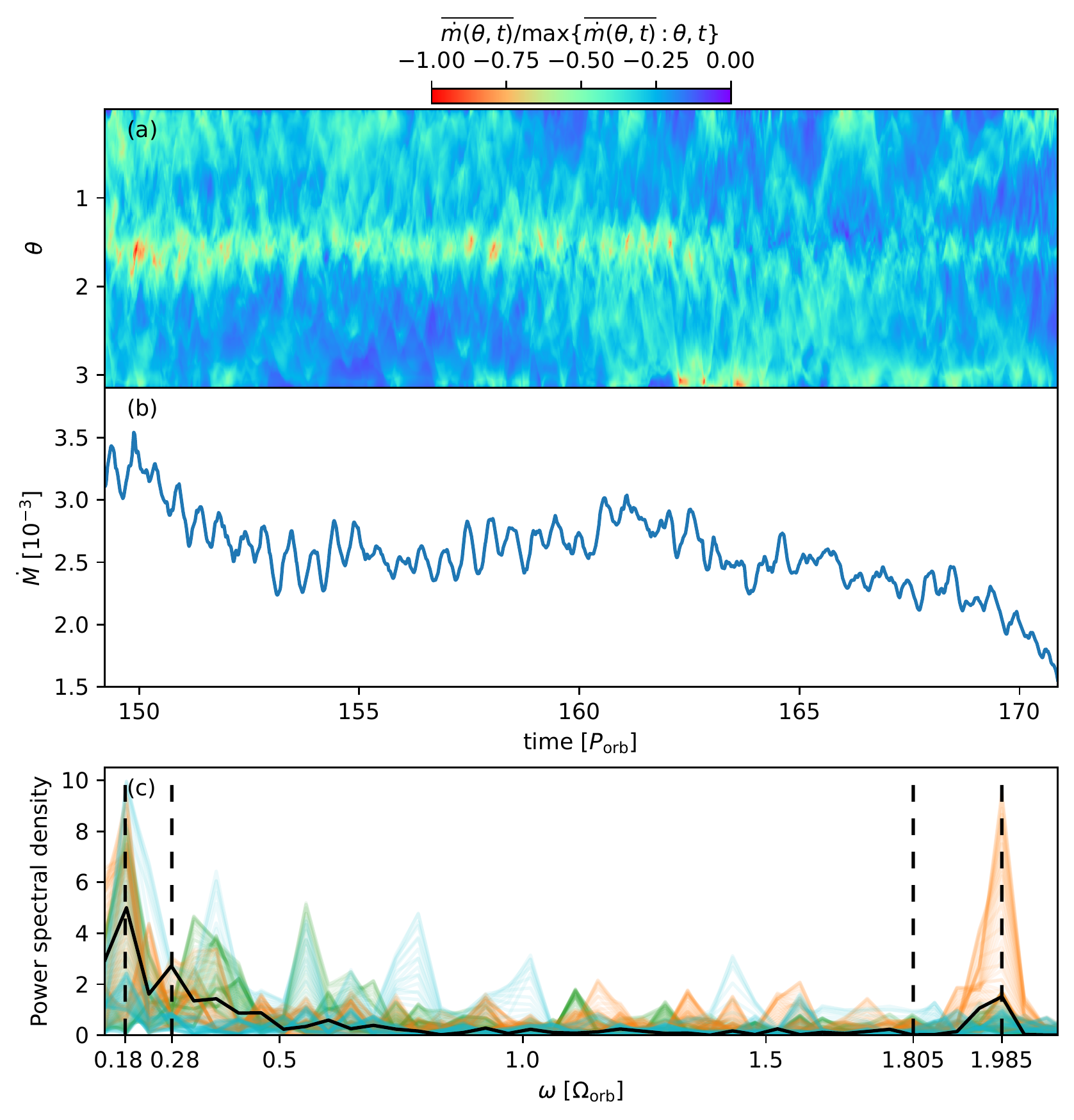}
   \caption{Detailed view on the variability of mass flux for model A. Panel (a): Space-time diagram of the local mass flux through the inner boundary. Panel (b): Time evolution of the mass accretion rate onto the binary. Panel (c): Power spectral density of the total mass accretion rate onto the binary (black line) and of the mass flux at each colatitude (colored lines). Green lines correspond to the range $0 \le \theta \le \pi/3$, blue lines to the range $2\pi/3 \le \theta \le \pi$, and orange lines to the range $\pi/3 \le \theta \le 2\pi/3$. }
\label{fig:fourierB}
\end{figure}

\begin{figure*}[t]
\centering
      \includegraphics[width=0.49\textwidth]{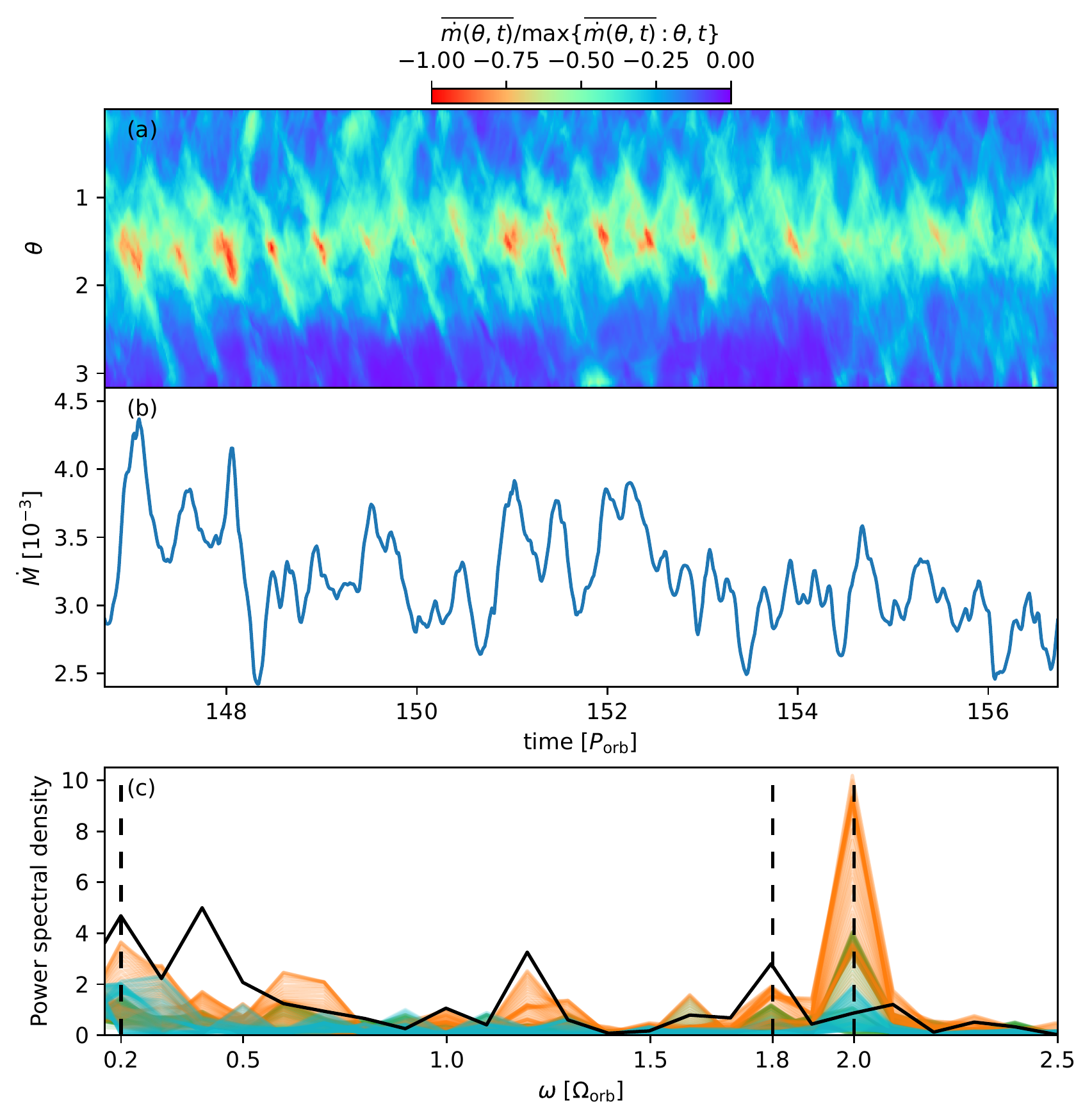}
            \includegraphics[width=0.49\textwidth]{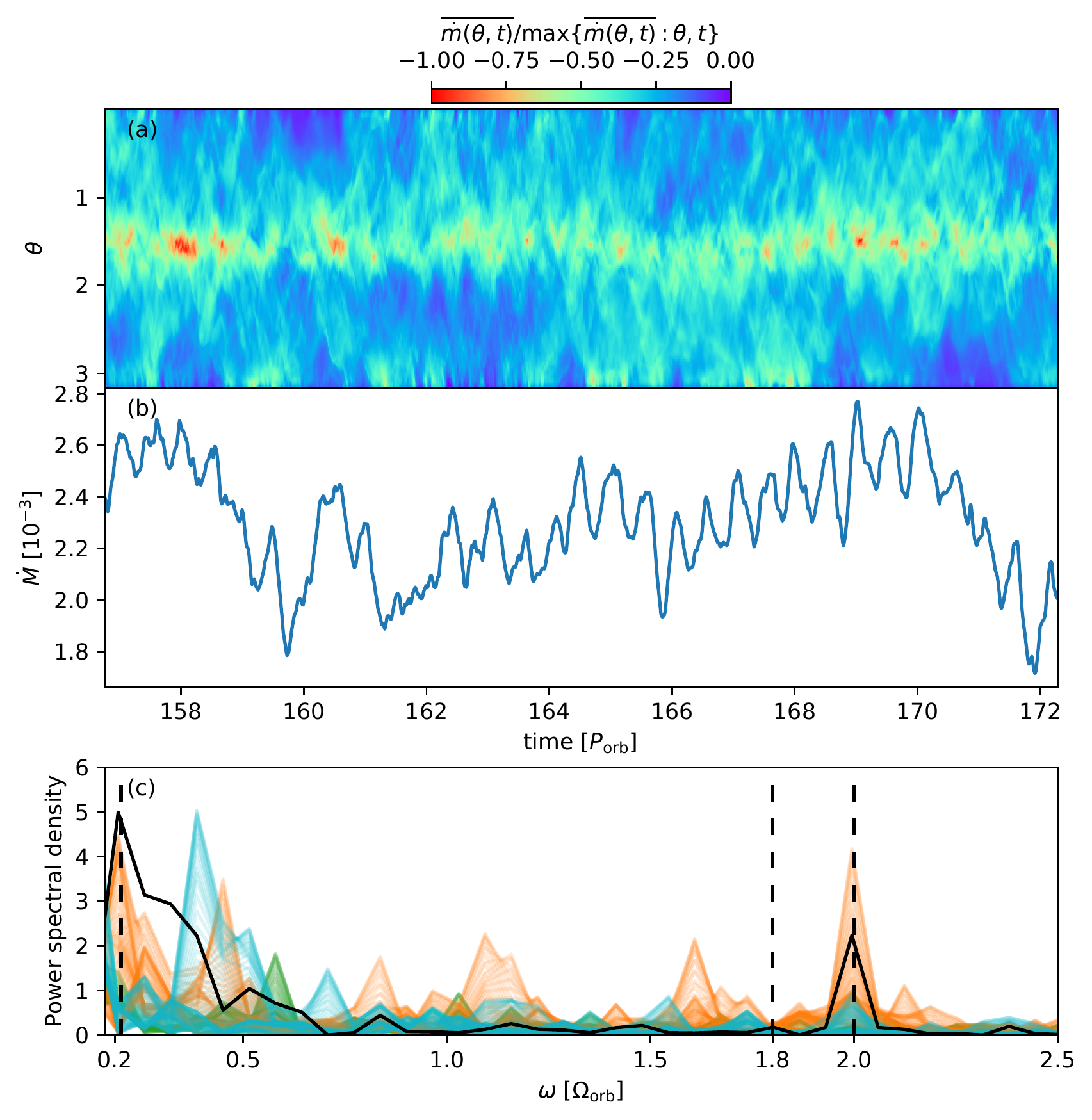}
   \caption{Detailed view on the variability of mass flux for model B and its two different accretion regimes (left and right panels). Meaning of symbols and lines in each panel is the same as in Fig.~\ref{fig:fourierB}. }
\label{fig:fourierA}
\end{figure*}

Now that we have investigated the secular binary separation evolution, we more thoroughly analyze the gas dynamics in the vicinity of the central binary, in particular, the time variability of mass and angular momentum accretion in the simulations that permit accretion. In Figs.~\ref{fig:spacetimeB} and \ref{fig:spacetimeA}, we show the latitudinal space-time diagram of the mass and angular momentum fluxes onto the binary, normalized by their maximum value in the considered time interval, for runs A and B. In the top panel of Figs.~\ref{fig:fourierB} and \ref{fig:fourierA}, we show a more detailed view of a shorter time interval. To construct these plots, we increased the simulation output rate to $80/P_\text{orb}$. We see that mass and angular momentum fluxes exhibit periodic variability at all colatitudes. This variability is manifold: we observe a high frequency variability that is modulated by a lower frequency, at least near the orbital plane.

To identify the modes associated with mass and angular momentum accretion variability, we use Fourier transform to compute the power spectral density of the mass accretion rate $\dot{M}$ and mass flux $\dot{m}(\theta)$. We distinguish four colatitude ranges: $0\le \theta \le \pi$, $0 \le \theta \le \pi/3$, $2\pi/3 \le \theta \le \pi$, and  $\pi/3 \le \theta \le 2\pi/3$ for the mass flux. We show our results in Figs.~\ref{fig:fourierB}c and \ref{fig:fourierA}c for simulation runs A and B.
In both simulations we identify two main peaks and their harmonics: one located at $\omega_\text{b} \simeq 2~\Omega_{\rm orb}$ and the other one at $\omega_\rho \simeq \Omega_{\rm orb}/5$. Here, $\omega_\text{b}$ corresponds to the forcing angular frequency of the quadrupolar moment contribution to the binary potential for a binary mass ratio $q=1$, which is the frequency at which material is pulled toward the central binary. 
The frequency $\omega_\rho$ is also often seen in CBDs, where it corresponds to an overdensity in the envelope orbiting at the angular frequency $\omega_\rho$.  This overdensity is often called ``lump'' and typically forms when some of the accreting material is strongly torqued in the vicinity of the inner boundary, which flings it back into the envelope where it locally accumulates. The interaction of the binary forcing frequency and the orbital angular frequency of the overdensity materializes as a modulation with beat angular frequency $\omega_{\rm beat} = \omega_\text{b} - \omega_\rho$.

We note a dramatic change in the latitudinal distribution from $t \simeq 156\ P_{\rm orb}$ in simulation run B  (Fig.~\ref{fig:spacetimeA}). For $t \lesssim 156\ P_\text{orb}$, we find that mass accretion shows inclined and periodic stripes spanning all latitudes. We interpret this as an indication of the presence of a tilted lump, successively feeding the individual binary components through accretion streams. We give more details in Appendix~\ref{App:tilt}.
In the context of CBDs, accretion onto the binary results exclusively from analogous accretion streams propagating in a low-density cavity encompassing the central binary. For $t \gtrsim 156\ P_\text{orb}$, the mass and angular momentum accretion becomes more isotropic, suggesting the absence of such well-structured latitudinally extended and tilted lump. Simultaneously, the complexity of the variability increases.

Interestingly, we see that while the $\omega_\text{b}$ mode appears in all three latitudinal regions in both runs and in both regimes of run B, such peak is not present in the power spectral density of the total mass accretion rate in the first regime of simulation run B. We can explain this by the asynchronocity of mass accretion between colatitudes, which results from the migrating accretion stream and which suggests that local latitudinal analysis is necessary when studying short-term evolution of accretion in CEE. The presence of a power spectral density peak at $\omega_\rho$ in the three latitudinal regions in the first phase of simulation run B suggests that there is a large latitudinal extent of an overdense region amplifying the accretion. However, this peak frequency is not present for the total mass accretion rate in simulation run A in the same time interval. This difference is likely due to the eccentric structure of overdensities above and below the orbital plane (see Sect.~\ref{sec:ecc}), which splits $\omega_\rho$ and its harmonics about their original value. Similar phenomenon was identified in CBD simulations \citep[e.g.,][]{Shi2012,Noble2012,Dorazio2013}.
In simulation run A, we can identify two peaks at around $0.18~\Omega_{\rm orb}$ and $0.28~\Omega_{\rm orb}$, which correspond to the splitting of $\Delta \omega \simeq 0.05$ about an unsplit lump angular frequency $ \omega_\rho \simeq 0.23$. For the second phase of simulation run B, additional peaks appear for $2\pi/3 \le \theta \le \pi$ and  $\pi/3 \le\theta\le 2\pi/3$ at around $0.13~\Omega_\text{orb}$ and $0.255~\Omega_\text{orb}$, which correspond to an angular frequency splitting of $\Delta \omega \simeq 0.0625$ about an unsplit lump angular frequency $ \omega_\rho \simeq 0.1925$.

\subsection{The lump}\label{sec:lump}

\begin{figure}[t]
\centering
  \includegraphics[width=0.49\textwidth]{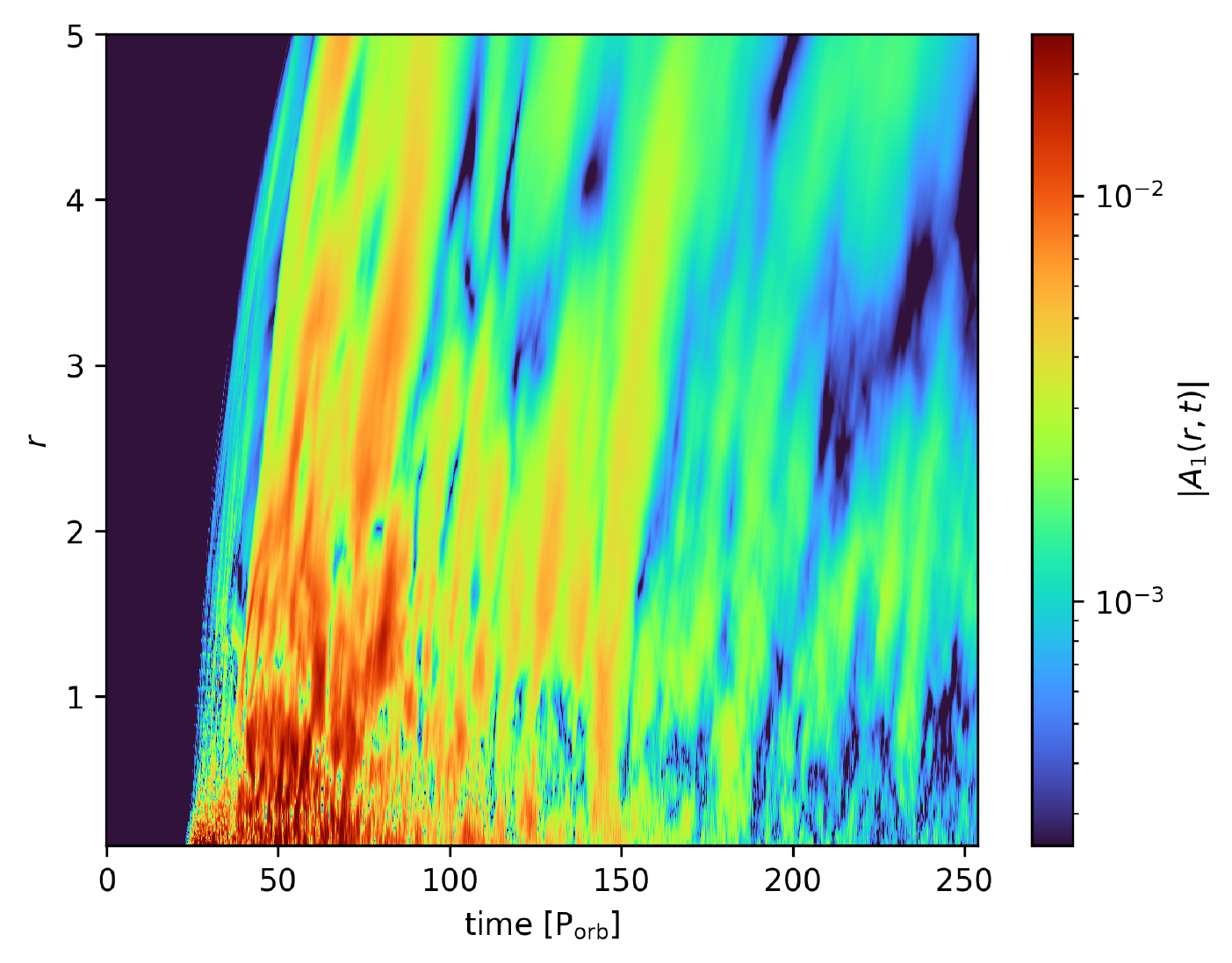}\\
  \includegraphics[width=0.49\textwidth]{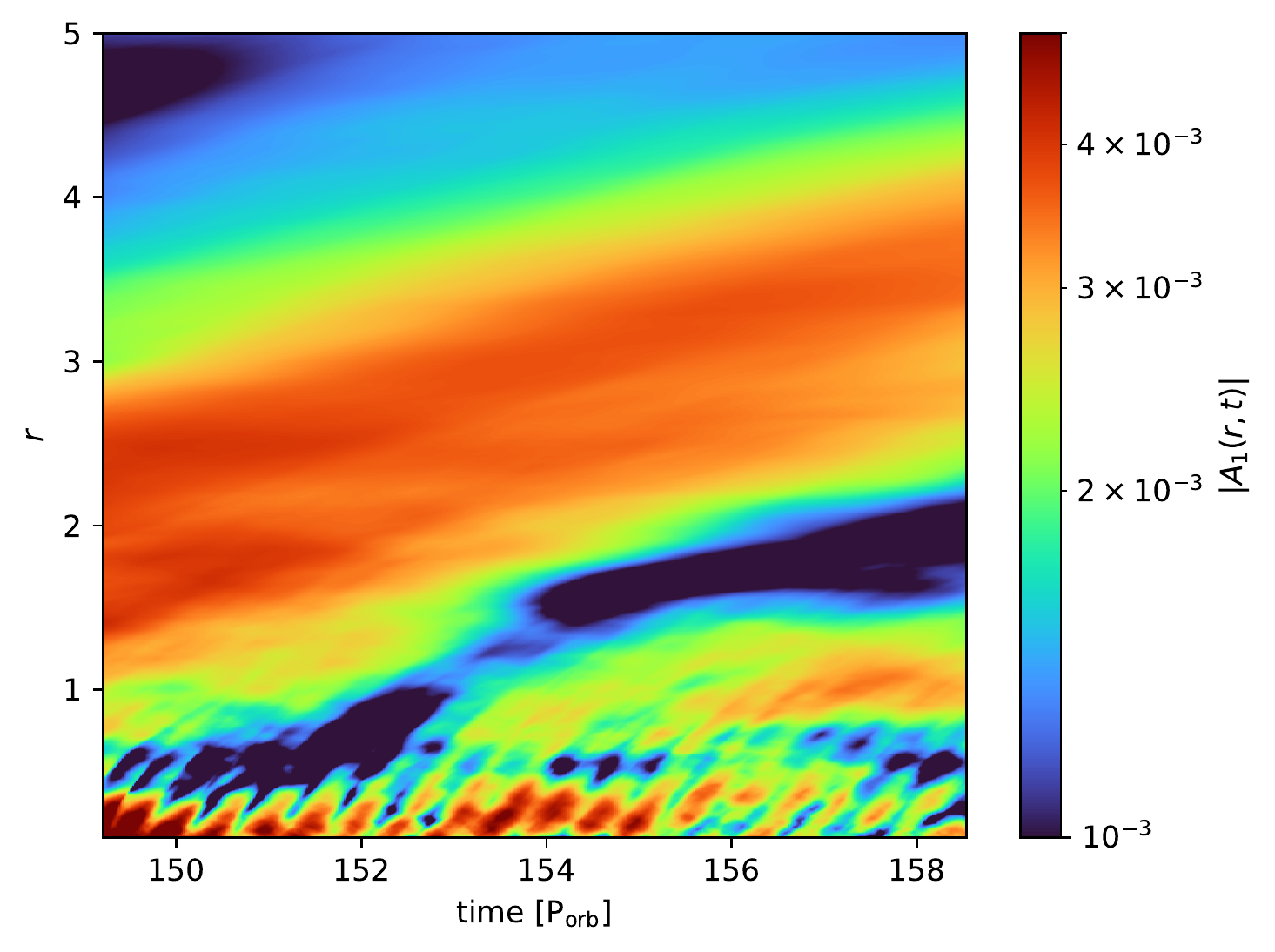}
   \caption{Space-time diagram of the $m = 1$ mode of the Fourier transform of the $\theta$-integrated density with respect to the azimuth $\varphi$ for model A. Panel (b) is obtained with a much larger time resolution than panel (a), such that high frequency fluctuations are well resolved. On panel (b), we see the overdensity generated at early time ($t \simeq 147~P_{\rm orb}$) propagating outward and expanding radially, and a new lump building up from  $t \simeq 152~P_{\rm orb}$. }
   \label{fig:A1B}
\end{figure}

\begin{figure}[t]
\centering
  \includegraphics[width=0.49\textwidth]{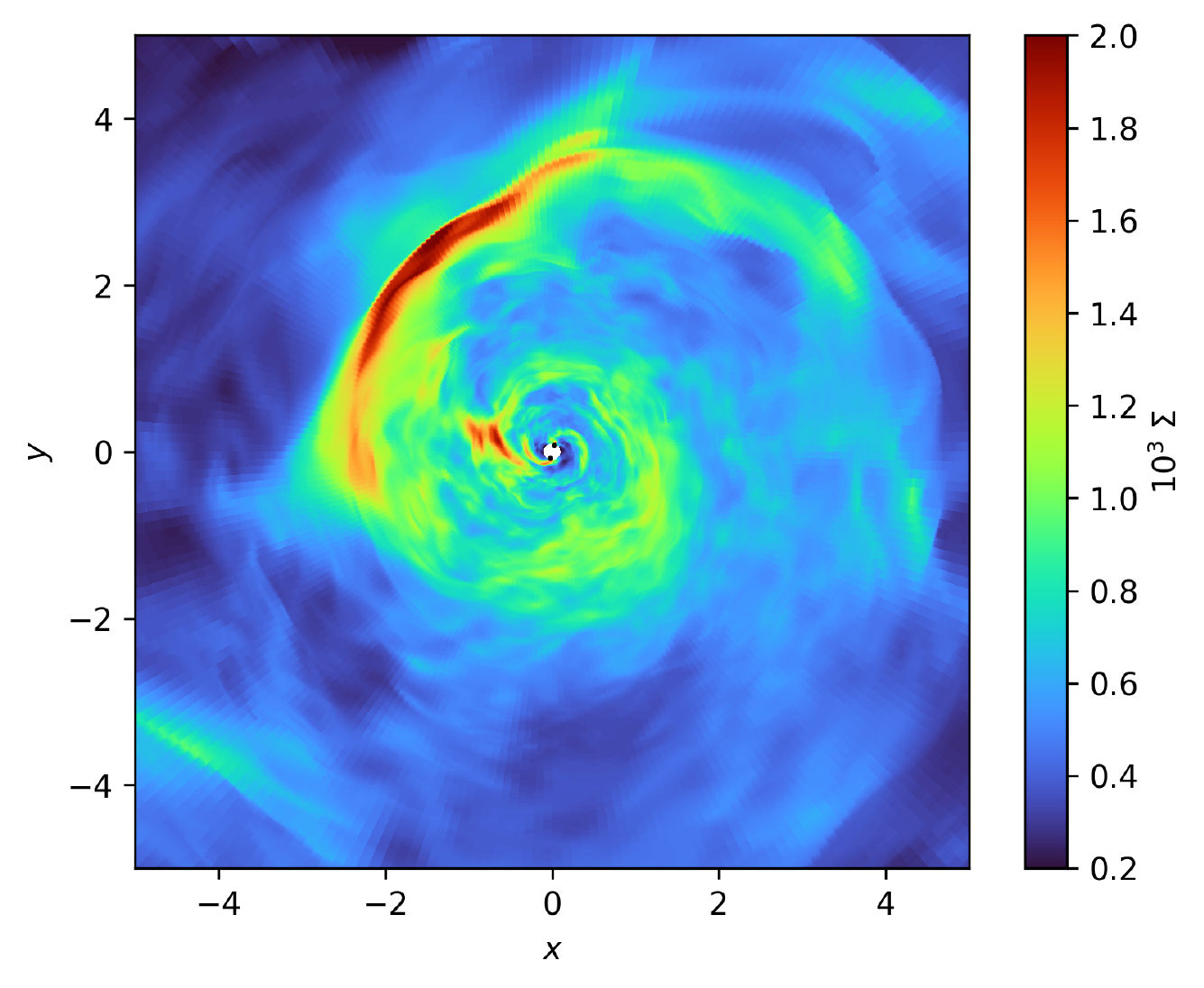} \\
   \includegraphics[width=0.49\textwidth]{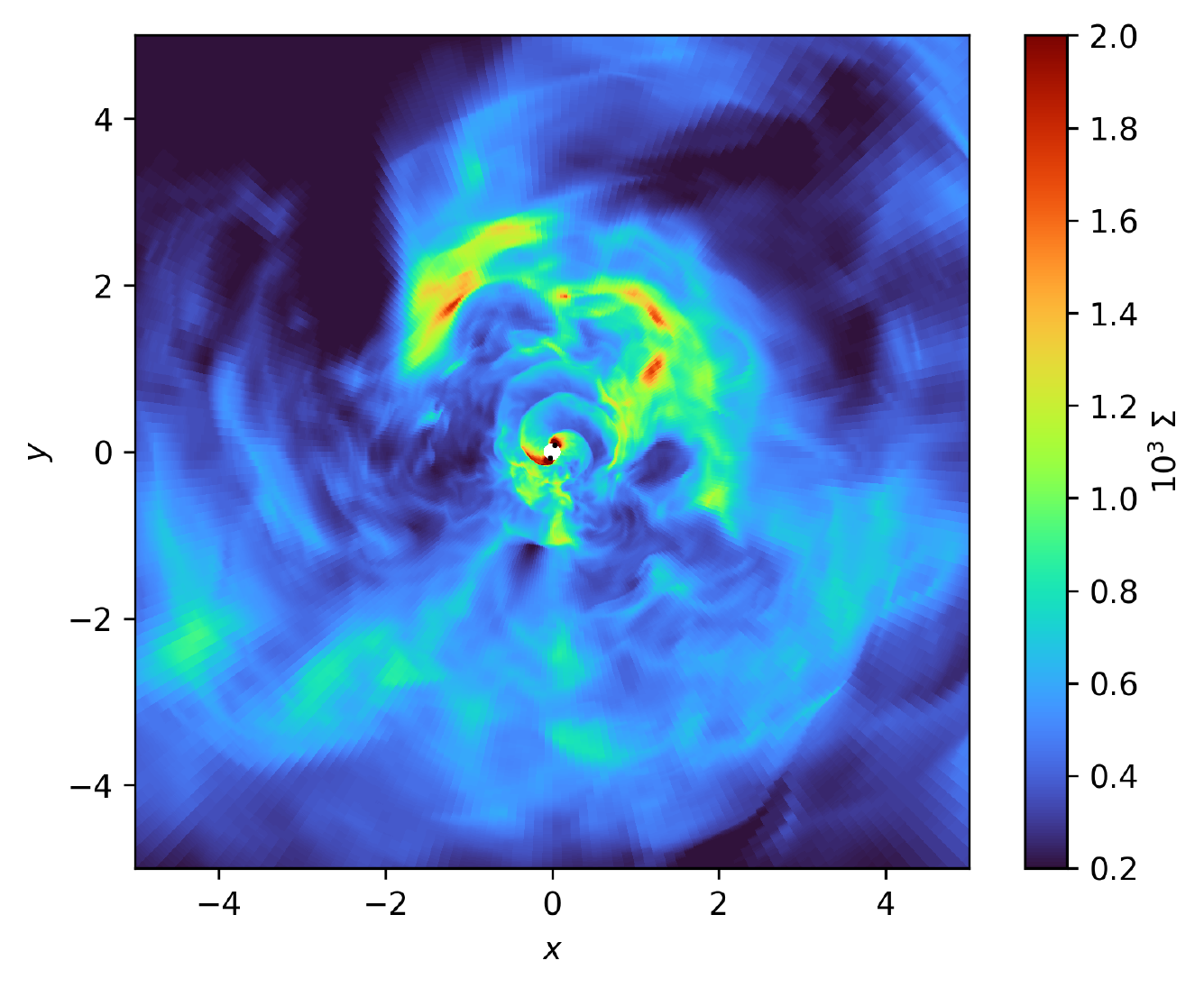}
   \caption{Surface density about the orbital plane (\ref{eq:sigma}) for model A (top) and A' (bottom) at $ t   = 158~P_{\rm orb}$. We see the outward propagation of a lump and the formation of a new one for model A, and the absence of a structured lump in model A' resulting from the absence of accretion streams. Black dots indicate the position of the two cores.}
   \label{fig:Sigma}
\end{figure}

Although we saw signatures of the lump in the power spectra, the density snapshots in Fig.~\ref{fig:snap} do not make the existence of a lump glaring. To better visualize the lump and to assess its potential effects on the inner envelope dynamics and accretion onto the central binary, in Fig.~\ref{fig:A1B} we examine the space-time evolution of quantity $A_{1}(r,t)$ \citep[e.g.,][]{Roedig2011,Shi2012,Roedig2012,Noble2012,Armengol2021}, which is the $\theta$-integrated $m=1$ mode of the Fourier transform of the density with respect to the azimuth $\varphi$,
 \begin{equation}
     A_{1}(r,t) = \int \rho e^{i\varphi}r \dd \theta \dd \varphi \ . 
 \end{equation}
First, we see a high-frequency variation of $A_1$ in the inner envelope ($ r\lesssim 0.8$), which is caused by the forcing angular frequency, $\omega_\text{b}$. A small fraction of such overdensities contribute to the increase of mass and angular momentum accretion shown in Figs.~\ref{fig:spacetimeB} and \ref{fig:spacetimeA} while the rest of the material is strongly gravitationaly torqued by the binary and is flung back into the envelope. These outflowing streams collide and accumulate in a large range of colatitudes starting from $r \simeq 0.8$. The resulting overdense region  dilutes and propagates radially far into the envelope, as we can see from the outward propagating overdensity in Fig.~\ref{fig:A1B}. The inner part of this overdense lump feeds the inner envelope, but eventually the lump propagates far enough into the envelope that it no longer interacts with the inner region and a new lump begins to form again. We illustrate this process in Fig.~\ref{fig:Sigma}, where we show the surface density averaged in the $z$ direction for a thin region of opening angles $\pm\pi/8$ about the orbital plane,
\begin{equation}\label{eq:sigma}
    \Sigma = \int_{7\pi/16}^{9 \pi/16} \rho r \sin \theta \dd\theta \ .
\end{equation}

While overdensities we see in our simulations are in many aspects remarkably similar to the lump present in CBDs simulations, they exhibit fundamental differences. In CBD simulations, $\omega_\rho$ is the orbiting frequency of a single lump that is fed by accreting material flung back into the envelope and that typically remains near the cavity edge. In contrast, in our CEE simulations, $\omega_\rho$ characterizes the formation frequency of nonaxisymmetric overdensities that propagate far into the envelope.

\subsection{Eccentricity growth and evolution}\label{sec:ecc}

\begin{figure}[t]
\centering
  \includegraphics[width=0.49\textwidth]{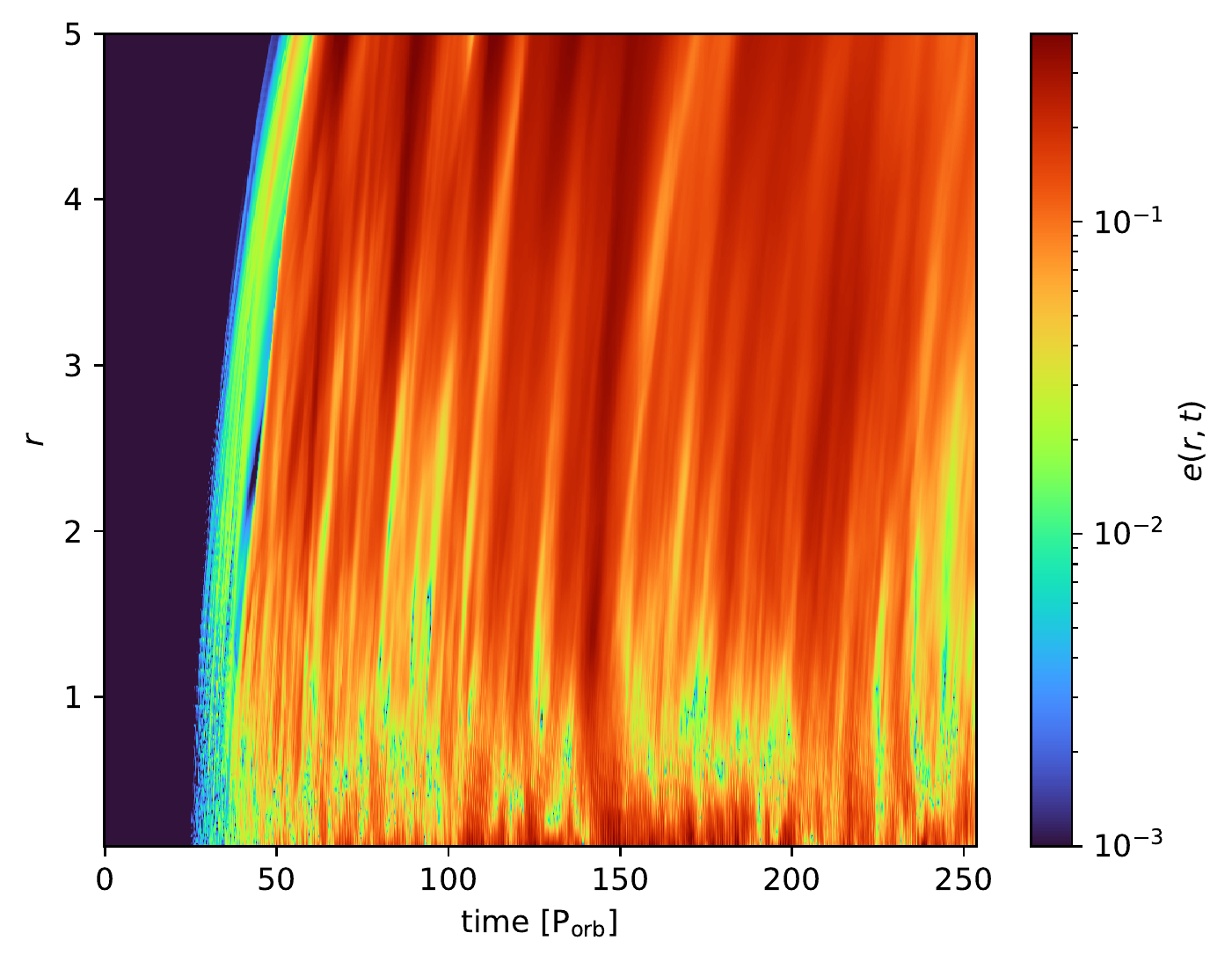}
       \includegraphics[width=0.49\textwidth]{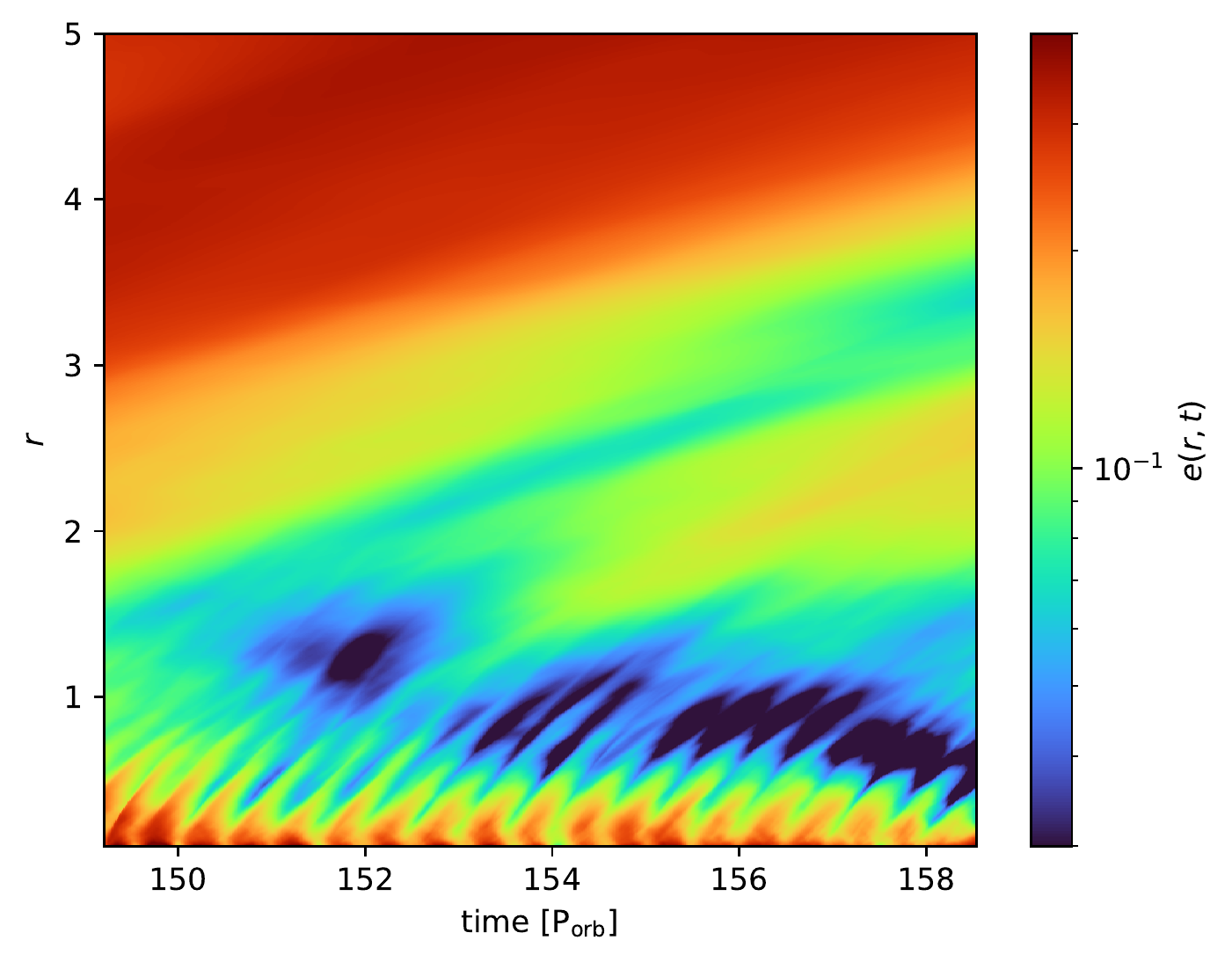}
   \caption{Space-time diagram of shell-averaged envelope eccentricity. The meaning of symbols is the same as in Fig.~\ref{fig:A1B}.}
   \label{fig:eB}
\end{figure}

\begin{figure}[t]
\centering
\includegraphics[width=0.49\textwidth]{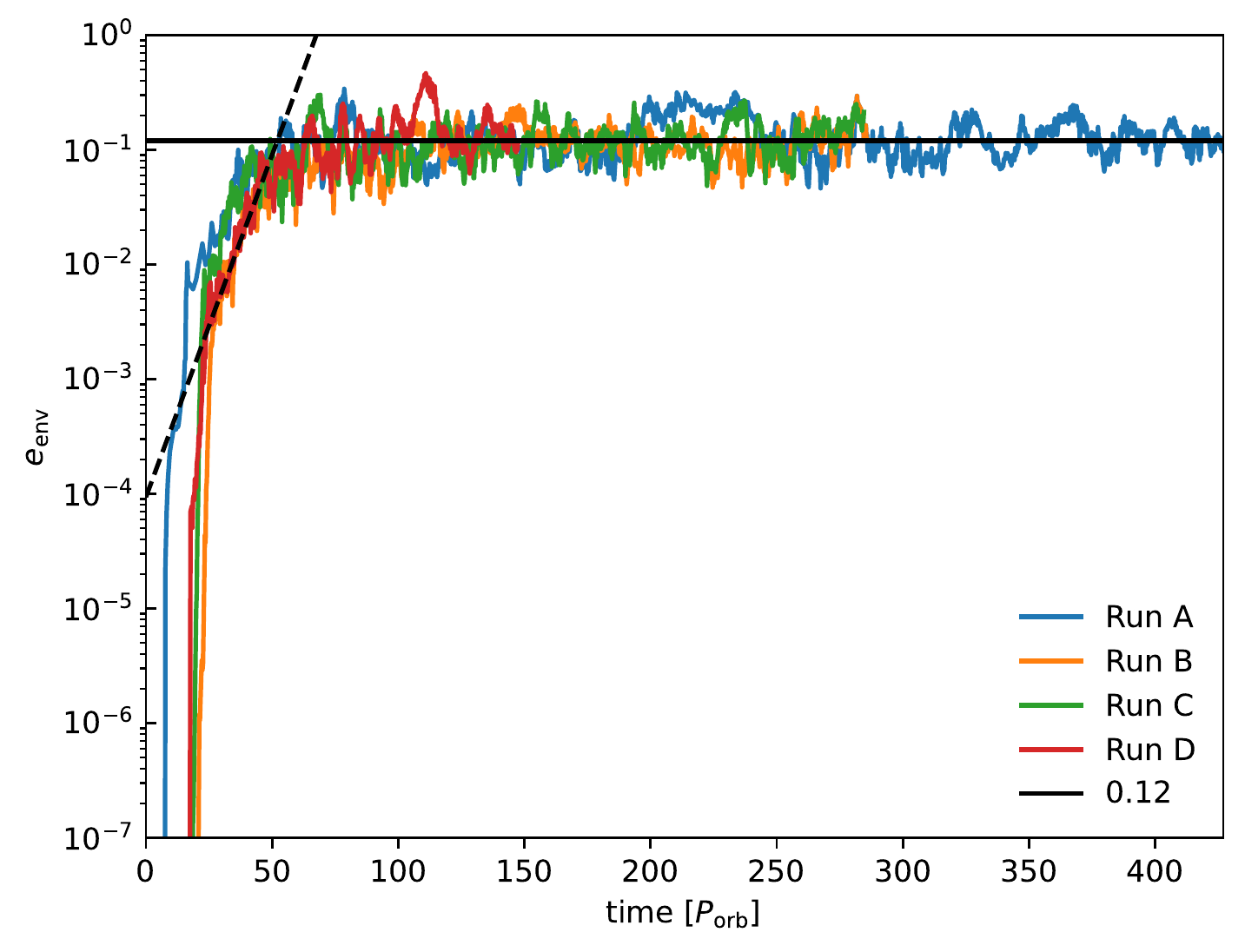}
\caption{Evolution of the mean envelope eccentricity within the numerical domain. The black dashed line shows  a linear fit yielding a growth rate $\lambda_\text{env} \simeq 0.022~\Omega_{\rm orb}$.}
\label{fig:eccev}
\end{figure}

Throughout this work, we assume that the binary orbit remains circular and fixed. As a result, we can only speculate about implications of our findings for binary eccentricity (see Sect.~\ref{sec:cee_impl}), but we can directly study the related eccentricity of the envelope. We would expect that the initially noneccentric envelope encompassing an equal-mass binary on a circular orbit will not become eccentric. However, the frequency splitting of $\omega_\rho$ observed in Figs.~\ref{fig:fourierB}  and \ref{fig:fourierA}  suggests that envelope eccentricity develops in our simulations. To illustrate this more thoroughly, in Fig.~\ref{fig:eB} we show the space-time diagram of shell-averaged envelope eccentricity
 \begin{equation}
     e(r,t) = \frac{\left\lvert\int \rho u_r e^{i \varphi} \dd S \right\rvert}{\int \rho u_\varphi \dd S} \ .
 \end{equation}
Similarly to $A_1$, $e$ is subject to a high frequency variation in the inner envelope ($ r\lesssim 1$) according to the forcing frequency $\omega_\text{b}$. The dynamics of accretion and of the lump is tightly linked to the generation and propagation of eccentricity in the envelope. Eccentricity is excited by the amplification of small asymmetries in the interaction between accretion flows and the central binary either by stream impact on the inner boundary \citep[e.g.,][]{Shi2012} or by resonant Lindblad excitation \citep[e.g.,][]{Lubow1991a,Lubow1991b,Papaloizou2001,Munoz2020}.  We see that while a fraction of the newly generated eccentricity is contained in the colliding outflowing streams forming lumps, the rest is trapped in between successively created lumps, where $e$ grows over time. Consequently, as the lumps propagate outward, the eccentricity follows. 

In Fig.~\ref{fig:eccev}, we show the evolution of the mean envelope eccentricity within our simulation domain, $e_\text{env}$, which is defined as
\begin{equation}
    e_{\rm env} = \frac{\left\lvert\int \rho u_r e^{i \varphi} \dd V \right\rvert}{\int \rho u_\varphi \dd V} \ .
\end{equation}
We find that similarly to CBDs \citep[e.g.,][]{Shi2012}, the $e_\text{env}$ initially increases very rapidly in response to the quadrupole perturbation associated with the replacement of the time- and latitude-averaged binary potential with the true expression. For  $25 \lesssim t/P_\text{orb} \lesssim  50$, eccentricity grows exponentially with a growth rate $\lambda_\text{env} \simeq 0.022~\Omega_{\rm orb}$. The growth rate does not depend on the initial angular momentum of the envelope nor on the presence or absence of viscosity. Interestingly, the value of $\lambda_\text{env}$ is of the same order as the eccentricity saturation growth rate of $\sim 0.018~\Omega_{\rm orb}$ obtained by \cite{Shi2012} in the context of CBDs, which could suggest a common physical origin. After $t \simeq  50~P_{\rm orb}$, the exponential growth saturates and $e_\text{env}$ reaches a statistically stationary state with a mean value $e_{{\rm env},f} \simeq 0.12$, which is independent of $\beta$ and of the presence or absence of viscosity. Such eccentricity saturation likely results from nonlinear effects, which suggests that eccentricity excitation and damping reach a quasi-equilibrium that may be maintained throughout the entire post-dynamical spiral-in phase \citep[e.g.,][]{Shi2012,Teyssandier2016,Miranda2017,Munoz2020}.

\subsection{Convective stability and angular momentum transport in the envelope}\label{sec:AMtrans}

During the post-dynamical CEE, the central binary interacts with the surrounding gas and a complex interplay between the torques and internal stresses continuously injects, removes, and redistributes angular momentum within the envelope. In this Section, we investigate the stability of the envelope and analyse its dynamics by characterizing the various angular momentum transport processes.

\subsubsection{Solberg–H{\o}iland criterion for convective stability}
\label{sec:solberg}
\begin{figure}[t]
\centering
      \includegraphics[width=0.5\textwidth]{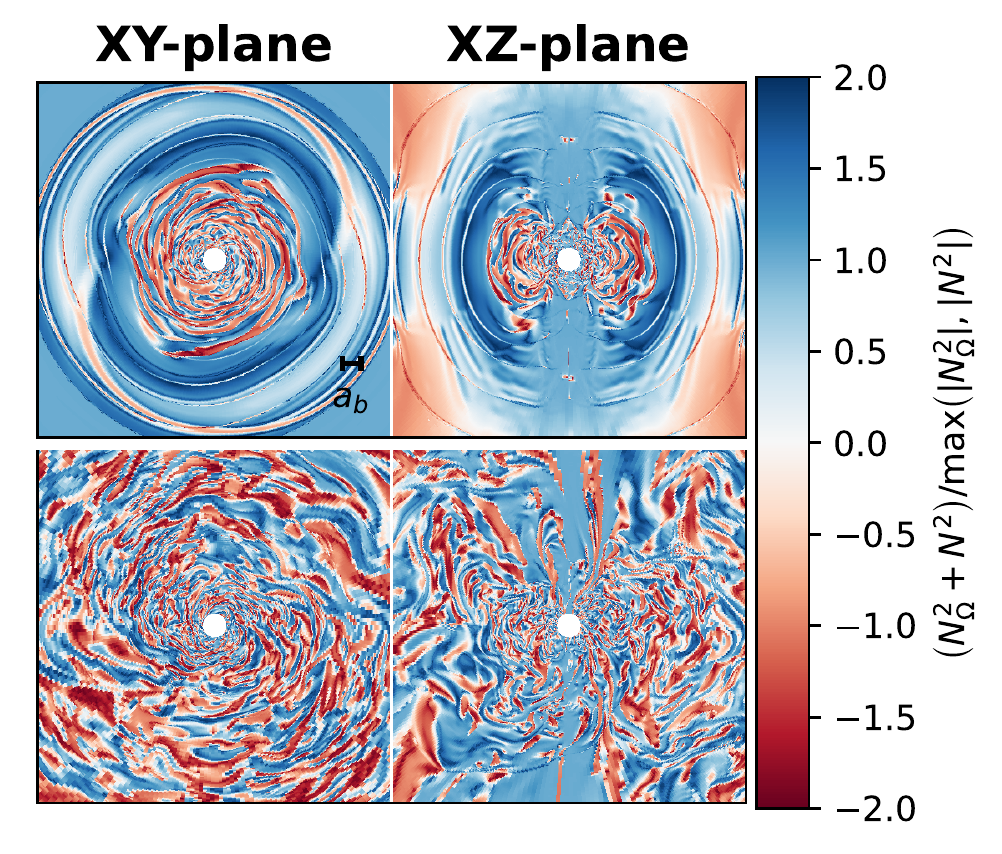}
      \caption{First Solberg–H{\o}iland criterion for convective stability for run B at $t= 20~P_{\rm orb}$ (top row), and $t= 250~P_{\rm orb}$ (bottom row). Negative values indicate convective instability according to Eq.~(\ref{eq:solb1}).}
\label{fig:solb1}
\end{figure}

In the absence of viscosity, thermal diffusion, and  radiation pressure, \cite{Solberg1936} and \cite{Hoiland1941} proposed the following necessary but not sufficient condition for convective stability, which for a stratified and rotating fluid with $\boldsymbol{\Omega} = \Omega\be_z$ reads
\begin{equation}\label{eq:solb1}
N^2_{\Omega} + N^2 > 0 \ ,
\end{equation}
where 
\begin{equation}
N^2_{\Omega} =      \frac{1}{s^3}\frac{\partial \ell_z^2 }{\partial s} \quad {\rm and} \quad N^2 =  -  \frac{1}{c_p}\bg \cdot \bnabla S \ ,
\end{equation}
$\ell_z=s^2 \Omega$ is the specific angular momentum, $S$ is the specific entropy, and $c_p$ is the heat capacity at constant pressure. In Fig.~\ref{fig:solb1}, we show the Solberg–H{\o}iland criterion for convective stability in the $xy$ and $xz$ planes soon after replacing the averaged binary potential with its full expression and at late-time. We see that as soon as they are present, gravitational perturbations from the central binary destabilize the flow according to the Solberg–H{\o}iland criterion. In practice, this translates into small scale turbulent mixing between spiral arms (see Fig.~\ref{fig:snap}, first row), which is initially not strong enough to destroy the spiral structure. As the envelope expands and the stabilizing effect of density stratification is reduced, the vertical size of the turbulent eddies increases and the spiral structure is partially destroyed. We observe behavior resembling the ab initio simulation of dynamical plunge-in from \cite{Ohlmann2016}, where the theoretically stable and unstable layers alternate in a geometrically thick disk-like structure about the orbital plane. The radial spatial frequency decreases outward as the stabilizing effect of stratification becomes weaker.

\subsubsection{Local torque balance}

While the various volume-integrated torques presented in Sect.~\ref{sec:torques} trace the evolution of the total angular momentum reservoir of the common envelope in our numerical domain, it is also important to examine the spatial variation of such torques. As we show in detail in Appendix~\ref{App:dotJz}, the local angular momentum transfer rate across the common envelope reads
\begin{equation}
    \dot{J}_z(r,t) = \dot{J}_{z, \rm adv}(r,t) + \dot{J}_{z, \rm grav}(r,t) + \dot{J}_{z, \rm visc}(r,t) \ ,
\end{equation}
where
    \begin{align}
    & \dot{J}_{z, \rm adv}(r,t) = - \int_{\partial r} \rho s u_\varphi u_r \dd S \ , \\
    & \dot{J}_{z, \rm grav}(r,t) = \int^{R_{\rm domain}}_r \left( \int_{\partial r} \rho \frac{\partial \Phi}{\partial \varphi} \dd S \right) \dd r \ , \\ 
    & \dot{J}_{z, \rm visc}(r,t) = - \int_{\partial r} \left[\left(   \boldsymbol{\br} \times \boldsymbol{T} \right) \cdot \be_z \right] \cdot \boldsymbol{n}_\perp \dd S \ .
    \end{align}

\begin{figure}[t]
      \includegraphics[width=0.5\textwidth]{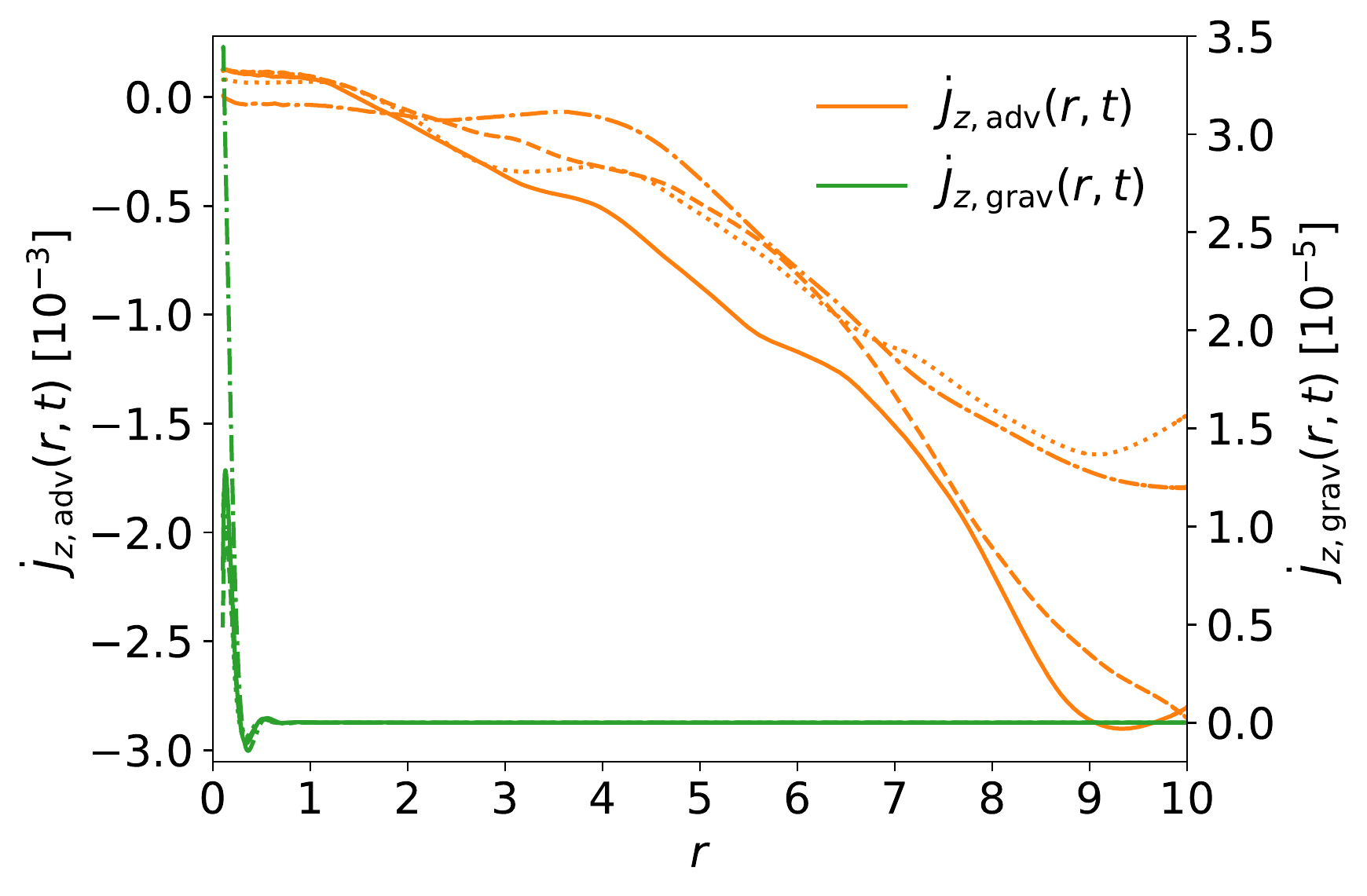} \\
      \caption{Advective and gravitational contributions to the local angular momentum transfer rate for models A (full lines), model A' (dash-dotted lines), model B (dashed lines), and model C (dotted lines). The quantities are averaged in time interval $250 \le t/P_{\rm orb} \le 275$.}
\label{fig:torqprofAB}
\end{figure}

In Fig.~\ref{fig:torqprofAB}, we show the contributions to $\dot{J}_z$ as a function of $r$  for simulation runs A, A', B, and C. We see that for accreting models A, B, and C, gravitational torque only plays a minor role in the redistribution of angular momentum in the inner envelope and essentially no role far from the central binary. This occurs because the density is globally a decreasing function of $r$ and because $\lim_{r \to \infty} \partial \Phi / \partial \varphi = 0$. Instead, it is the advective torque that transports angular momentum. Up to $r \simeq 1.5$, the angular momentum is transported inwards $\dot{J}_{z, \text{adv}} > 0$, but for larger $r$ the angular momentum flows outward. However, when accretion is prevented by reflecting boundary conditions (simulation run A'), gravitational torque dominates advective torque up to $r \simeq 0.2$, which  results in inward angular momentum transport. At larger $r$, the advective torque transports angular momentum outward. The main differences in the advective torque profiles between the four models result from  different contributions of turbulence and  different mean-flow angular structure.

\subsubsection{Turbulent transport of angular momentum in the envelope}\label{sec:turbtransport}
 
\begin{figure*}[t]
\includegraphics[width=\columnwidth]{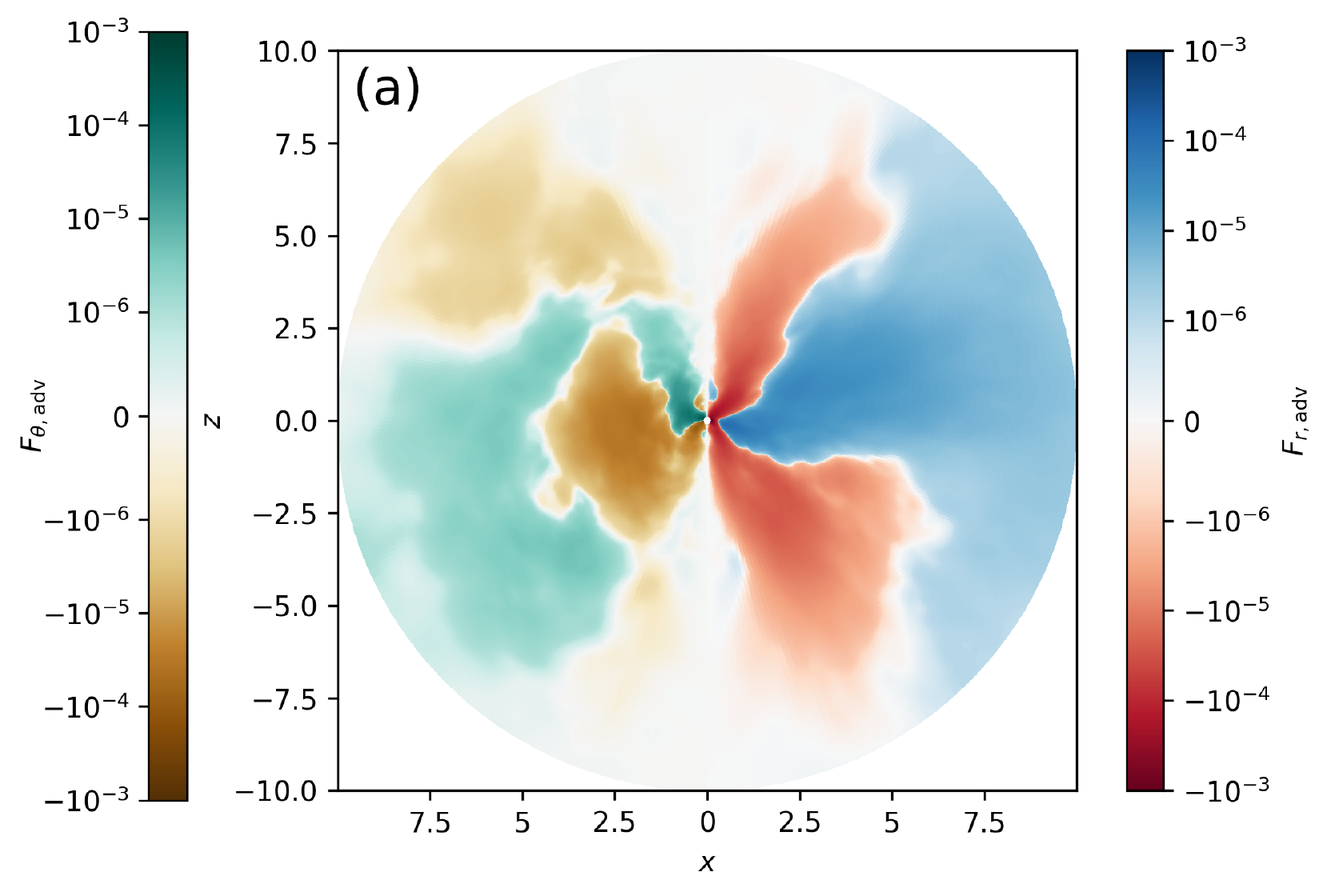} 
\includegraphics[width=\columnwidth]{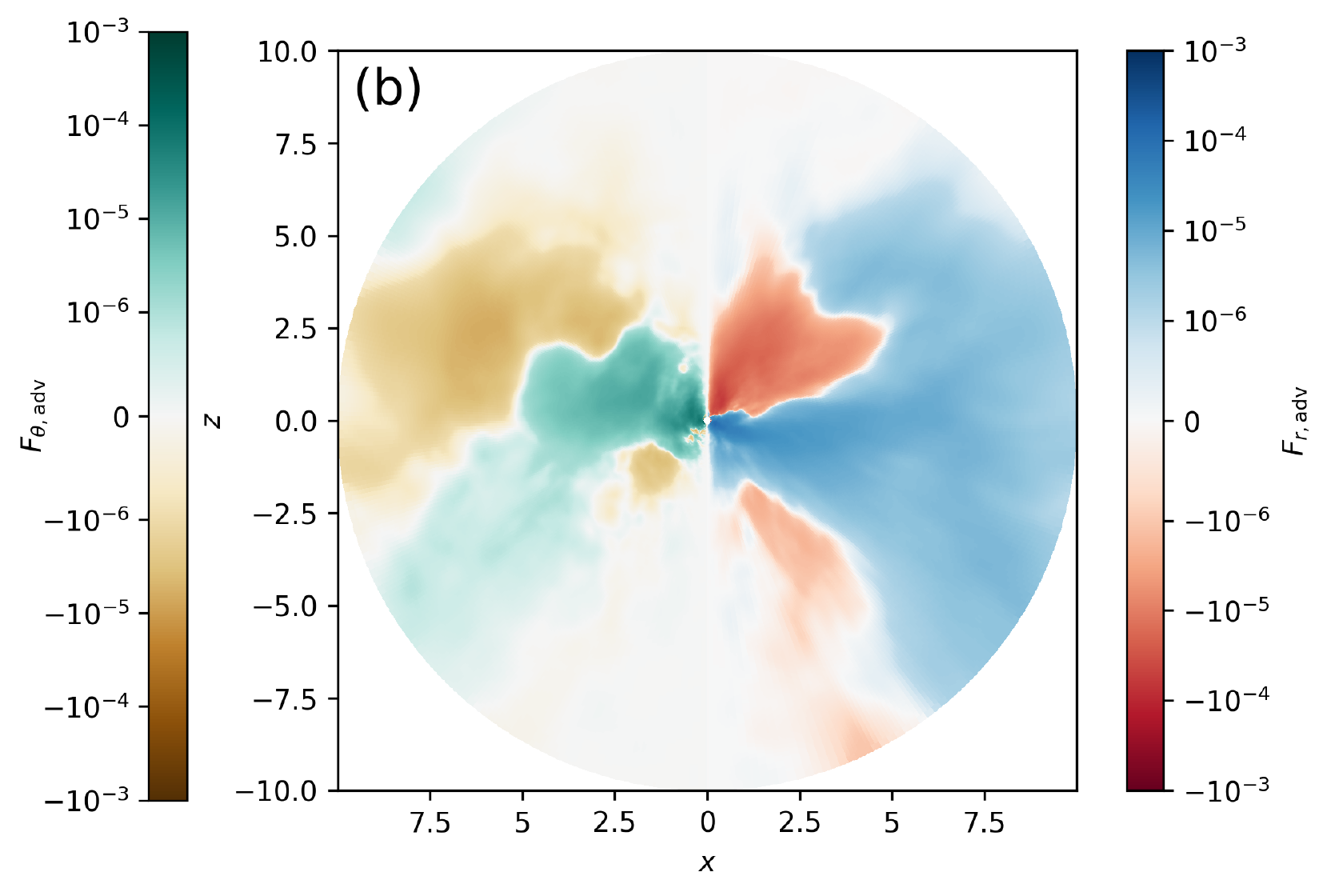} \\
\includegraphics[width=\columnwidth]{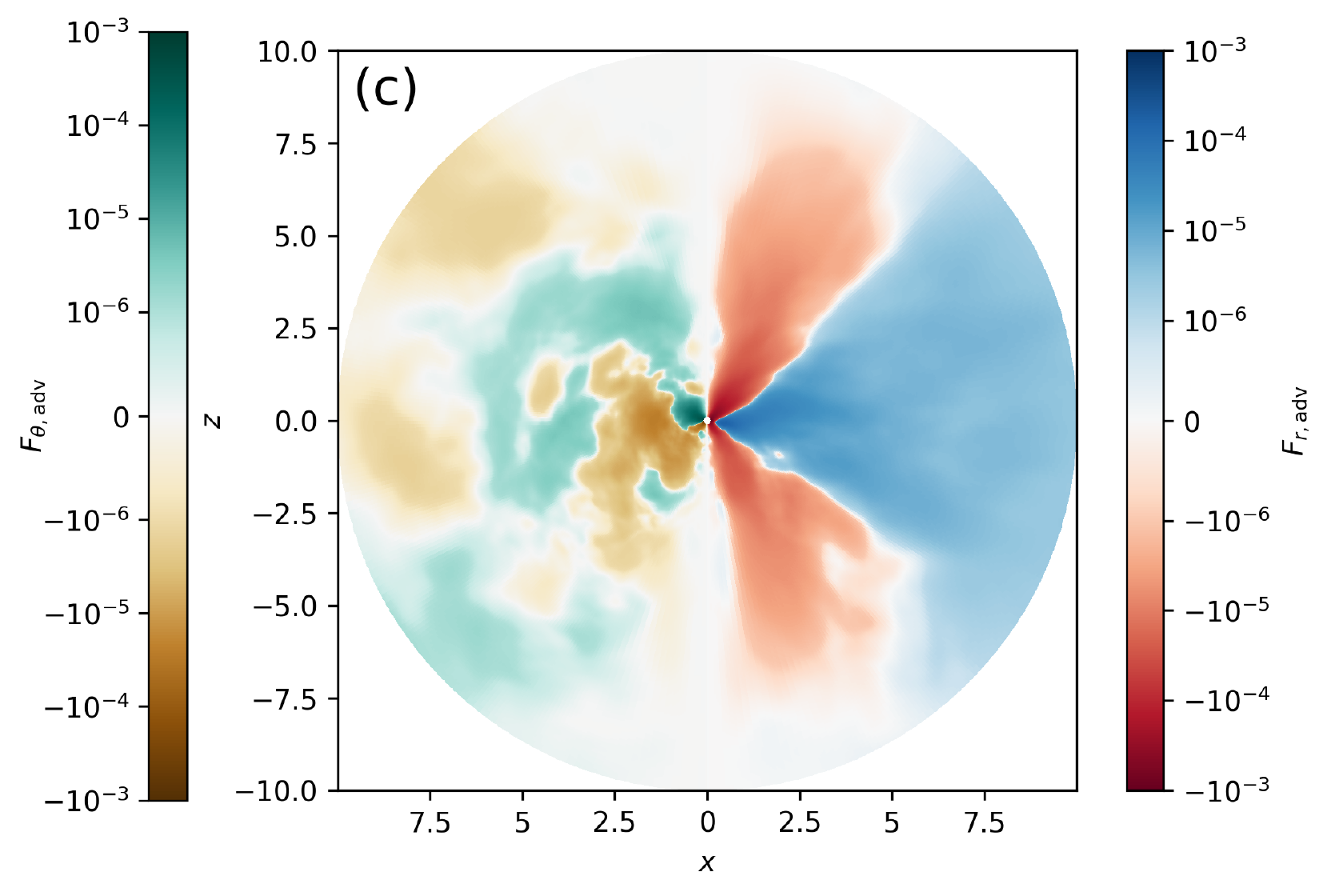} 
\includegraphics[width=\columnwidth]{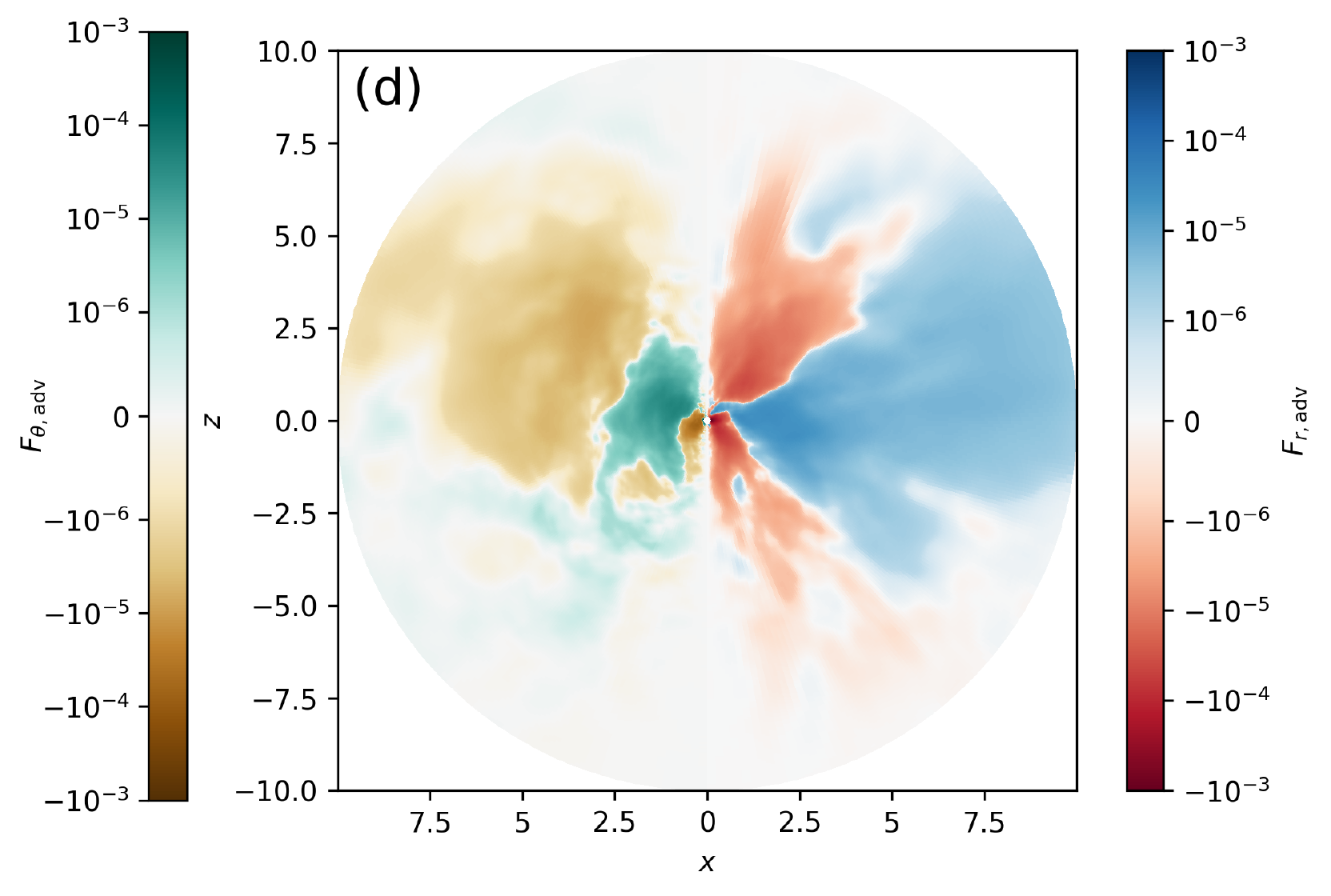} 
   \caption{Azimuthally averaged advective radial and latitudinal fluxes of angular momentum, $ F_{r, \rm adv} = s \overline{\rho}\,\overline{u_r u_\varphi}$ and $F_{\theta, \rm adv} = s \overline{\rho}\,\overline{u_\theta u_\varphi}$ averaged over time interval $250 \le t/P_{\rm orb} \le 275$ for models A (panel a), A' (panel b), B (panel c), and C (panel d).}
\label{fig:F_A1}
\end{figure*}

\begin{figure*}[t]
\includegraphics[width=\columnwidth]{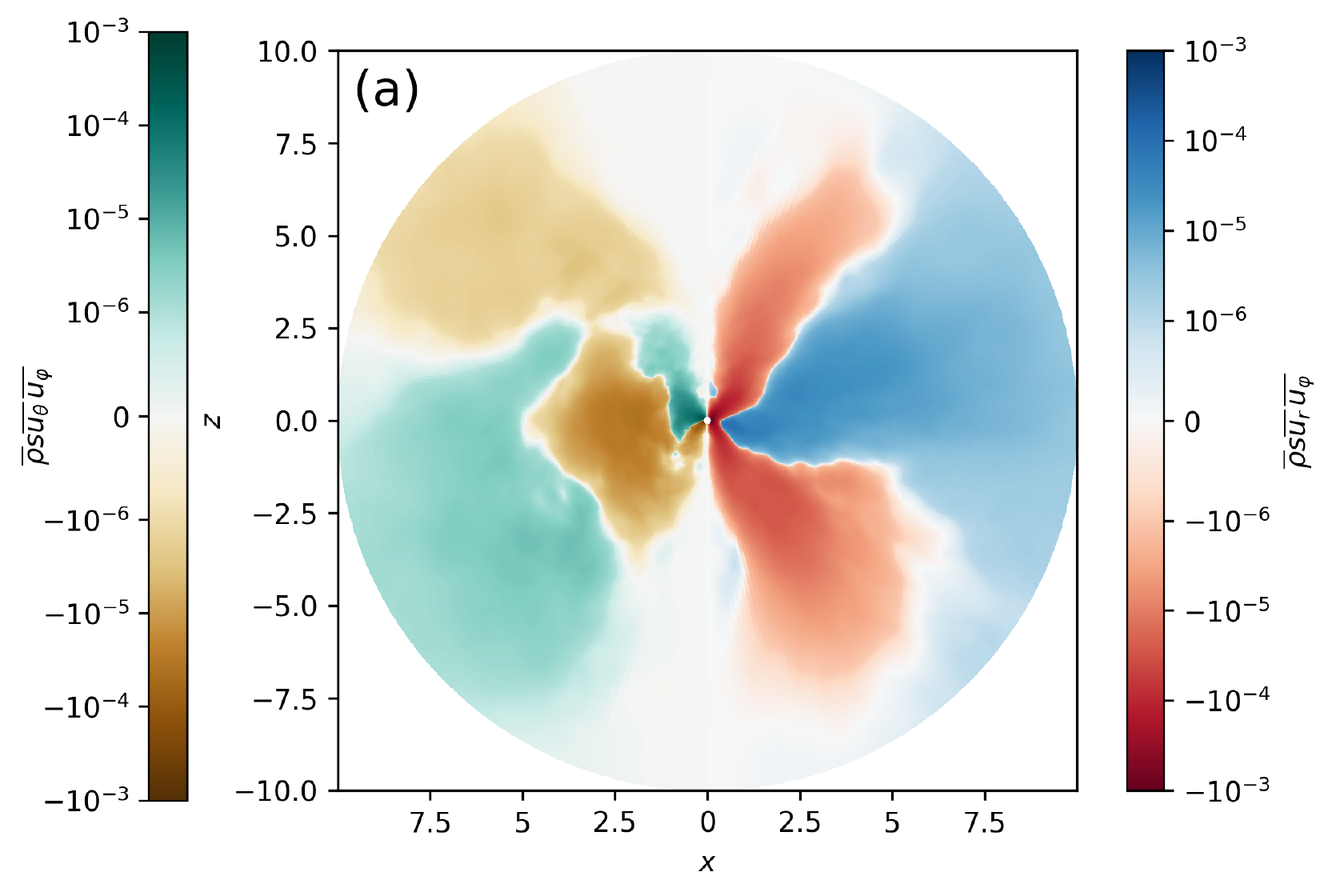} 
\includegraphics[width=\columnwidth]{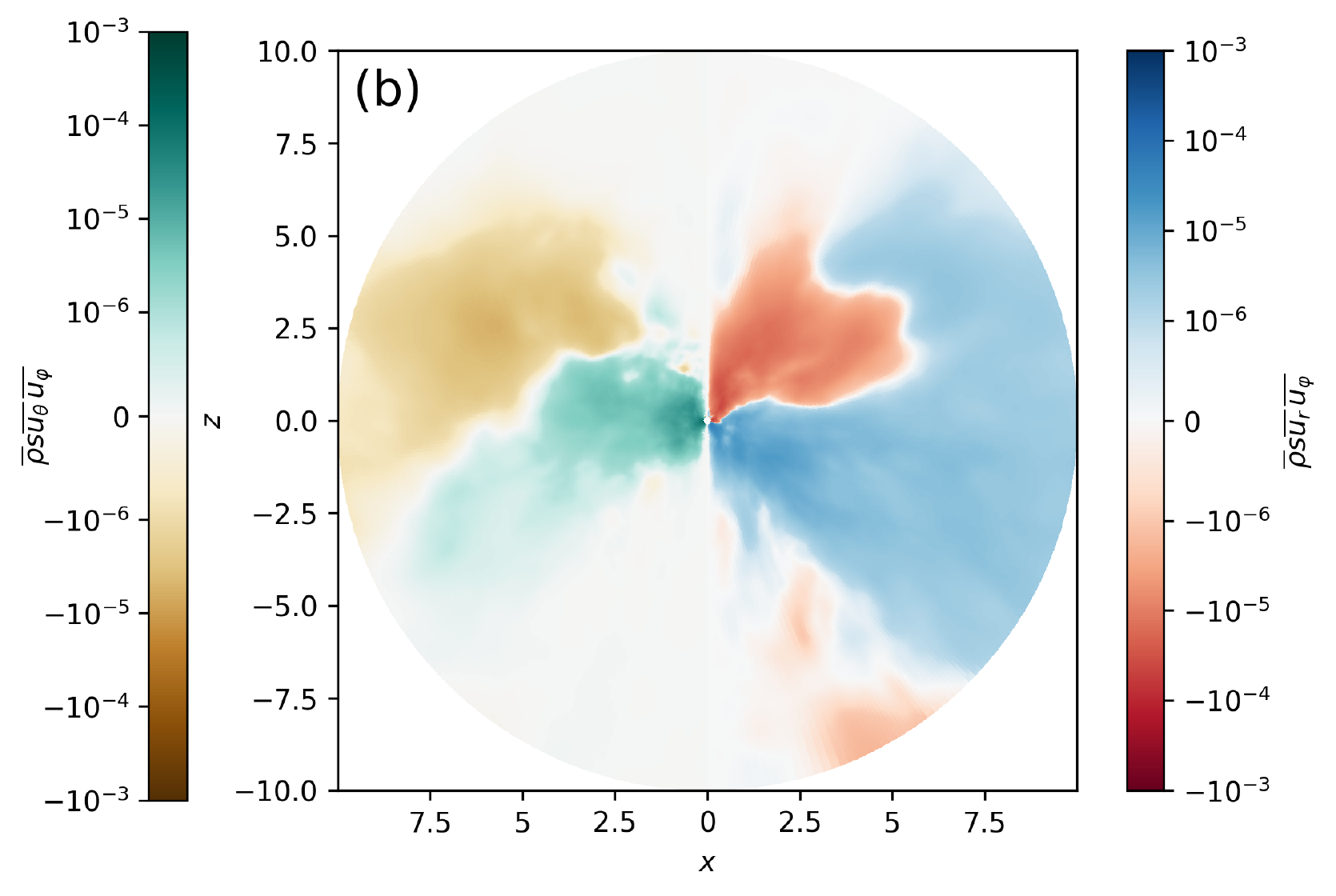} \\
\includegraphics[width=\columnwidth]{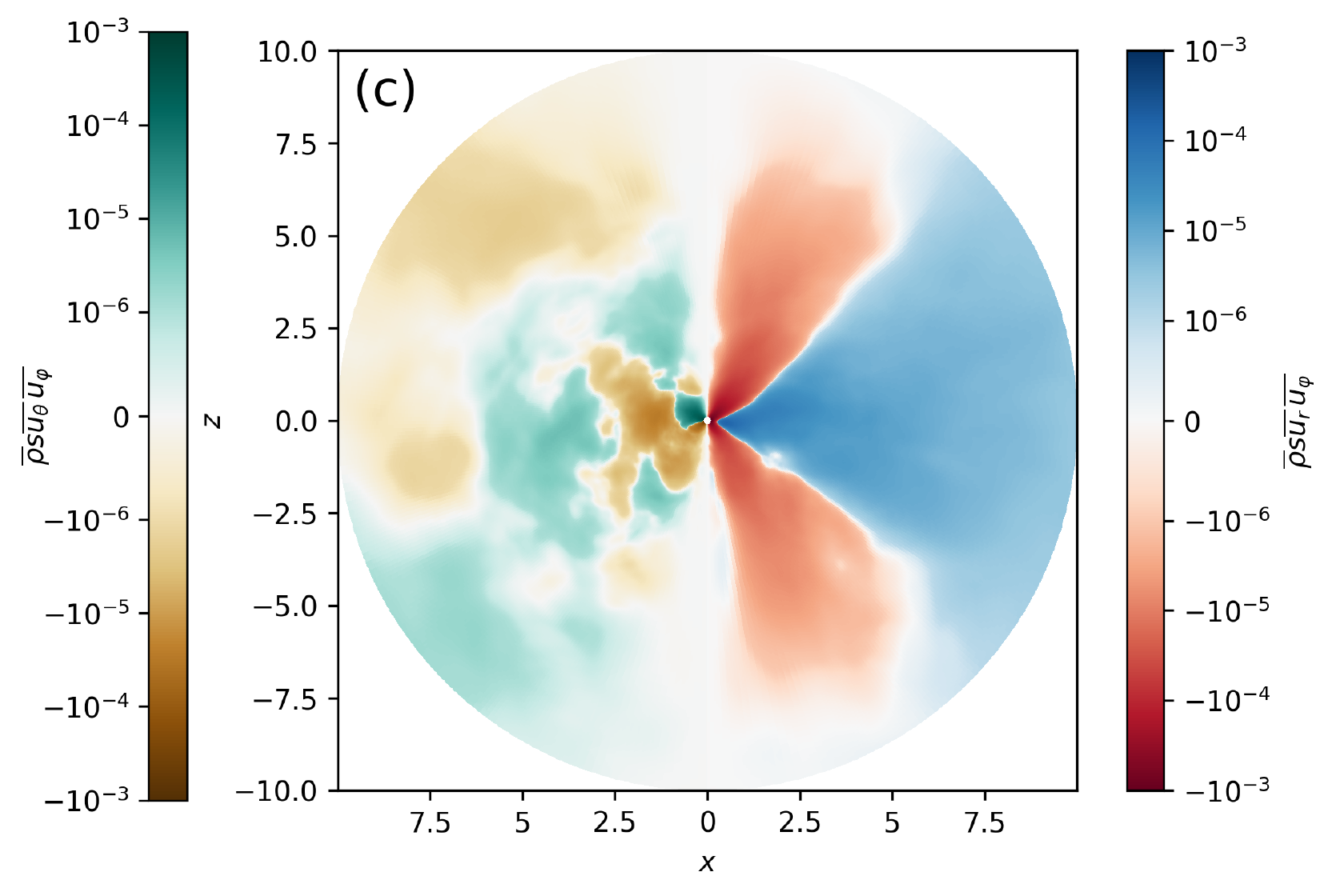} 
\includegraphics[width=\columnwidth]{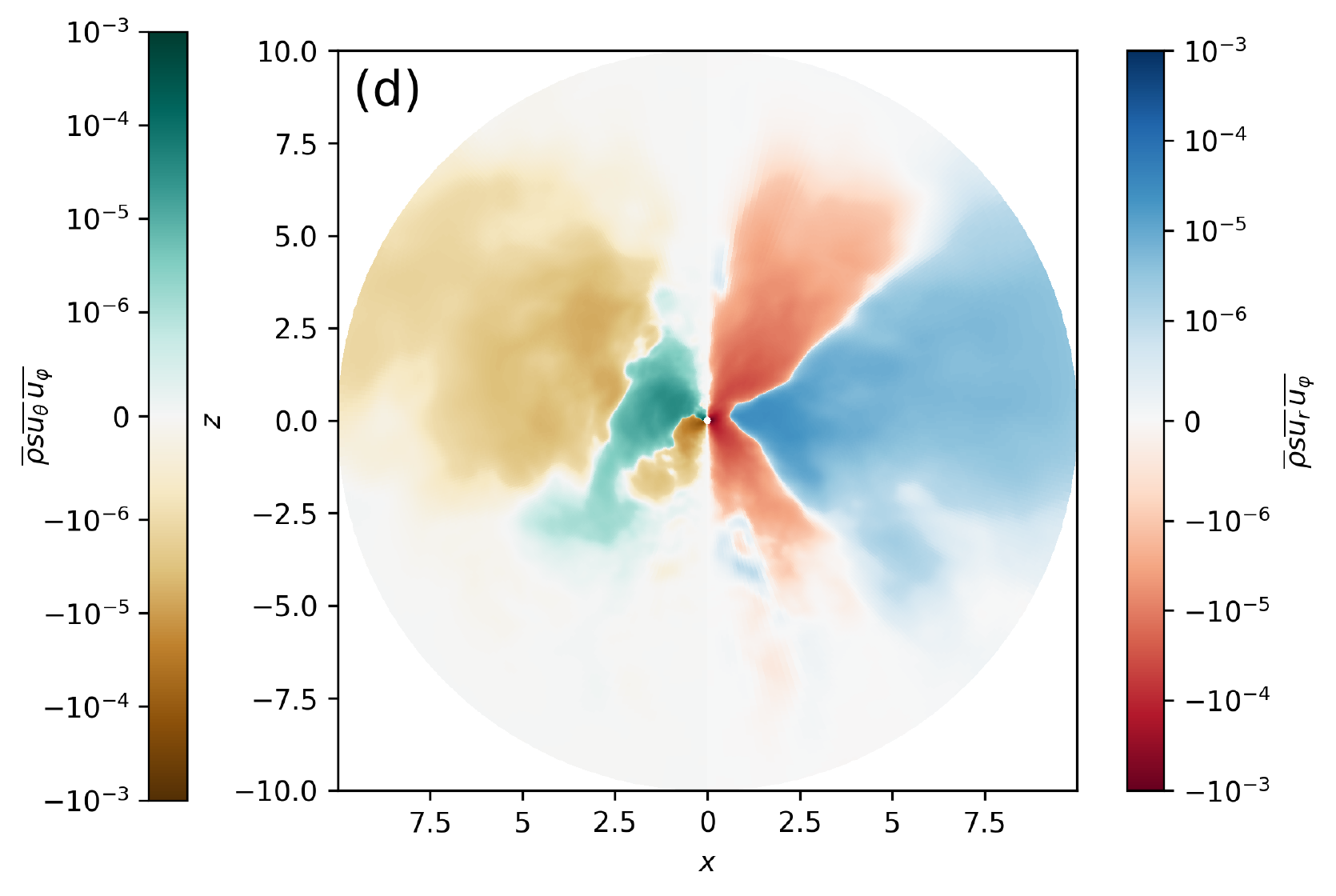}
   \caption{Mean flow contribution to the azimuthally averaged advective angular momentum radial and latitudinal fluxes averaged over time interval $250 \le t/P_{\rm orb} \le 275$ for models A (panel a), A' (panel b), B (panel c), and C (panel d).}
\label{fig:F_A2}
\end{figure*}

\begin{figure*}[t]
\includegraphics[width=\columnwidth]{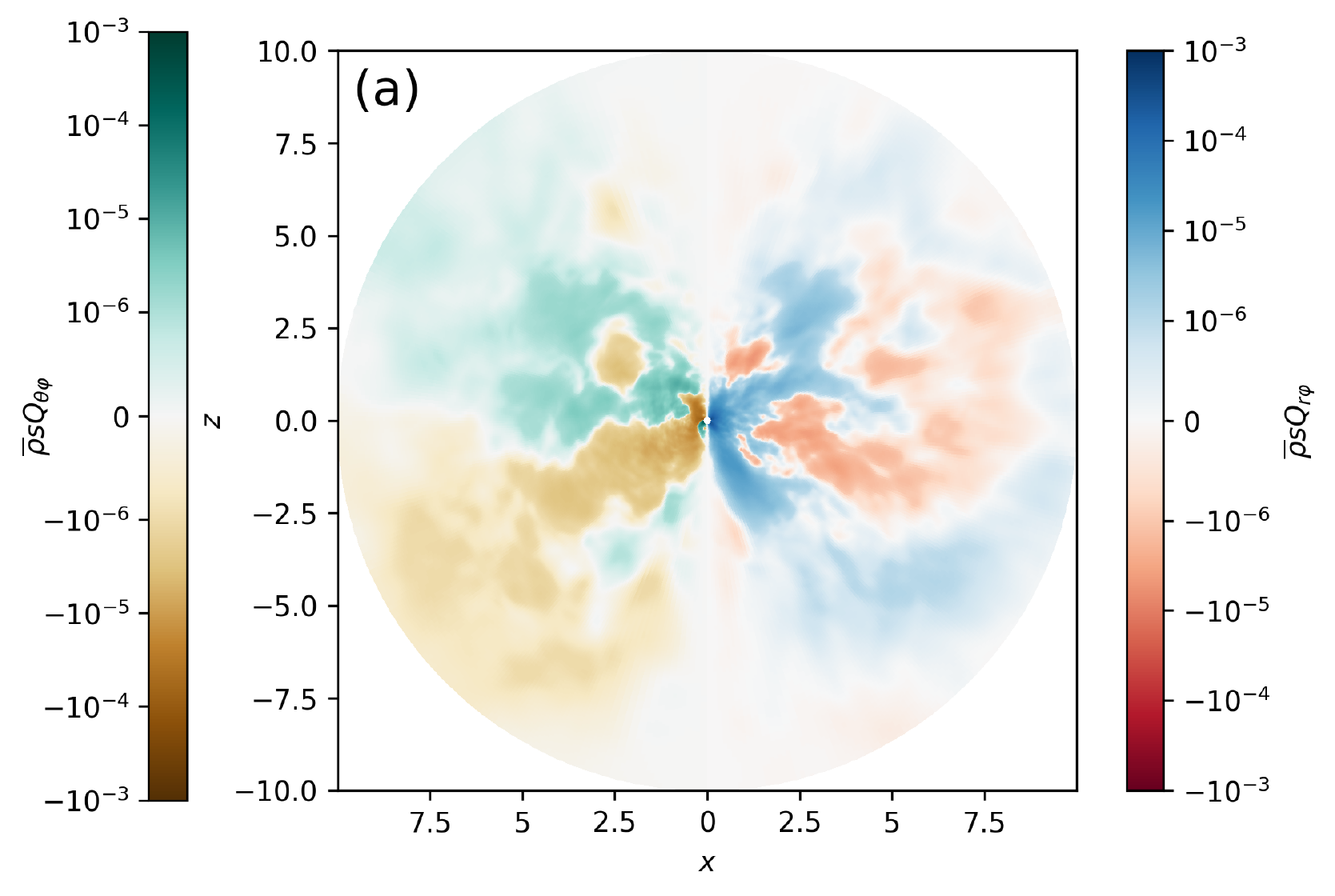} 
\includegraphics[width=\columnwidth]{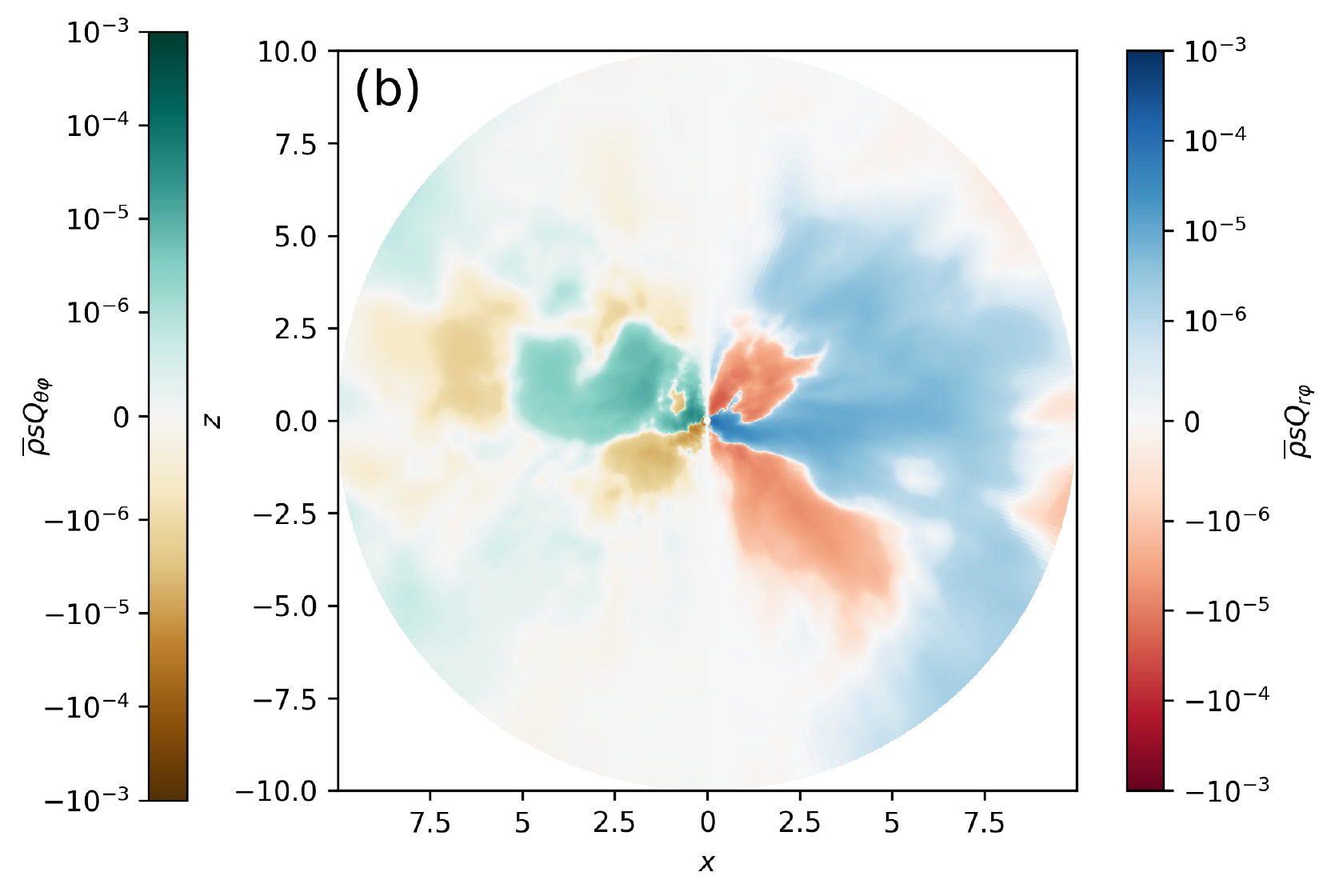} \\
\includegraphics[width=\columnwidth]{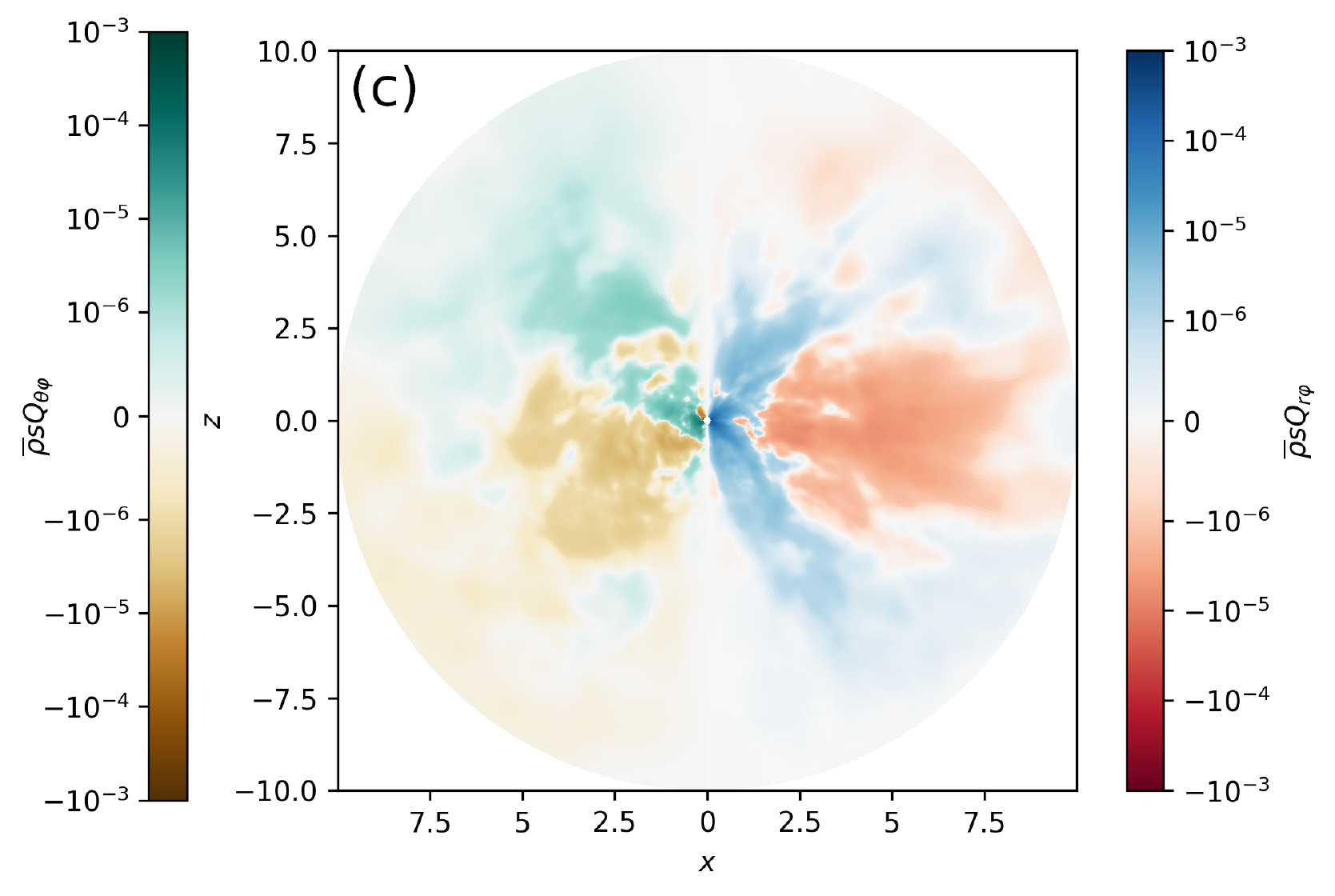} 
\includegraphics[width=\columnwidth]{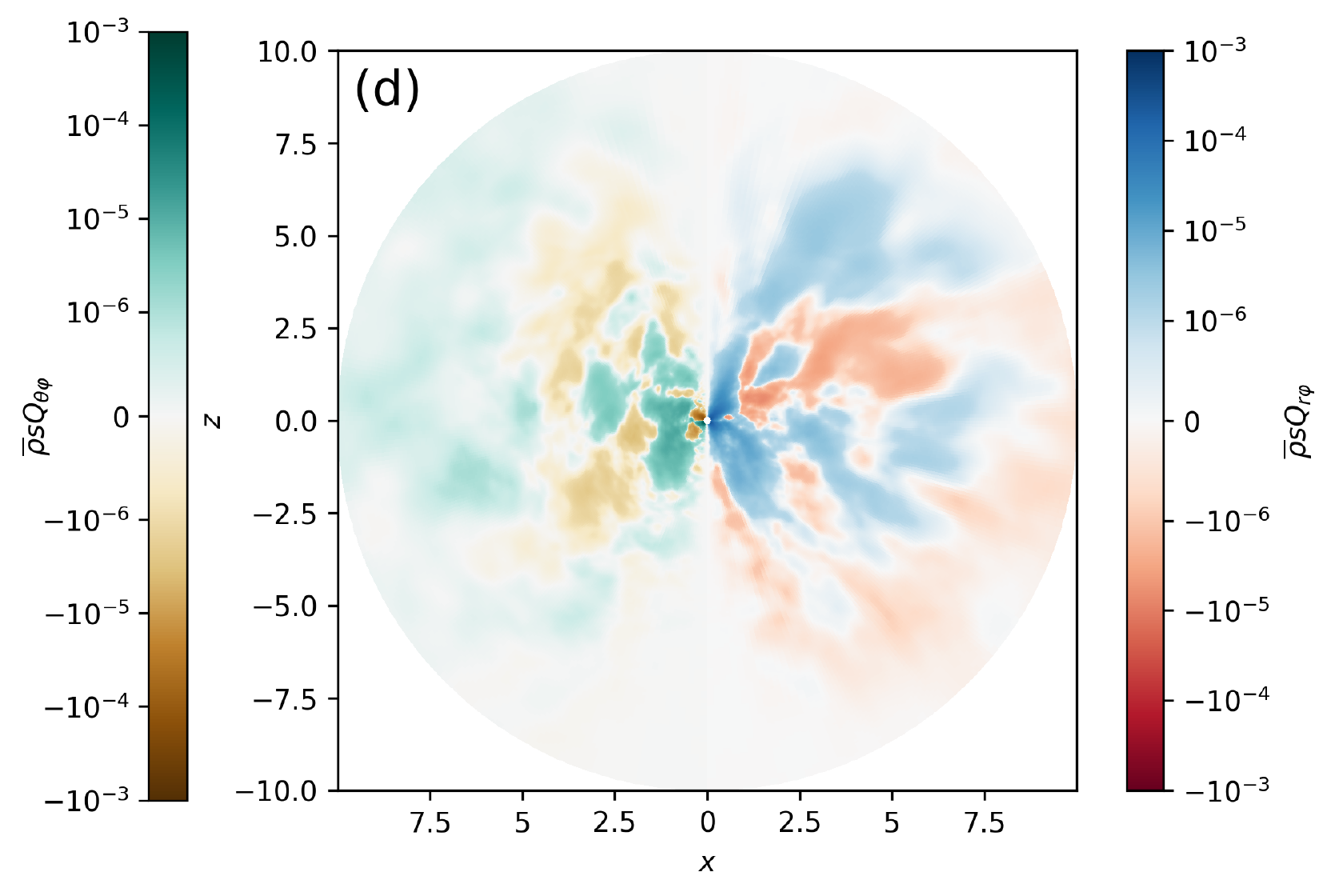}
\caption{Turbulent flow contribution to the azimuthally averaged advective angular momentum radial and latitudinal fluxes averaged over time interval $250 \le t/P_{\rm orb} \le 275$ for models A (panel a), A' (panel b), B (panel c), and C (panel d).}
\label{fig:F_A3}
\end{figure*}

We can use hydrodynamic mean-field theory to asses the turbulent fluxes of angular momentum \citep[e.g.,][]{Kapyla2019,Rudiger2022}. Taking the azimuthal average of the angular momentum equation in the $z$-direction, using Reynolds decomposition to define velocity fluctuations  about  their  averages as $ u_i' = u_i - \overline{u_i}$, and ignoring density fluctuations from its mean value, we obtain the conservation law  
\begin{equation}\label{eq:AMcons}
\begin{aligned}
    \frac{\partial (\overline{\rho s u_\varphi})}{\partial t}  \simeq &  - \bnabla \cdot \left[ s \left(  \overline{\rho}\,\overline{\buu_\perp} \overline{u_\varphi}  + \overline{\rho} \overline{\buu' u_\varphi'}  - 2 \left( \overline{\rho \nu}\boldsymbol{\overline{S}} + \overline{\rho \nu \boldsymbol{S}' }\right) \cdot \be_\varphi   \right)\right] \\ & - \overline{\rho \frac{\partial \Phi}{\partial \varphi}} \ ,    
\end{aligned}
\end{equation}
where  $\overline{\buu_\perp} = ( \overline{u_r}, \overline{u_\theta})$ is the meridional mean velocity, overlined quantities indicate azimuthal average, and $\overline{\boldsymbol{S}}$ and $\boldsymbol{S}'$ are the rate-of-strain tensors of the mean flow and of the fluctuating flow defined as
\begin{align}
\overline{S_{ij}} &= \frac{1}{2}  \left( \partial_i \overline{u_j }+ \partial_j \overline{u_i} - \frac{2}{3} (\bnabla \cdot \overline{\buu}) \delta_{ij} \right)\ ,  \\  \quad S'_{ij}&= \frac{1}{2}  \left( \partial_i u_j'+ \partial_j u_i' - \frac{2}{3} (\bnabla \cdot \buu') \delta_{ij} \right) \ .
\end{align}
The first term on the right-hand side of Eq.~(\ref{eq:AMcons}) corresponds to the advective transport by large scale meridional flow, the second represents the turbulent meridional flux, and the third term is the viscous transport. Turbulent angular momentum fluxes are often described using Reynolds stress,
\begin{equation}\label{eq:Qij}
    Q_{ij} = \overline{ u_i'  u_j'} \ .
\end{equation}
This stress tensor is often separated into nondiffusive (``$\Lambda$-effect'')  and diffusive contributions described by turbulent viscosity,
\begin{equation}
    Q_{ij} = Q_{ij}^{\Lambda} + \mathcal{N}_{ijkl} \frac{\partial \overline{u_k}}{\partial x_l} \ ,
\end{equation}
where $ Q_{ij}^{\Lambda}$ is the nondiffusive part and $\mathcal{N}_{ijkl}$ is the (turbulent) viscosity tensor \citep[e.g.,][]{Kitchatinov1994,Kitchatinov1995,Kapyla2011,Rudiger2022}. Even in the case where we do not prescribe subgrid viscosity, an effective (convective) turbulent viscosity still exists and can be derived from the expression of the turbulent viscosity tensor $\mathcal{N}_{ijkl}$. Conversely, when we prescribe $\nu > 0$ (simulation run D), an additional effective viscosity associated with the simulation's intrinsic turbulence still exists. In this case, the total effective viscosity is given by the sum of the two contributions. The disentangling of the two contributions to the Reynolds stress and the measurement of the associated simulation's intrinsic effective turbulent viscosity is however beyond the scope of this work. Instead, we focus on the total stress. 

We assume that the turbulent velocity $\buu'$, which is the deviation from the mean velocity, is the nonaxisymmetric component of the fluid flow velocity. However, because the large scale flow resulting from the gravitational torque exerted by the binary orbit is itself nonaxisymmetric, the contribution of turbulence to angular momentum transport is likely overestimated, especially in the close vicinity of the binary. Unfortunately, there is no straightforward way to establish what the mean flow is in our simulations. This is an issue also in the context of accretion and CBDs, where \cite{Hawley2000}, \citet{Hawley2001}, \citet{Shi2012}, and \cite{Armengol2021} use departure from density weighted shell average to compute velocity perturbations. Still, one could extract the actual turbulent flow with reasonable accuracy by filtering out the large scale flow using Fourier and inverse transforms \citep[e.g.,][]{Kapyla2011}. This is however beyond the scope of this work and we refer to the nonaxisymmetric perturbation $\buu'$ as the turbulent fluid flow velocity, though one has to keep in mind that this may be inaccurate in the binary close vicinity. 

In Fig.~\ref{fig:F_A1}, we show the meridional components of the azimuthally averaged advective fluxes of the total angular momentum $\boldsymbol{F}_{\rm adv} =  s\overline{\rho}\,\overline{\buu_\perp u_\varphi} = s \overline{\rho} ( \overline{\buu_\perp}\,\overline{u_\varphi} + \overline{\buu' u_\varphi'})$. In Figs.~\ref{fig:F_A2} and \ref{fig:F_A3}, we show the mean and turbulent flow contributions, $s \overline{\rho}\,\overline{\buu_\perp}\,\overline{u_\varphi}$ and  $s \overline{\rho}\,\overline{\buu' u_\varphi'}$. We see that for accreting models, the mean axisymmetric flow results in outward angular momentum advective flux in a equatorial disk-like structure with an opening angle that is smaller for higher initial angular momentum. Outside of the disk-like structure, the angular momentum advective flux points inward. The morphology of the radial turbulent transport of angular momentum is more complicated, because it changes sign in both cylindrical directions $s$ and $z$. Such disk-like structure is not present in our nonaccreting model A'. Indeed, the inward flow is deflected by the inner boundary and any polar mass flux asymmetry between northern and southern hemispheres, however small, is amplified and can even lead to a polar outflow in one of the hemispheres.

\begin{figure}[t]
      \includegraphics[width=0.5\textwidth]{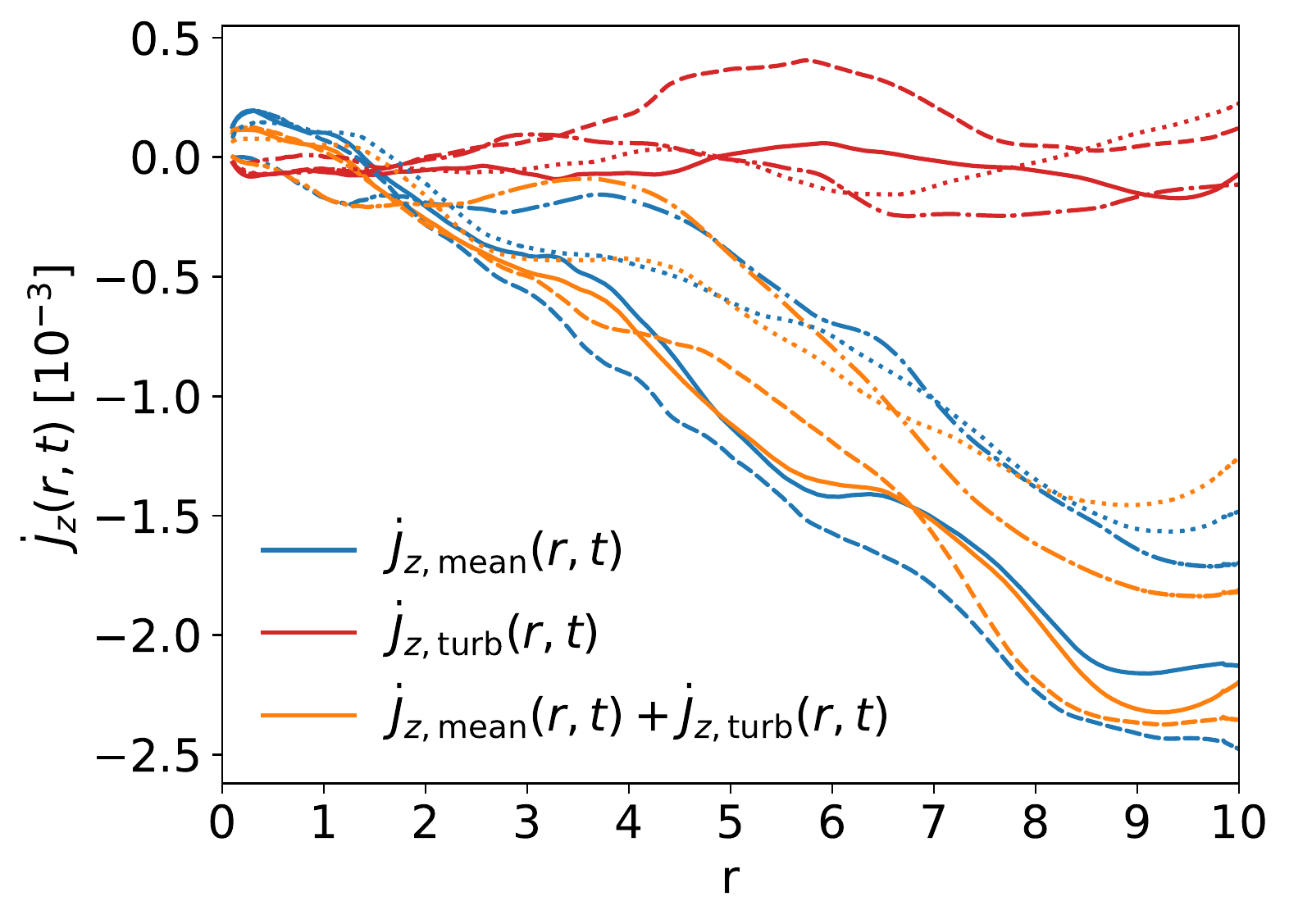} \\
      \caption{Mean and turbulent components of the advective contribution to local angular momentum transfer rate for models A (full lines), A' (dash-dotted lines),  B (dashed lines), and C (dotted lines). The quantities are averaged over a time interval $250 \le t/P_{\rm orb} \le 275$. Neglecting density perturbations, $\dot{J}_{z, \rm adv} = \dot{J}_{z, \rm mean}+\dot{J}_{z, \rm turb}$.}
\label{fig:turbprofAB}
\end{figure}

In Fig.~\ref{fig:turbprofAB}, we show the radial profile of the mean-flow and turbulent contributions to the angular momentum transfer in terms of Reynolds stress,
\begin{align}
  \dot{J}_{z,\rm mean} &= - \int_{\partial r} \overline{\rho} s \overline{u_r}\overline{u_\varphi} \dd S\ ,\\  \dot{J}_{z,\rm turb} &= - \int_{\partial r} \overline{\rho} s Q_{r\varphi} \dd S \ .
\end{align}
We find that for our three inviscid and accreting runs, the net radial angular momentum transport is essentially dominated by the contribution from the mean flow, which is directed inward for $r \lesssim 1.5$ and outward for $r \gtrsim 1.5$.

The mean axisymmetric flow also leads to angular momentum advective transport in the $\theta$ direction. Specifically, such mean flow advects angular momentum toward the orbital plane in the inner part of the envelope and away from it further out. Conversely, turbulent flow advects angular momentum away from the midplane in the close vicinity of the binary and toward it in the rest of the envelope. Overall, the structure of the total angular momentum flux follows the mean flow contribution, where the angular momentum is advected toward the orbital plane in the inner envelope and away from the orbital plane far from the binary.

\subsubsection{Vertical eddy scales}
\label{sec:vertical_eddy_scales}

Since we are interested in the ability of turbulent structures to transport angular momentum radially in the envelope, we aim to estimate the typical vertical scale of turbulent convective eddies exchanging angular momentum with one another. To make sure that we properly isolate turbulent flow, we focus on its latitudinal component in the orbital plane. This is because the nonaxisymmetric contribution of the large-scale mean-flow, which pollutes the inferred turbulent velocity, results from the envelope's response to the gravitational perturbations exerted by the binary and is zero in the $\theta$-direction in the orbital plane.  Let us first introduce the normalized auto-correlation of the turbulent latitudinal velocity on the orbital plane, which we azimuthally average and we  integrate over an arbitrary radial domain $[r_{\rm min},r_{\rm max}]$,
\begin{equation}\label{eq:autocorr}
     R_{\theta \theta}(r',t) = \frac{ \int_{r_{\rm min}}^{r_{\rm max} - r'} \overline{u_\theta'(r,\pi/2,\varphi) u_\theta'(r+r',\pi/2,\varphi) }\dd r}{\int_{r_{\rm min}}^{r_{\rm max} - r'} \overline{u_\theta'(r,\pi/2,\varphi) u_\theta'(r,\pi/2,\varphi) }\dd r}  \ .
\end{equation}
We also introduce the integral radial scale \citep[e.g.,][]{Townsend1976,Oneill04,Mora2020},
\begin{equation}\label{eq:lamb}
    \Lambda(t) = \int_{0}^{\infty}  R_{\theta \theta}(r',t) \dd r' \ .
\end{equation}
Because the low amplitude tail of Eq.~(\ref{eq:autocorr}) may contain information extraneous to turbulent motion,  we integrate Eq.~(\ref{eq:lamb}) up to the first zero-crossing of $R_{\theta \theta}(r',t)$, as is commonly done in experimental and numerical fluid dynamics \citep[][]{Oneill04}. In homogeneous turbulence, $\Lambda$ can be interpreted as the typical radial scale of the energy-containing turbulent eddies. Because the mean density stratification becomes weaker as the distance from the central binary increases, turbulence is not homogeneous in our simulations, and the interpretation of $\Lambda$ is more ambiguous. Here, $\Lambda$ represents the weighted average of all turbulent radial scales in the flow with a dominant contribution from eddies containing higher energy. In an effort to mitigate this ambiguity, we integrate the auto-correlation of the azimuthally averaged turbulent latitudinal velocity on the orbital plane in three arbitrary regions of strong, moderate, and weak stratification labeled  I ($r \leq R_{\rm domain}/10$), II ($R_{\rm domain}/10 < r \leq R_{\rm domain}/2 $), and III  ($r > R_{\rm domain}/2 $). We interpret the resulting integral scale as the typical eddy scale in each region.

\begin{figure}[t]
\centering
\includegraphics[width=0.5\textwidth]{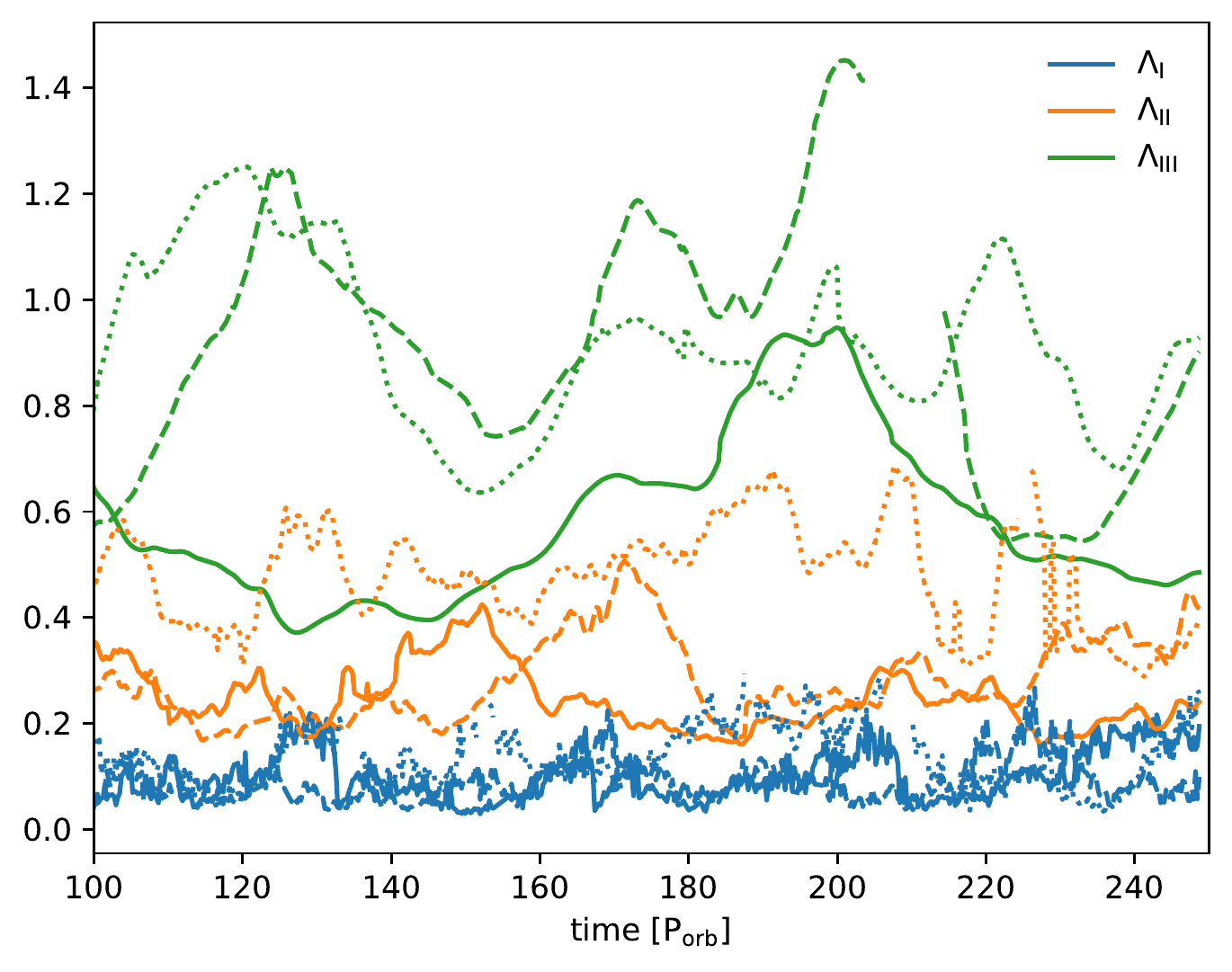} 
   \caption{Time evolution of the integral radial scales $\Lambda$ (Eq.~(\ref{eq:lamb})) in three regions  I ($r \leq R_{\rm domain}/10$), II ($R_{\rm domain}/10 < r \leq R_{\rm domain}/2 $), and III  ($r > R_{\rm domain}/2 $) for simulation runs A (full lines), B (dashed lines), and C (dotted lines). The discontinuity of $\Lambda_{\rm III}$ for run A is a result of $u_\theta'$ not changing sign on the same radial scales for all $\varphi$, that is  $R_{\theta \theta}(r',t)$ does not cross zero during a few orbital periods. }
\label{fig:lamba_l}
\end{figure}

In Fig.~\ref{fig:lamba_l}, we show the time evolution of $\Lambda$ in the three regions for our three inviscid simulation runs. In Table~\ref{tab:runs}, we provide the time-averaged values, $\Lambda^{t_\text{min}-t_\text{max}}$, where $t_\text{min}$ and $t_\text{max}$ denote the time interval of averaging in the units of $P_\text{orb}$. We see that in all three models the vertical extent of turbulent eddies decreases outward as the stabilizing effect of density stratification decreases. However, depending on the amount of angular momentum present in the envelope at the onset of the post-dynamical inspiral phase (parameter $\beta$), the radial dependence of turbulent eddy vertical scales varies substantially. The more angular momentum the secondary star injects into the shared envelope during dynamical inspiral, the more the envelope gets deformed by centrifugal forces. Hence, the injected angular momentum modifies density stratification by expanding the envelope anisotropically and affects envelope stability through the sign of angular momentum gradient (Eq. (\ref{eq:solb1})). As a result, we find that envelopes with lower angular momentum content at the onset of the post-dynamical phase end up being less effectively stratified, which leads to a reduction of their ability to limit the vertical extent of turbulent eddies. However, it is important to note that in region I, $\Lambda_{\rm I}^{\rm 140-250} \simeq 0.6~a_\text{b}$ for both simulation runs A and B while $\Lambda_{\rm I}^{\rm 140-250} \simeq 0.877~a_\text{b}$ for run C. The fact that we observe similar eddy scales for runs A and B with different initial total angular momentum in this inner region is due to our initial spin-up setup where the innermost layers are spun-up to critical rotation, while they remain subcritical during spin-up for run C. Additionally, we further note that far away from the binary in region III, simulation runs B and C yield similar vertical eddy scale. This roughly constant value likely constitutes a limit at very low stratification that may depend on the numerical size of the domain \citep[see also][]{Garaud2017}. This limit formally implies that the assumption of a vertical eddy scale that is proportional to the local pressure scale height \citep[e.g.,][]{Vitense1953,Zahn89} may fail in the limit of low stratification as the effective viscosity would locally tend to infinity.

\subsubsection{The role of viscosity}
\label{sec:visc}

\begin{figure}[t]
      \includegraphics[width=0.5\textwidth]{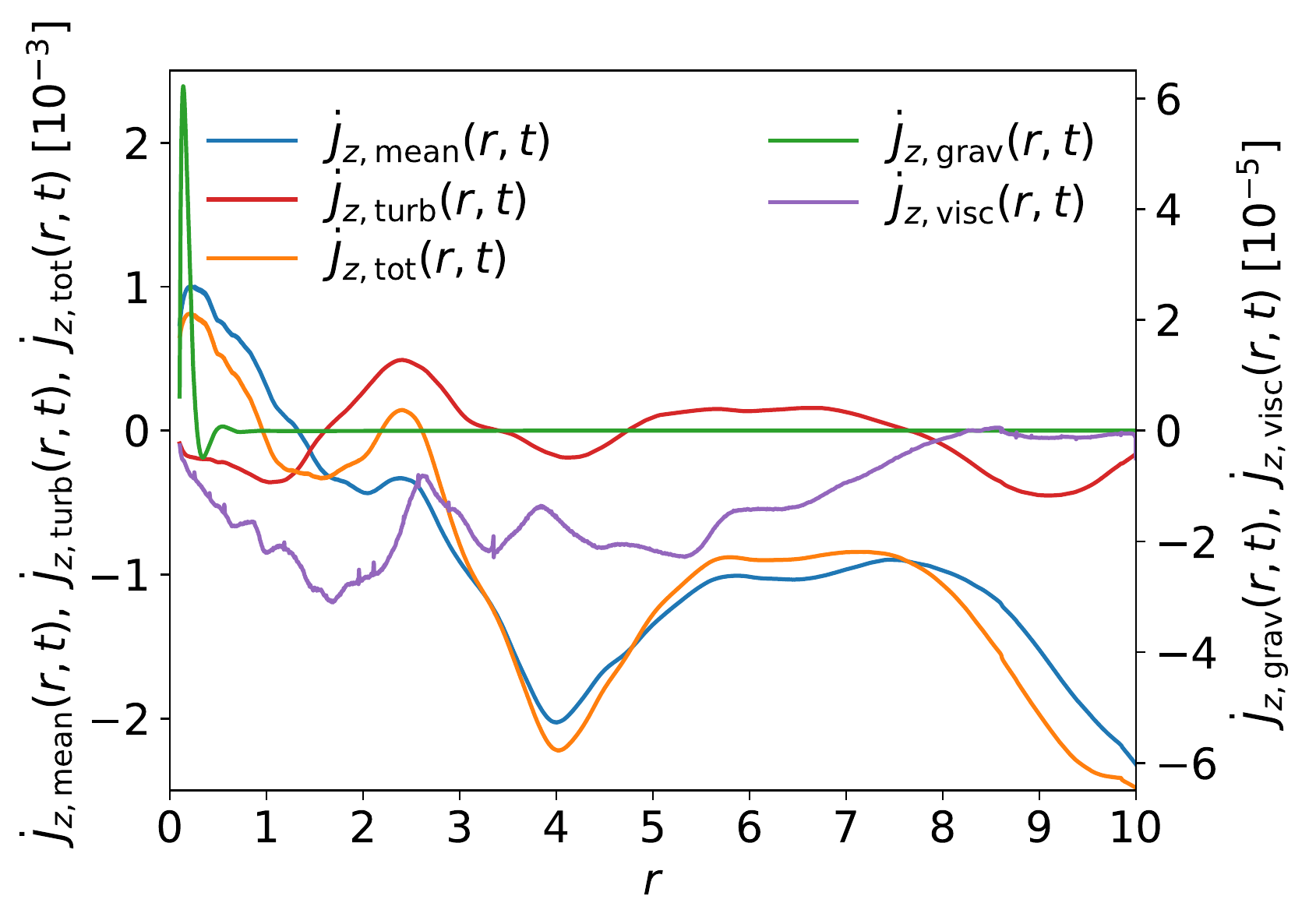} \\
      \caption{Mean, turbulent, gravitational, and viscous contributions to local angular momentum transfer rate across the common envelope for model D, averaged in time from $t = 140~P_{\rm orb}$ to $t = 150~P_{\rm orb}$.}
\label{fig:torqprofD}
\end{figure}

\begin{figure}[t]
\includegraphics[width=0.49\textwidth]{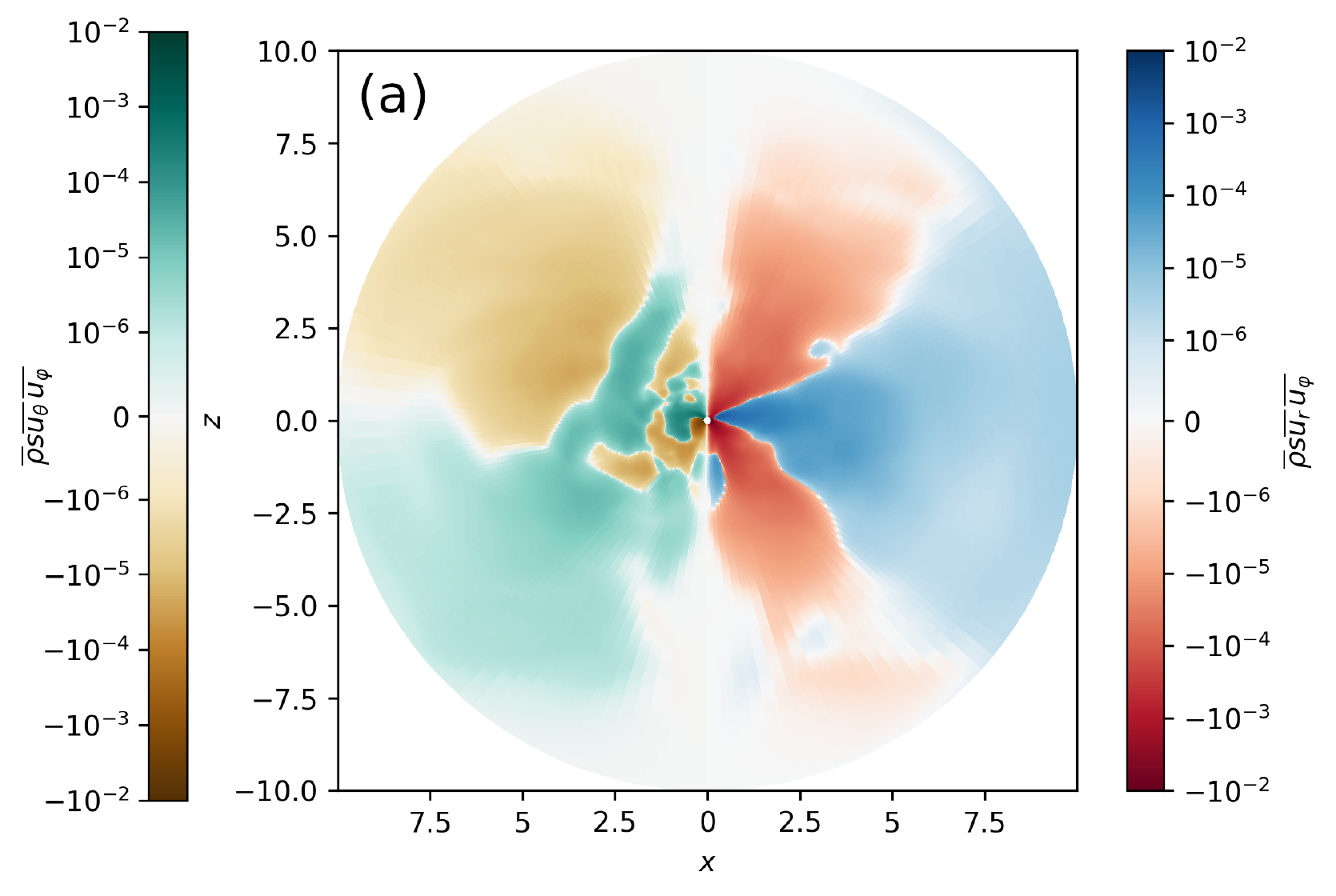} \\
\includegraphics[width=0.49\textwidth]{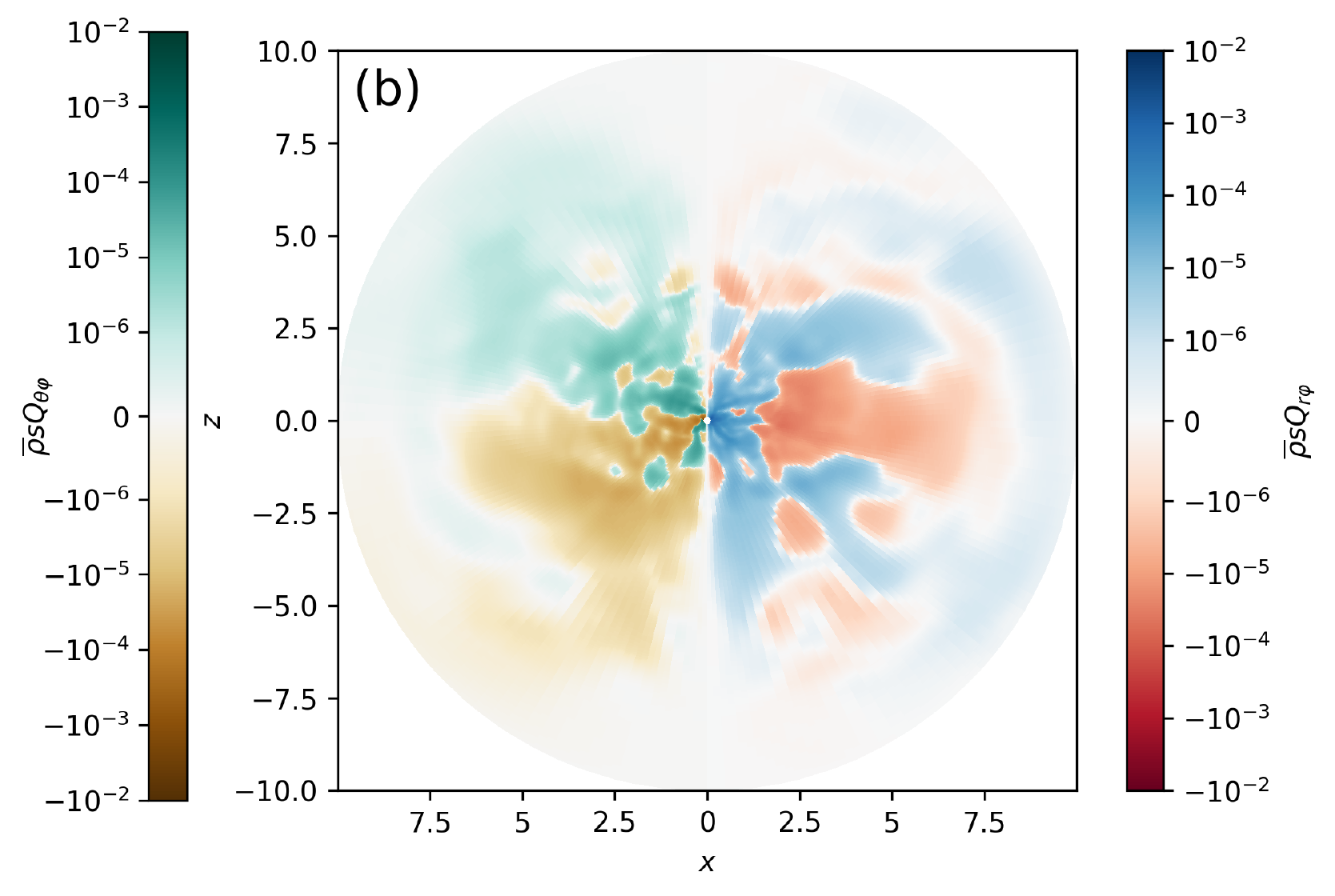} \\
\includegraphics[width=0.49\textwidth]{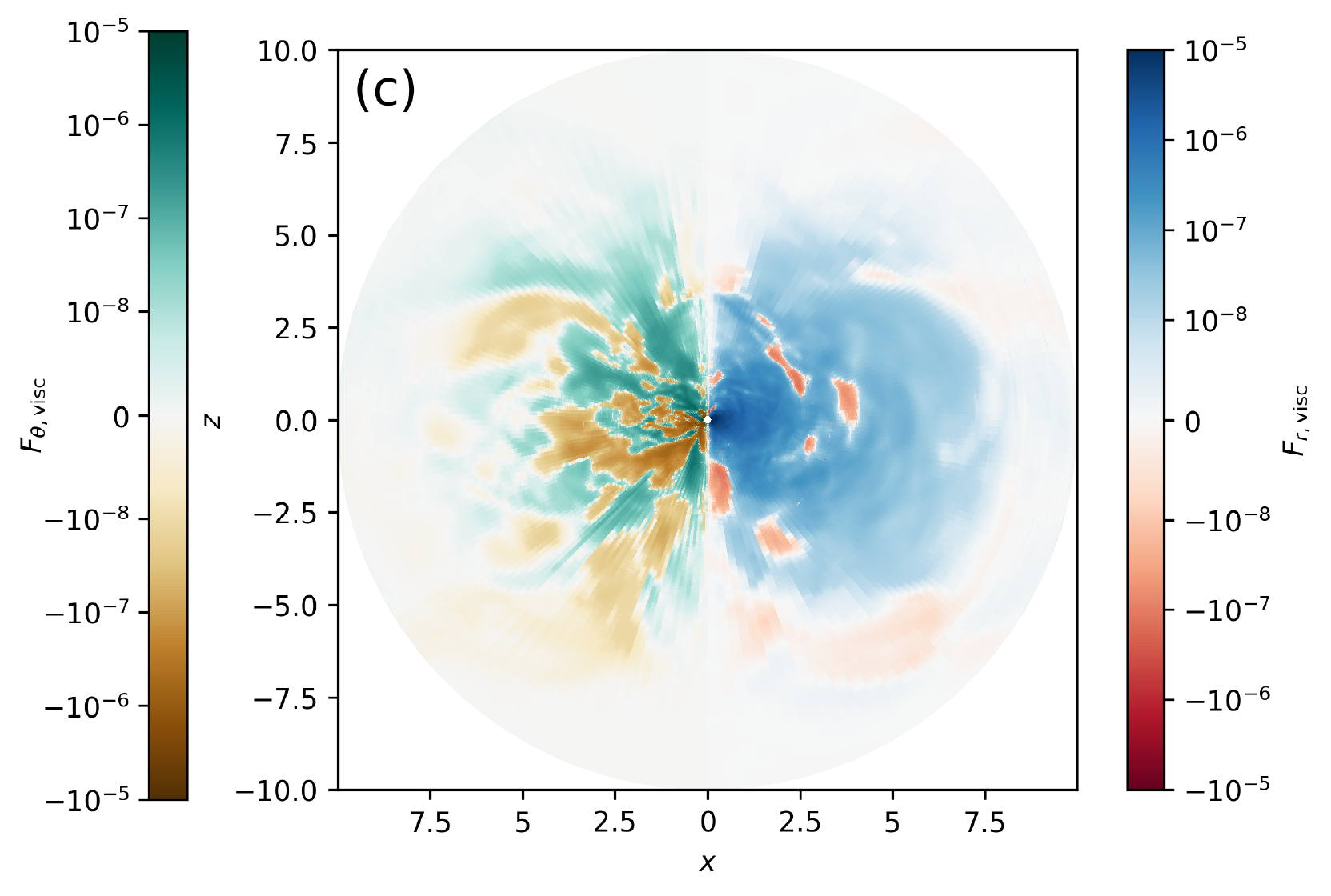}
   \caption{Mean flow (a), turbulent flow (b), and viscous (c) contributions to the azimuthally averaged advective radial and latitudinal angular momentum fluxes for the viscous model D. We averaged the quantities over ten orbital periods for  $140 \le t/P_{\rm orb}  \le 150$.}
\label{fig:Fvisc}
\end{figure}

In simulation run D, we prescribed an isotropic viscosity whose effects on the envelope dynamics add to those from the effective viscosity associated with Reynolds stresses.
In Fig.~\ref{fig:torqprofD}, we show the mean, turbulent, and viscous contributions to local angular momentum transfer rate across the common envelope for model D and in Fig.~\ref{fig:Fvisc} we show the mean flow, Reynolds stress, and viscous contributions to the azimuthally averaged advective radial and latitudinal angular momentum fluxes.
We see that for our  value of $\alpha_\nu$, the azimuthally averaged total angular momentum viscous flux is essentially directed outward, and only plays a minor role in the radial and latitudinal transport of angular momentum in the shared envelope. Hence, simulation run D retains the same angular momentum transport features as our inviscid runs.
This result reflects the efficiency of the transport by the mean flow and by the Reynolds stress transport and suggests that the  effective viscosity associated with the latter is likely several orders of magnitude larger than $\nu$.
Still, despite being small, the prescribed viscosity has a stabilizing effect on the shear flow in the inner envelope, and the ratio between turbulent and mean components of the advective angular momentum transfer rate is, at a given time, smaller in simulation run D. Such stabilizing effect of viscosity also locally delay the onset of turbulence.

\section{Discussions}\label{sec:discussions}

\subsection{Comparison with CBDs}

In this work, we draw many analogies between simulations of CBDs and post-dynamical stages of CEE and it is thus instructive to briefly discuss differences and similarities between these situations. First, there are significant differences in the origin of the gas surrounding the binary. CBDs can often occur as remnants of star formation out of molecular clouds or they are thought to accompany orbiting super-massive black holes \citep{Begelman1980,Bate1997,Milosavljevic2005,Matsumoto2019}. In this situation, the density distribution and angular momentum content of the disk depends not only on the properties of the binary, but perhaps more significantly on the accretion for larger distances. Conversely, CEE can often be regarded as an isolated object, where the formation process inextricably links together the distribution of density, energy, and angular momentum in the envelope with the properties of the central binary. Second, CBDs are often observed and simulated as relatively optically and geometrically thin, 2D objects. Instead, the post-dynamical shared envelope in CEE contains large amount of mass, which prevents cooling and keeps the geometry strictly three-dimensional. Ultimately, the shared envelope disperses and whatever gas remains should cool to a thin disks, as is observed in post-AGB binaries \citep[e.g.,][]{Dermine2013,Kluska2022}. The transition between these two regimes of post-dynamical CEE should be a subject of future study.
Finally, thermal convection is weak or even absent in CBDs, where it is the magnetorotational instability \citep[MRI,][]{Balbus1991} that is instead often recognized as the main source of turbulence \citep[e.g.,][]{Cabot1996,Stone1996,Balbus1998} modeled with a turbulent effective viscosity using the $\alpha$ ansatz \citep[][]{SS73}. Conversely, common envelopes are expected to be vigorously convective, making thermal convection inevitable \citep[e.g.,][]{Soker1993,Ohlmann2016,Sabach2017,Grichener2018,Wilson2019}.

Despite these fundamental differences, we have shown that there are similarities and even commonalities between these two systems. We found that mass and angular momentum accretion onto the central binary (when allowed) has the same temporal variability with two characteristic frequencies. The first frequency is associated with the quadrupolar moment contribution to the binary potential, while the second one with the formation and propagation of overdensities, which share many common characteristics with the lump located near the cavity edge in CBD simulations. Because of the complicated geometry of common envelopes,  we have shown that a local analysis of the accretion flux is necessary to understand its short-term variability. The behavior of the orbital separation evolution is dictated by the same condition for the two problems, specifically, $j < 3/8$ gives orbital contraction when $q=1$ and $e_\text{b}=\dot{e}_\text{b}=0$. This condition  suggests predominant orbital contraction in CEE simulations, while  orbital expansion is possible for a wide orbital parameter range in CBD simulations \citep[e.g.,][]{Miranda2017,Munoz2019}.
Finally, we found that the shared envelope develops eccentricity, which grows with an exponential growth rate that is of the same order as that obtained by \cite{Shi2012} in the context of CBDs and which saturates to reach a statistically stationary value as is also seen in CBD simulations \citep[e.g.,][]{Miranda2017,Munoz2020}. To summarize, the abundant literature and ongoing work in the field of accretion and CBDs can be of precious help to better understand the post-dynamical inspiral phase of CEE.

\subsection{Implications for CEE}
\label{sec:cee_impl}

\begin{figure}
    \centering
    \includegraphics[width=0.48\textwidth]{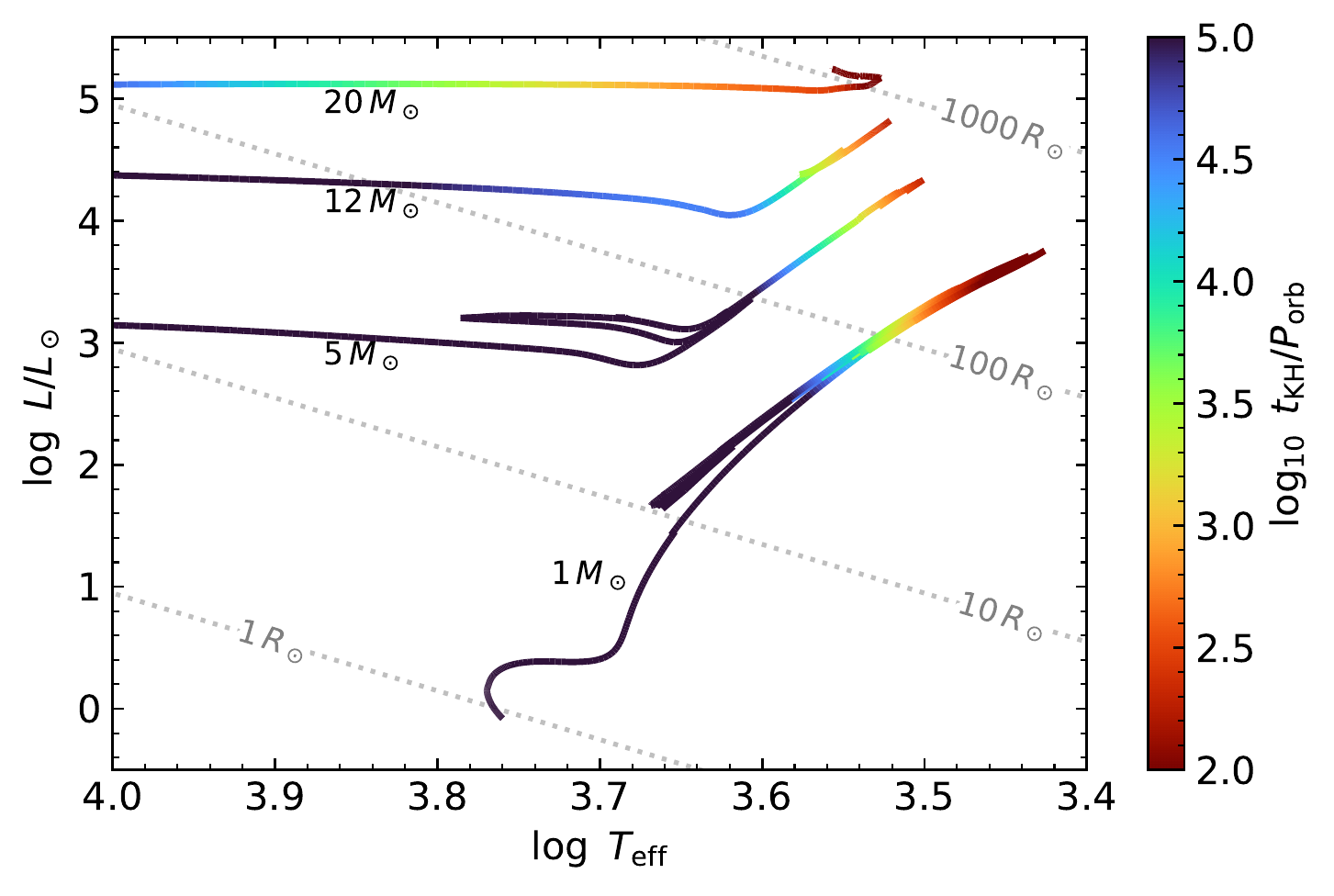}
    \caption{Ratio of thermal timescale $t_\text{KH}$ to the orbital period of the binary $P_\text{orb}$ inside the shared envelope in the Hertzprung--Russel diagram constructed for four solar-metallicity nonrotating evolutionary tracks from the MIST database \citep{Dotter2016,Choi2016}. Here, $t_\text{KH} = G(M_1+\menv)^2/(2RL)$, where $L$ is the luminosity of the star.}
    \label{fig:timescales}
\end{figure}

Our findings have a number of implications for CEE. We find that the orbital evolution of the central binary does not stall even when the gas in the immediate vicinity of the binary corotates. Instead of the commonly assumed drag, the binary transfers angular momentum to the envelope by generating spiral waves and turbulence. We find that the associated timescale of orbital contraction is $\tau_\text{b} = |a_\text{b}/\dot{a}_\text{b}| \sim 10^3$ to $10^4$ orbital periods of the binary when accretion is allowed or  slowly decreases to $\tau_\text{b} \sim 10^5 P_\text{orb}$ at $t \simeq 450\,P_\text{orb}$ when accretion is prevented. These timescales are similar to what is typically found in CBDs \citep[e.g.][]{Artymowicz1991}. We emphasize that this timescale refers to the inner binary orbit, which is much smaller than the outer extent of the envelope. Our results suggest that while there is gas in the shared envelope, the binary should continue to spiral in due to nonlocal interactions with the nearby gas, albeit much slower than in the preceding dynamical plunge-in phase. Many ab initio works on CEE find that the central binary orbits continue to slowly shrink at the end of the simulations. Based on our results, we suggest that the orbital contraction rate should not gradually approach zero but always remains at a rather small but finite value. We also suggest that achieving complete envelope ejection and final orbital separations compatible with expectations could be possible simply by following the evolution for much longer time than what is currently done.

No thermal energy transfer through the envelope is required to reduce the orbital separation. However, at some point in time, the energy diffusion timescale through the envelope should become comparable to $\tau_\text{b}$. When that happens, it is possible that thermal coupling between the binary and the envelope is established, which might affect the orbital decay. For example, efficient removal of energy deposited in the vicinity of the binary could increase the envelope density around the binary, which would lead to higher torques and smaller $\tau_\text{b}$. In order for a thermal ``self-regulating'' process to have a chance to accelerate the orbital decay, the thermal timescale of the envelope of the primary star, $t_\text{KH}$, has to be shorter, or of the order of $\tau_\text{b}$.  

In order to evaluate $t_\text{KH}/\tau_\text{b}$ for different stars, we need to make three approximations. First, we make use of the fact that $\tau_\text{b}/P_\text{orb} \sim 10^3$ to $10^5$ and we study instead quantity $t_\text{KH}/P_\text{orb}$. Second, $P_\text{orb}$ is evaluated assuming that parameters of our simulation described in Sect.~\ref{sec:model}, $a_\text{b}/R = 0.16$ and $M_1/M = 0.2$, are applied uniformly to all progenitor primary stars. Third, stellar quantities $R$, $M$, and $L$ represent the values of the stellar model before the binary interaction. In Fig.~\ref{fig:timescales}, we show the ratio $t_\text{KH}/P_\text{orb}$ evaluated along evolutionary tracks of four single stars with masses $1$, $5$, $12$, and $20\,\msun$. We see that for low-mass red giants without fully developed convective envelopes, $R \lesssim 100\,\rsun$, the ratio is $t_\text{KH}/P_\text{orb} \gtrsim 10^5$, which implies that the envelope would not have enough time to thermally couple to the inspiralled binary before its orbit significantly decays. For these stars, the time window for any additional processes acting to remove the envelope on long timescales might be severely restricted, because the lifetime of the envelope is set by the fast orbital decay timescale rather than the thermal timescale. An example of such possible long-lasting processes are strong pulsations or dust-driven wind \citep{Clayton2017,Glanz2018}. For AGB and high-mass stars, the thermal coupling between the inspiralled binary and the envelope seems more likely. However, unless $t_\text{KH}/P_\text{orb} \lesssim 10^3$, which occurs only near the maximum expansion of the stars, the orbital decay could still remain unaffected by the thermal processes if the binary is able to efficiently accrete from the envelope due to a ``pressure valve'' such as launching of jets \citep{Soker1994,Chamandy2018,Shiber2019}.

Naturally, our simple estimates in Fig.~\ref{fig:timescales} have a number of caveats. For example, the binary might dynamically plunge-in to much lower values of $a_\text{b}/R$, as was seen in several recent simulations \citep{Ohlmann2016,Lau2022a}, or the binary might relatively quickly shrink its orbit early in the post-dynamical phase, as Fig.~\ref{fig:mmcrit} suggests. Furthermore, CEE is often preceded by strong thermal-timescale mass transfer, which leads to a significant decrease of primary's luminosity due to thermal restructuring of the envelope. The luminosity could decrease by a up to a factor of $10$, which would enlarge $t_\text{KH}$ \citep[e.g.,][]{Blagorodnova2021}. All of these effects would tend to increase $t_\text{KH}/P_\text{orb}$ and make thermal timescale influences less likely. Conversely, as the envelope expands, $t_\text{KH}$ and the diffusion timescale decrease. Since there are number of effects working in opposite directions, the overall importance of thermal effects on the inspiral is not immediately obvious. This long-term evolution cannot be easily studied by direct multidimensional simulations, but some insight can be obtained with 1D models with prescriptions calibrated to include physical processes studied in this work. In the future, we aim to enlarge our grid of parameters to make such parameterization possible.

In the light of our results, it is also interesting to discuss the energy-conserving $\alpha_\text{CEE}$ formalism that is commonly used to predict CEE outcomes. Our results suggest that much of the orbital decay during the post-dynamical phase might be over before thermal effects become important, which lends support to the energy-conserving formalism. At the same time, the orbital decay is clearly separated into two regimes: a fast dynamical plunge-in and much slower post-dynamical inspiral. Even if energy is conserved in both of them, the value of the $\alpha_\text{CEE}$ parameter might be different, because both types of orbital decay depend differently on binary properties and envelope structure. Furthermore, the post-dynamical inspiral depends on the efficiency of accretion, which could be influenced, among other effects, by jets. All of this could lead to different effective values of $\alpha_\text{CEE}$ for different populations of binary stars, which is not surprising, but perhaps also to a spread of $\alpha_\text{CEE}$ among a single population. In any case, our results generally motivate the development of two-step or multistep CEE formalisms \citep[e.g.,][]{Hirai2022}.

Although one of our original motivations for this work was to see whether the binary could reaccrete some of its angular momentum and expand its orbit, our results suggest that this is unlikely. The shared envelope is very thick and accretion near the polar regions brings in gas with very low specific angular momentum. The situation could change at later phases when the remaining envelope is able to cool to a thinner disk. Investigating this transition should be a subject of future study.

Finally, in our work we have made the assumption that the binary orbital motion has completely circularized after the dynamical plunge-in. This is in agreement  with 3D hydrodynamical simulations, which typically find quasi-circular orbits at the end of this phase provided that the initial eccentricity is low, \citep[e.g.,][]{Ricker2012,Passy2012,Ohlmann2016,Glanz2021}. However, this does not necessarily imply that the orbit remains circular throughout the post-dynamical inspiral phase. A variety of physical processes can lead to the growth or decrease of binary eccentricity, such as accretion streams impact on binary components or the gravitational interaction between the central binary and its nonaxisymmetric eccentric envelope such as the one we find in Sect.~\ref{sec:ecc}. While such phenomena can potentially lead to the binary eccentricity growth, binary eccentricity may generate new resonances that can in turn damp eccentricity \citep[e.g.,][]{Lubow1991a,Lubow1991b}. Orbital eccentricity may therefore be nonzero during the post-dynamical spiral-in phase, and stabilize or oscillate about a fixed value, similarly to what is seen in CBDs \citep[e.g.,][]{Roedig2011,Zrake2021}, and could help to explain nonzero eccentricities seen in some post-AGB and post-CEE binaries \citep[e.g.,][]{Dermine2013,Kruckow2021}.

\section{Conclusions}
\label{sec:conclusions}

In this work, we performed a series of 3D hydrodynamic numerical simulations of the post-dynamical inspiral phase of CEE. We used the procedure of \cite{Morris2006,Morris2007,Morris2009} to mimic the outcome of the preceding dynamical plunge-in and to establish controlled initial conditions for our simulations (Fig.~\ref{fig:snap}). Our first aim was to determine the timescale of binary separation evolution  in response to the various torques acting on the system when accretion is turned on or off. We have computed the various torques acting on the binary and we found that they always result in the contraction of the orbit, regardless of whether accretion is allowed or not (Fig.~\ref{fig:mmcrit}, Sect.~\ref{sec:bin_evol}). When accretion is allowed, mass and angular momentum accretion drive the orbital contraction and the orbital contraction timescale rapidly reaches a quasi-steady value of $\mathcal{O}(10^3-10^{4}~P_{\rm orb})$. Without accretion, orbital contraction is solely driven by the gravitational torque. Because of the envelope expansion, the amplitude of the gravitational torque slowly decreases, leading to a slow increase of the orbital contraction timescale. After 450~$P_{\rm orb}$, this timescale reaches a value of $\mathcal{O}(10^5~P_{\rm orb})$. 

Our results imply that while the binary is embedded in gas, the orbit contracts even if the gas immediately surrounding the binary is corotating. The orbital decay timescale is much slower than what is seen during the dynamical plunge-in. This suggests a significant reduction of orbital separation and more efficient envelope ejection is possible even after the dynamical plunge-in and that current simulations have not been carried out over a sufficiently long period to observe this effect. Since the orbital separation is very small compared to the outer extent of the envelope, the post-dynamical inspiral timescale is much shorter than the thermal timescale for primary stars with radius $\lesssim 100\,\rsun$. Even for larger stars, the post-dynamical decay does not have to be significantly influenced by the thermal response of the envelope and thermal ``self-regulation'' does not seem to be unavoidable, but this is contingent on the dynamics very close to the binary such as the presence or absence of accretion or jets. The short inspiral timescales lend support to adiabatic treatment of CEE, but motivate viewing CEE as an (at least) two-step process.

Our second aim was to find the typical frequencies associated with the short-term variability of mass accretion onto the binary and to compare the results to CBDs. We found that the main features of mass accretion variability in the context of post-dynamical CEE are similar to that of CBDs. Specifically, the variability is connected to the forcing angular frequency of the quadrupolar moment contribution to the binary potential $\omega_\text{b} = 2~\Omega_{\rm orb}$ and to the frequency associated with the formation of nonaxisymmetric overdensities in the inner part of the envelope $\omega_\rho = \Omega_{\rm orb}/5$. Such overdensities result from accreting material being flung back into the envelope, and accumulating at a distance of roughly six binary separations from the binary center of mass (Figs.~\ref{fig:spacetimeB}--\ref{fig:A1B}, Sect.~\ref{sec:tdepacc}). We found that the resulting overdense and eccentric ``lump'' then propagates far into the envelope, contrary to the case of CBDs, and feeds and enhances mass accretion (Fig.~\ref{fig:Sigma} and Sect.~\ref{sec:lump}). Because of the spherical shape of the CEE problem contrasting with flat CBDs, such frequencies do not necessarily characterize the global mass accretion rate. Instead, the presence or absence of latitudinally  migrating accretion streams leads to the synchronicity or asynchronocity of the mass flux at all colatitudes. When asynchronous, time variability of the latitudinal integrated mass flux may be smoothed out. A local analysis of the accretion flux is therefore necessary to understand its short-term variability. Finally, we found that envelope eccentricity is excited in the vicinity of the binary and propagates outward within and in-between successive lumps. During this process, the eccentricity amplifies and builds up in the envelope, which leads to the splitting of the accretion frequencies (Fig.~\ref{fig:A1B} and Sect.~\ref{sec:ecc}). The envelope mean eccentricity grows exponentially with a growth rate $\lambda_e \simeq 0.022~\Omega_{\rm orb}$ until it reaches a statistically stationary value of $e_{{\rm env},f} \simeq 0.12$ independent of the initial angular momentum of the envelope and of the presence or absence of viscosity (Fig.~\ref{fig:eccev}). Growth of eccentricity in the envelope could in turn excite eccentricity in the inner binary.

Our third aim was to understand how angular momentum is transported within the envelope (Sect.~\ref{sec:AMtrans}).  We showed that, similarly to previous dynamical plunge-in simulations, gravitational perturbations from the orbiting binary during post-dynamical in-spiral phase trigger the destabilization of the envelope. This destabilization results in turbulent convection contributing to the transport of energy and angular momentum throughout the shared envelope (Fig.~\ref{fig:solb1} and Sect.~\ref{sec:solberg}). However, we showed that this contribution is rather small. Instead, the angular momentum flux is dominated by large scale axisymmetric fluid flows, which consist of an outward transport in a geometrically thick disk-like structure about the orbital plane, except in the close vicinity of the central binary where the flux points inward. 
However, because the net (latitudinally integrated) radial angular momentum transport by the mean flow at a given radius is rather weak, turbulent transport measured by Reynolds stresses in fact is also important and can have an effect of the same order as that of the mean flow on the global transport of angular momentum. In particular, we showed that Reynolds stresses can locally strongly damp or enhance the outward transport of angular momentum by the mean flow (Figs.~\ref{fig:torqprofAB}--\ref{fig:turbprofAB}). 

Our final aim was to characterize the role of viscosity originating both from turbulent motions and from unspecified processes acting on subgrid scales. We showed that because the stabilizing effect of density stratification decreases outward, vertical convective eddy scales increase with radial distance from the central binary. We showed that envelopes with higher initial angular momentum content are more strongly stratified and thus contain smaller convective cells, however, their typical size is limited to a fraction of the domain radius in the limit of low stratification, suggesting that $\alpha$-type viscosity models would fail in outer layers (Figs.~\ref{fig:lamba_l}--\ref{fig:Fvisc}). We further found that prescribing a background kinematic viscosity with $\alpha_\nu = 10^{-3}$ does not significantly affect the binary separation evolution, nor the transport of angular momentum within the shared envelope (Sect.~\ref{sec:visc}).

Our new way of studying late stages of CEE has its limitations. For example, it is important to keep in mind that we have excised a central region encompassing the binary and that we only considered two extreme regimes of accretion (maximum or none). We have further considered fixed binary orbit, assuming that the contraction or expansion timescale is much longer than the duration of our simulation. Our estimates of the orbital contraction timescale suggest that such assumption might not be completely valid when accretion is allowed. This limitation cannot be easily lifted: if we allowed the orbit to shrink or expand we would have to change the position of the inner boundary with time in order to not discard the flow dynamics in the vicinity of the binary or to prevent the binary from entering the numerical domain. This would affect  the conservations of  mass and angular momentum and the numerical cost of our simulations. Another solution would be to use Cartesian grid, but in such a case we would  lose the advantages of spherical geometry.
We have further assumed that binary eccentricity remains zero throughout our simulations. However, a variety of physical processes may lead to the binary eccentricity growth or decrease, although they may balance each other out. Finally, we do not include gas self gravity which could affect the binary-envelope interaction.

More sophisticated initial parameters could complicate the our results and need to be explored in future works. We plan to investigate eccentric binary orbits, binaries with mass ratios different from unity, more sophisticated inner boundary conditions, or the effect of magnetic fields. Changing these parameter could dramatically impact the binary-envelope interaction, resulting in very different variability, amplitude, and angular distribution of  mass and angular momentum accretion onto the binary, binary separation evolution, and angular momentum transport within the envelope.

\begin{acknowledgements}
We thank the anonymous referee for comments that improved this paper. We thank Kengo Tomida for discussions about Athena++. The research of DG and OP has been supported by Horizon 2020 ERC Starting Grant `Cat-In-hAT' (grant agreement no. 803158).
This work was supported by the Ministry of Education, Youth and Sports of the Czech Republic through the e-INFRA CZ (ID:90140). OP thanks the KITP program ``Bridging the Gap: Accretion and Orbital Evolution in Stellar and Black Hole Binaries'' for hospitality and inspiration: this research was supported in part by the National Science Foundation under Grant No. NSF PHY-1748958.
\end{acknowledgements}

\bibliographystyle{aa}
\bibliography{bibnew}

\onecolumn
\begin{appendix}
\section{Angular momentum conservation}\label{App:dotJz}

We take the cross product of $\br$ with the momentum Eq.~(\ref{eq:eq}), multiply by $\be_z$ and integrate it over the domain's volume, 
\begin{equation}\label{eq:sphi}
    \int \frac{\partial \rho s u_\varphi}{\partial t} \dd V + \int \bnabla \cdot (\rho s u_\varphi \buu) \dd V = - \int \rho \frac{\partial \Phi}{\partial \varphi} \dd V  - \int  \be_z \cdot \left(\br \times \bnabla \cdot \boldsymbol{T} \right) \dd V \ .
\end{equation}
Using the fact the viscous stress tensor $\boldsymbol{T}$ is symmetric, and writing $\br = x_j \be_j$,
\begin{equation}
        \br \times \bnabla \cdot \boldsymbol{T} =  \be_i \epsilon_{ijk} x_j \partial_l T_{kl} 
          = \be_i  \epsilon_{ijk}\partial_l \left( x_j T_{kl} \right) - \be_i  \epsilon_{ijk}\delta_{lj}T_{kl} 
          = \bnabla \cdot \left(\br \times \boldsymbol{T} \right) \ ,
\end{equation}
where $\epsilon_{ijk}$ is the Levi-Civita tensor  of rank three, and 
\begin{equation}
   \be_z \cdot \bnabla \cdot \left( \boldsymbol{\br} \times \boldsymbol{T} \right) =   \be_z \cdot {\rm div} \left( \boldsymbol{\br} \times \boldsymbol{T} \right)^{T}
    =\bnabla \cdot \left[\left(   \boldsymbol{\br} \times \boldsymbol{T} \right) \cdot \be_z \right] - \Tr{\left[ \left(   \boldsymbol{\br} \times \boldsymbol{T} \right) \bnabla\be_z \right]}
   =\bnabla \cdot \left[\left(   \boldsymbol{\br} \times \boldsymbol{T} \right) \cdot \be_z \right] \ ,
\end{equation}
Eq.~(\ref{eq:sphi}) can be rewritten as
\begin{equation}\label{eq:sphi2}
        \int \frac{\partial \rho s u_\varphi}{\partial t} \dd V + \int \bnabla \cdot (\rho s u_\varphi \buu) \dd V = - \int \rho \frac{\partial \Phi}{\partial \varphi} \dd V  - \int_{\partial R} \bnabla \cdot \left[\left(   \boldsymbol{\br} \times \boldsymbol{T} \right) \cdot \be_z \right]  \dd V \ .
\end{equation}
Finally, using Gauss divergence theorem, Eq.~(\ref{eq:sphi2}) can be rewritten as
\begin{equation}\label{eq:Jz}
\dot{J}_z  = - \int_{\partial R} \rho s u_\varphi\buu \cdot \boldsymbol{n}_\perp \dd S   - \int \rho \frac{\partial \Phi}{\partial \varphi} \dd V - \int_{\partial R} \left[\left(   \boldsymbol{\br} \times \boldsymbol{T} \right) \cdot \be_z \right] \cdot \boldsymbol{n}_\perp \dd S 
 = \dot{J}_{z,\rm adv} + \dot{J}_{z, \rm grav} +  \dot{J}_{z,\rm visc}   \ ,
\end{equation}
where $\boldsymbol{n}_\perp$ is is the outward-pointing unit vector at the boundaries' surface, $\dot{J}_z$ is the time derivative of the $z-$component of the gas angular momentum, and $\dot{J}_{z,\rm adv}$, $T_{z, \rm grav}$,  and $\dot{J}_{z,\rm visc}$  are the advective, gravitational, and viscous torques on the system, respectively. Replacing the viscous stress tensor by its expression finally yields,
\begin{equation}
    \dot{J}_{z, \rm visc} = - \int_{\partial R_{\rm out}} r_{\rm out} \sin \theta T_{r \varphi} \dd S  + \int_{\partial R_{\rm in}} r_{\rm in} \sin \theta T_{r \varphi} \dd S \ ,
\end{equation}
where 
\begin{equation}
         T_{r\varphi} = T_{\varphi r} =-  \rho \nu \left( \frac{1}{s} \frac{ \partial u_r}{\partial \varphi} + r \frac{ \partial }{\partial r} \frac{u_\varphi}{r}\right)  \ .
 \end{equation}

 Similarly, local torques are obtained by taking the cross product of $\br$ with the momentum Eq.~(\ref{eq:eq}), multiplying it by $\be_z$, taking the radial derivative, integrating the result over the volume comprised between $r$ and $r_{\rm out}$, and using the fact that the gravitational torque vanishes at the outer boundary, that is $\partial \Phi / \partial \varphi$ tends to zero far from the central binary \citep[see e.g.,][]{Miranda2017}.  
 
\section{Effect of boundary conditions on the gravitational torque}\label{App:dotJzgrav}

\begin{figure}[t]
\centering
\includegraphics[width=0.33\textwidth]{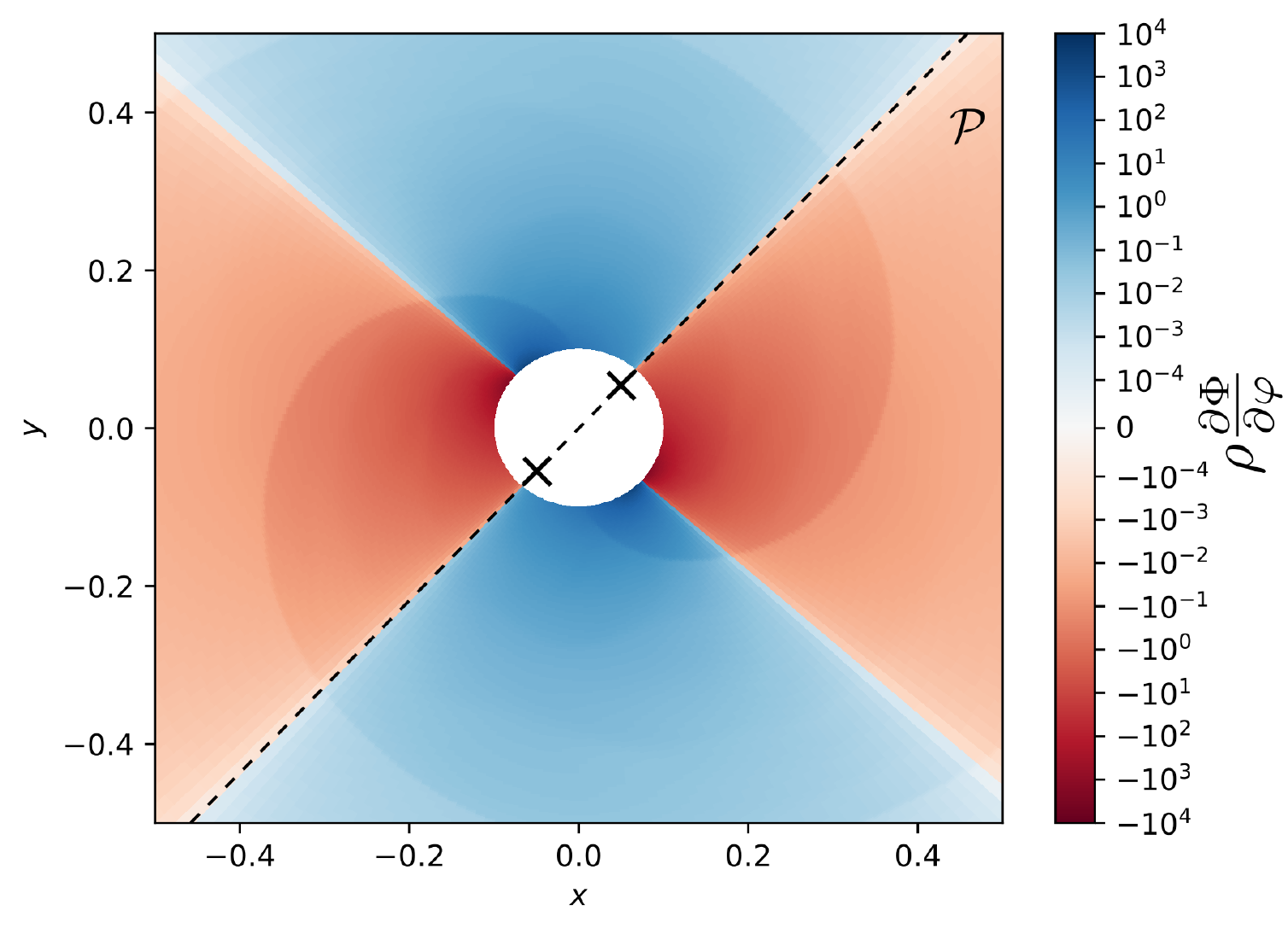} 
\includegraphics[width=0.33\textwidth]{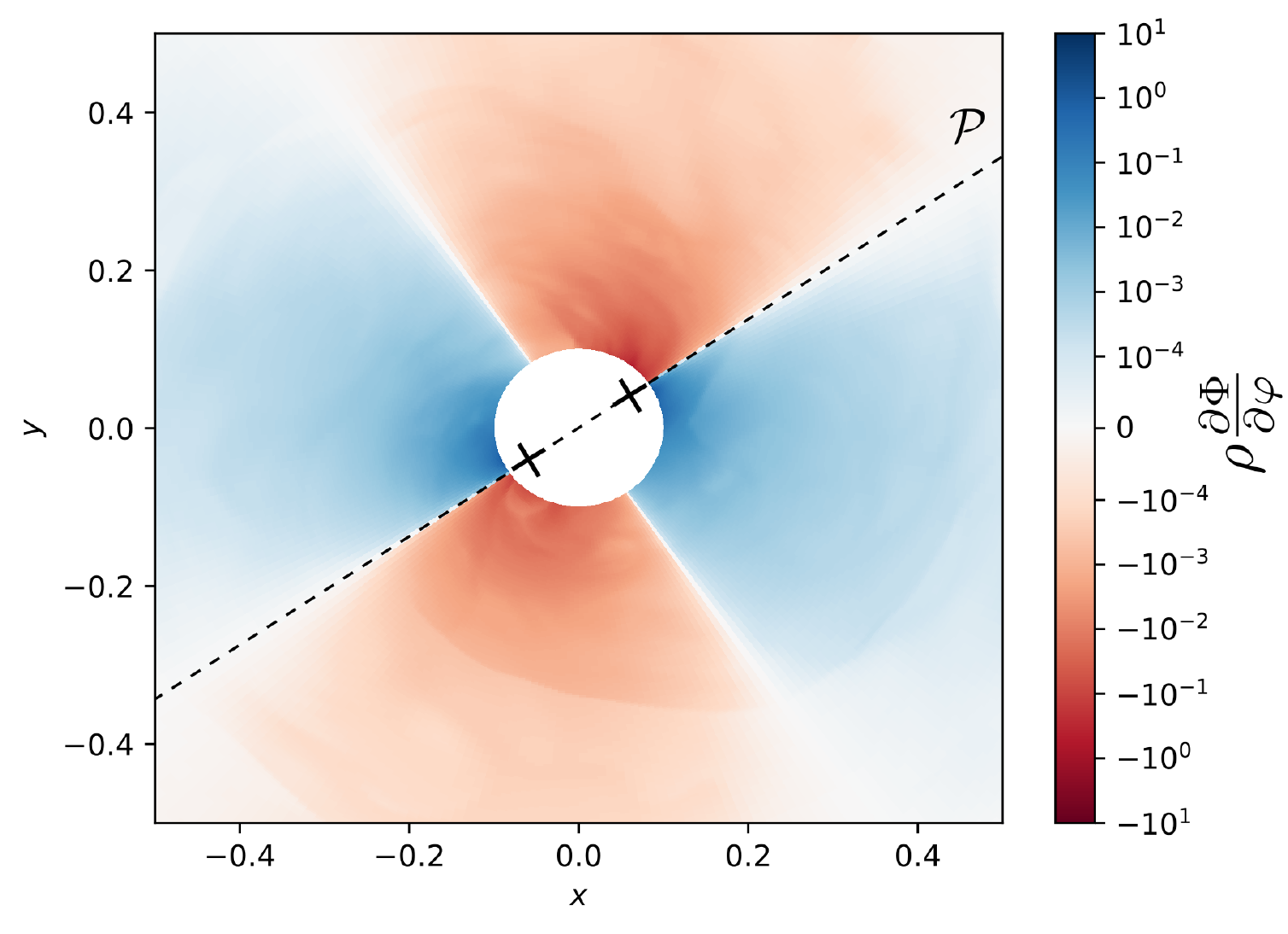} 
\includegraphics[width=0.33\textwidth]{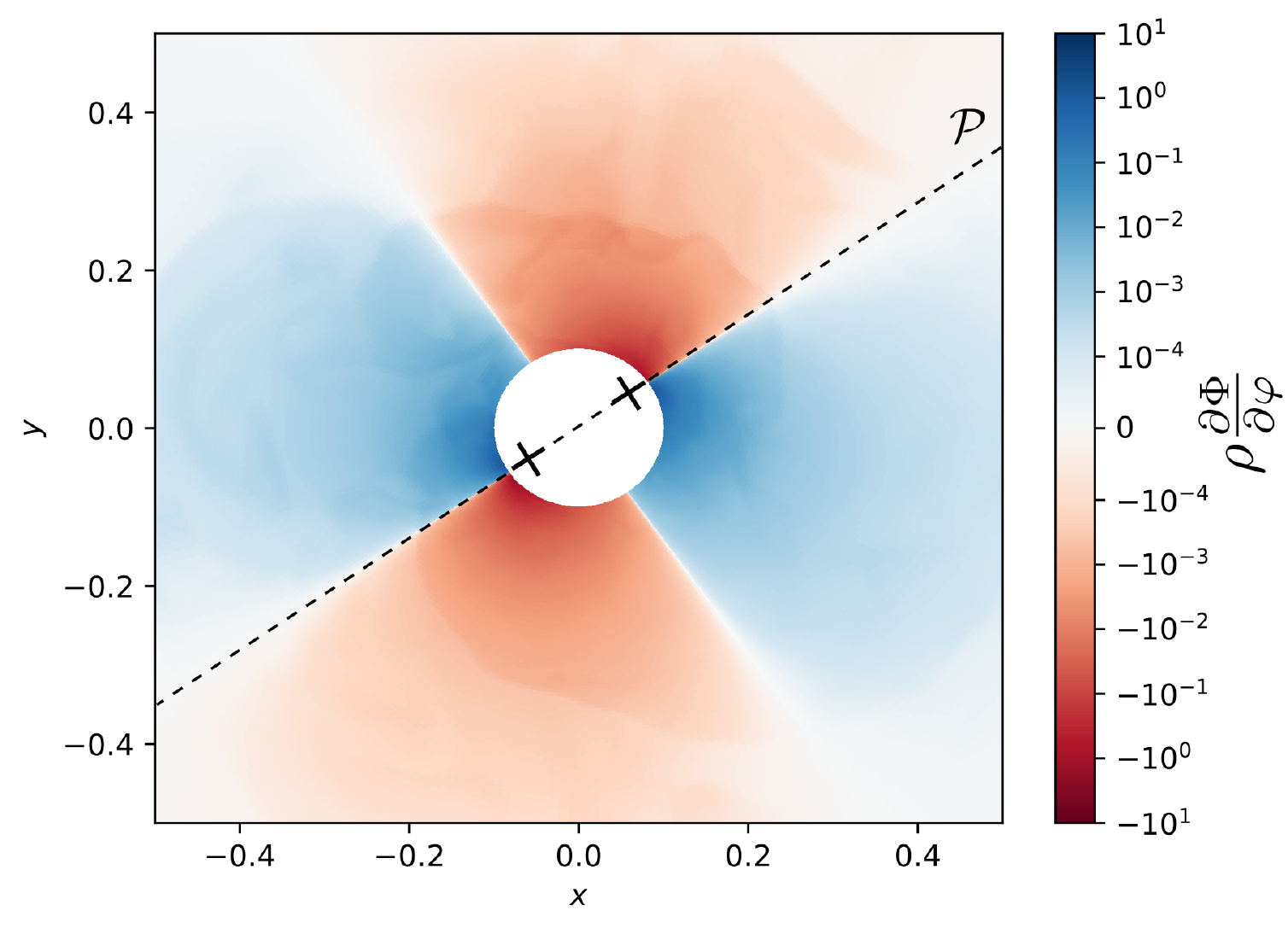} 
   \caption{Specific gravitational torque for simulation run A at $t \simeq 25~P_{\rm orb}$ (left) and $t \simeq 250~P_{\rm orb}$ (middle), and for run A' at $t \simeq 250~P_{\rm orb}$ (right). Black crosses indicate the location of the two cores, and the dashed black line indicates the intersection of $\mathcal{P}$ with the orbital plane.}
\label{fig:B1}
\end{figure}

Fig.~\ref{fig:mmcrit}(b) shows that the amplitude of the gravitational torque largely depends on the enforced inner boundary conditions. In particular, while $\dot{J}_{z,\rm grav}$ rapidly settles to statistically (almost) zero when the inner boundary is open to mass and angular momentum flow toward the binary, $\dot{J}_{z,\rm grav}$ decreases much slower and thus drives the orbital contraction of the binary when accretion is turned off.  To understand this fundamental difference, let us first investigate the conditions for $\dot{J}_{z,\rm grav}=0$. For the gravitational torque to be zero, mass distribution must satisfy the symmetry property of $|\partial \Phi / \partial \varphi|$ with regard to the plane $\mathcal{P}$ defined as
\begin{equation}\label{eq:symplane}
    \boldsymbol{e}_s \cdot (\br_1 - \br_2) = 0 \ ,
\end{equation}
that is the plane orthogonal to the orbital plane and intersecting the binary semi-major axis. In our simulations, the gravitational torque arises from the quadrupolar moment of the binary potential generating a two-armed spiral density wave and breaking the mass distribution symmetry with respect to the plane $\mathcal{P}$ (see Fig.~\ref{fig:B1}, left). We decompose the amplitude of the gravitational torque as
\begin{equation}\label{eq:Jzg}
    |\dot{J}_{z, \rm grav}| = \epsilon \rho_{\rm eff} \int \left\vert \frac{\partial \Phi}{\partial \varphi} \right\vert \dd V \ ,
\end{equation}
where $\int |\partial \Phi / \partial \varphi| \dd V$ is constant in our simulations, 
\begin{equation}
    \epsilon = \frac{|\int \rho \frac{\partial \Phi}{\partial \varphi} \dd V|}{\int \rho |\frac{\partial \Phi}{\partial \varphi}| \dd V} \ ,  \quad {\rm and} \quad  \rho_{\rm eff} = \frac{\int \rho |\frac{\partial \Phi}{\partial \varphi}| \dd V}{\int |\frac{\partial \Phi}{\partial \varphi}| \dd V} \ .
\end{equation}
Here, $0 \le \epsilon \le 1$  quantifies the asymmetry of the mass distribution with respect to $\mathcal{P}$, and $\rho_{\rm eff}$ is the weighted-averaged density with a dominating contribution from the gas subject to larger $|\partial \Phi / \partial \varphi|$. 

\begin{figure}[t]
\centering
\includegraphics[width=0.75\textwidth]{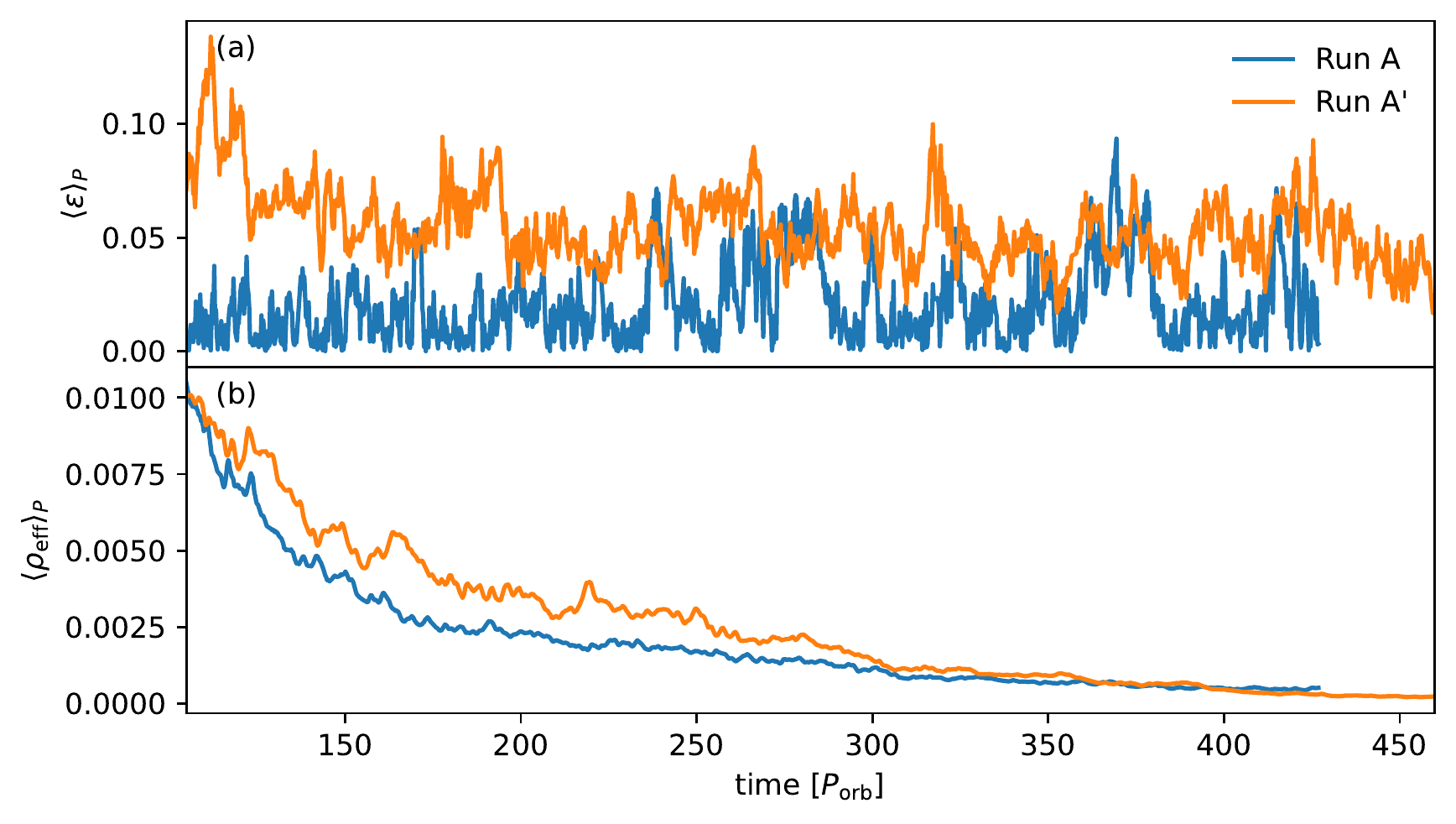} 
   \caption{(a)$\langle \epsilon \rangle_P = \int_t^{t+P_{\rm orb}}  \epsilon dt'$ quantifying the asymmetry of the mass distribution with respect to the plane $\mathcal{P}$ as a function of time. (b) Weighted-averaged density with a dominating contribution from the gas subject to larger $|\partial \Phi / \partial \varphi|$, $\langle \rho_{\rm eff} \rangle_P = \int_t^{t+P_{\rm orb}} \rho_{\rm eff} dt'$, as a function of time.}
\label{fig:B2}
\end{figure}

Fig.~\ref{fig:B2} shows the time evolution of $\langle \epsilon \rangle_P$ and $\langle \rho_{\rm eff} \rangle_P$ for models A and A', where $\langle \cdot \rangle_P$ indicates a temporal smoothing over one orbital period. We see that closing the inner boundary to mass and angular momentum flow toward the binary typically leads to larger $\epsilon$ and $\rho_{\rm eff}$. Turbulence-induced alteration of the asymmetric spiral density wave structure likely explains the difference in $\epsilon$ between the two models, simulation run A being more turbulent in the vicinity of the binary (see Fig.~\ref{fig:B1}). Reflecting boundary conditions indeed lead to the accumulation of material at the inner boundary leading to both larger $\rho_{\rm eff}$ and increased stabilizing density stratification. However, because the energy transferred from the binary orbit to the envelope as kinetic and internal energy is higher for higher $\rho_{\rm eff}$, the envelope expands more rapidly in simulations run A'. This  leads to a more rapid decrease of $\rho_{\rm eff}$ which in turn leads to the reduction of the stabilizing effect of density stratification and thus to a slow decrease of $\epsilon$. Inevitably, the gravitational torque in simulation run A' eventually also reaches statistically (almost) zero (see Fig.~\ref{fig:mmcrit}(b)). Because the gravitational torque is exclusively responsible for the orbital contraction when the binary does not accrete, $|a/\dot{a}|$ does not reach a steady value in simulation run A', and keeps increasing.

\section{Migrating accretion streams}\label{App:tilt}

\begin{figure}[t]
\centering
\includegraphics[width=\textwidth]{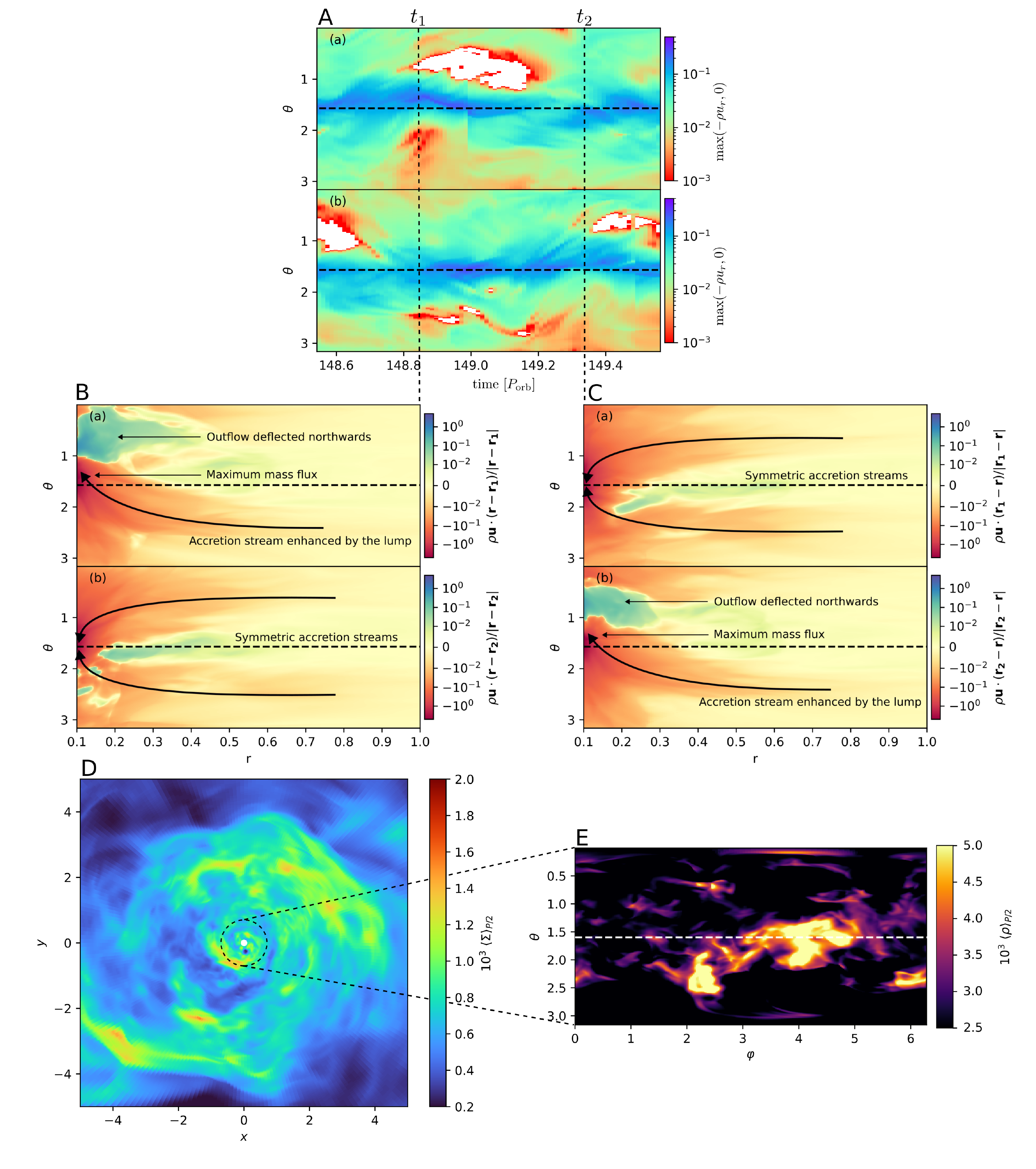} 
   \caption{A: Space-time diagram of the mass flux onto the individual components of the binary for model A, azimuthally averaged in the range $\varphi_1 - \pi/12 \leq \varphi_1 \leq \varphi_1 + \pi/12$ (a) and in the range $\varphi_2 - \pi/12 \leq \varphi_2 \leq \varphi_2 + \pi/12$ (b). The horizontal dashed lines indicate the orbital plane $\theta = \pi/2$. The two vertical dashed lines are separated by $P_\text{orb}/2$ and indicate two successive times, $t_1$ and $t_2$, when the mass flux is maximum at the highest latitude. B: Mass flux toward and away from the individual cores, azimuthally averaged in the range $\varphi_1 - \pi/12 \leq \varphi_1 \leq \varphi_1 + \pi/12$ (a) and in the range $\varphi_2 - \pi/12 \leq \varphi_2 \leq \varphi_2 + \pi/12$ (b) as a function of colatitude and radius at $t = t_1$. C: same as B but at $t = t_2$. D: Surface density about the orbital plane (\ref{eq:sigma}) averaged over $t_1 \leq t \leq t_2$. The dashed circle indicates the approximate radial location of the overdensity responsible for the migrating accretion streams ($r = 0.7$). E: Density cross section in the $\theta\varphi$ plane at $r = 0.7$ averaged over $t_1 \leq t \leq t_2$.}
\label{fig:C}
\end{figure}

In Sect.~\ref{sec:tdepacc} and Fig.~\ref{fig:spacetimeA}, we discuss that for $t\lesssim 156\ P_\text{orb}$, mass accretion onto the binary exhibits inclined and periodic stripes spanning a wide range of latitudes. To understand the cause of this latitudinally migrating accretion, we show a space-time diagram of the mass flux onto the individual components of the binary for model A during this time period in Fig.~\ref{fig:C} (A). We see that the mass flux peaks at a maximum latitude successively on core 1 (at $t=t_1$) and core 2 (at $t=t_2$), with a period of $P_\text{orb}/2 = t_2 - t_1$, that is, when the two cores switch position. This suggests that the source of the high-latitude accretion orbits much slower than the binary, that is $\Omega \ll \Omega_\text{orb}$. We show the  mass flux toward and away from the individual cores, azimuthally averaged within a $\pi/6$ opening angle about the position of the cores at $t=t_1$ and $t=t_2$, respectively in  Figs.~\ref{fig:C} (B) and (C). We see that high-latitude accretion through the inner boundary results from accretion streams originating from low-latitude regions. Such streams initially propagate radially, then toward the orbital plane when approaching the binary. Accretion streams approaching the orbital plane carry a certain amount of momentum. Their inertia allows them to travel a certain distance above the orbital plane before the material is pulled back toward it by the gravitational pull of the core. In addition, in the vicinity of the binary, the enhanced accretion streams deflect the outflow associated with spiral density waves to the north.

We show the surface density about the orbital plane (\ref{eq:sigma}) averaged over $t_1 \leq t \leq t_2$ in Fig.~\ref{fig:C} (D). We see that the overdensity located around $r \simeq 0.7$ is characterized by an angular frequency $\Omega_\text{lump} \ll \Omega_\text{orb}$ and is therefore a candidate for the source of high-latitude accretion. Finally, we show the density cross section in the $\theta\varphi$ plane at $r = 0.7$ averaged over $t_1 \leq t \leq t_2$ in Fig.~\ref{fig:C} (E). We find that this overdensity is spatially extended over a wide range of $\theta$. Its presence and its shape explains the migrating accretion streams shown in Fig.~\ref{fig:spacetimeA} as follows. As individual orbiting cores pass in front of the overdensity, they first pull its material from high colatitudes, enhancing accretion streams (Figs.~B(a) and C(b)). Due to their high inertia, such accretion streams reach the inner boundary in the northern hemisphere. The material impacting the inner boundary is partially accreted by the binary, the rest is pulled back toward the orbital plane. As the core now faces the part of the overdensity that is located close to the orbital plane, mass accretion becomes symmetric about the orbital plane. Finally, the core eventually approaches the northern hemisphere end of the overdensity, analogously the accretion streams pulled from this part of the lump reach the inner boundary in the southern hemisphere. As one core orbits away from the lump and becomes subject to symmetric mass accretion (Figs.~B(b) and C(a)), the companion approaches it, and the phenomenon is repeated.

\end{appendix}
\end{document}